\documentclass[12pt,reqno,amssymb]{article}
\usepackage{graphicx}
\usepackage{url}
\usepackage{multirow}
\usepackage{color}
\usepackage{amsmath}
\usepackage{subcaption}
\usepackage{pgfplots}

\usepackage{tikz}
\usetikzlibrary{shapes.geometric,arrows}
\usepackage{braket}
\usepackage{adjustbox} 
\usepackage{booktabs}

\usepackage{geometry}
\usepackage{array}
\usepackage{multirow}
\usepackage{amsfonts}
\usepackage{qcircuit}
\usepackage{hyperref}      

\usepackage[linesnumbered,ruled,vlined]{algorithm2e}

\usepackage{tikz}
\usetikzlibrary{shapes.geometric, arrows, positioning}
\usetikzlibrary{positioning,calc,shapes.geometric,arrows.meta}
\usepackage{qcircuit}
\usepackage{comment}
\usepackage{enumitem}

\usepackage{array,booktabs,tabularx,makecell} 

\usepackage{tikz}
\usetikzlibrary{shapes.geometric, arrows}
\usepackage{qcircuit}

\newcommand*{\E}{\mathbb{E}}

\includecomment{sectiontocompile}

\usepackage{pgfplots}
\pgfplotsset{compat=newest}
\usepgfplotslibrary{groupplots}
\usepgfplotslibrary{dateplot}

\numberwithin{equation}{section}

\begin{document}

\title{Genetic Transformer-Assisted Quantum Neural Networks for Optimal Circuit Design}

\author{Haiyan Wang\\
School of Mathematical and Natural Science\\
Arizona State University, Phoenix, AZ 85069, USA\\
haiyan.wang@asu.edu
}

\maketitle

\begin{abstract}
We introduce \emph{Genetic Transformer–Assisted Quantum Neural Networks}
(GTQNNs), a hybrid learning framework that combines a transformer encoder with
a shallow variational quantum circuit and automatically fine-tunes the circuit
via the NSGA-II multi-objective genetic algorithm.  The transformer reduces
high-dimensional classical data to a compact, qubit-sized representation,
while NSGA-II searches for Pareto-optimal circuits that \textbf{(i)} maximize
classification accuracy and \textbf{(ii)} minimize primitive-gate count—an
essential constraint for noisy intermediate-scale quantum (NISQ) hardware.
Experiments on four benchmarks (Iris, Breast-Cancer, MNIST, and
Heart-Disease) show that GTQNNs match or exceed state-of-the-art quantum
models while requiring much fewer gates for most cases.  A hybrid Fisher
information analysis further reveals that the trained networks operate far
from barren plateaus; the leading curvature directions increasingly align with
the quantum subspace as the qubit budget grows, confirming that the
transformer front-end has effectively condensed the data.  Together, these
results demonstrate that GTQNNs deliver competitive performance with a quantum
resource budget well suited to present-day NISQ devices.
\end{abstract}



\section{Introduction}\label{sec1}
Quantum machine learning (QML) has grown into a vibrant research area, promising to extend classical machine-learning techniques by exploiting exponentially large Hilbert spaces over the past decade ~\cite{Schuld, abbas2021power,Farhi2018,pointing2024simplicity,Sagingalieva2023,chen2024,maccormack2022branching}.  Quantum computers could therefore process high-dimensional data more efficiently than their classical counterparts~\cite{NielsonChuang,qml1,qml2,qml16,qml15}.  Early efforts centred on extending familiar algorithms to the quantum computing, for example, quantum neural networks (QNNs)~\cite{qml6,qml9,qml11,qml12,qml13,qml14} and quantum support-vector machines~\cite{qml16,qml10}.  Yet those prototypes revealed serious obstacles to scalability and real-world deployment.

Consequently, attention has shifted toward algorithms tailored to today’s noisy intermediate-scale quantum (NISQ) hardware~\cite{Wan2017Sep,Beer2020Feb,Cong2019Dec,Sharma2022May,Zhou2023May}.  This effort has spawned a new generation of  QML techniques, explicitly designed to tolerate noise and limited qubit counts while still offering quantum-enhanced performance~\cite{Preskill2018Aug,Cerezo2022,Beer2020}. The number of qubits required in QNNs tends to increase linearly with the number of data features, quickly exceeding the limited qubit capacity of current quantum hardware. Consequently, this restricts their applicability to smaller datasets instead of utilizing the exponential scaling of Hilbert space dimension with the number of qubits \cite{alam2022deepqmlp,liu2024training,Singh2024}. This scaling is problematic on NISQ devices due to increased error rates and circuit depth, leading to a higher likelihood of decoherence and computational inefficiency. The combination of these scalability issues and gate complexity challenges significantly hinders the practical implementation of QNNs on existing quantum platforms, making the handling of complex, feature-rich datasets a formidable task and posing a significant bottleneck in fully leveraging quantum computing for advanced machine learning applications \cite{pointing2024simplicity,schuld2021kernel}.

Genetic algorithms have also proven highly adaptable in quantum-computing settings, where they are employed to search large, rugged design spaces \cite{Ji2020,Lahoz-Beltra2016,Acampora2021,ARG21}. In particular, multi-objective genetic approaches that automatically synthesise quantum circuits \cite{ARG21,Wang2025,Wang2025-1} help overcome common obstacles such as trapping in local minima and the barren-plateau phenomenon \cite{baran2021,Chivilikhin2020}.

Quantum computing offers the potential for exponential speed-ups in certain computational tasks, while transformer architectures have revolutionized natural language processing and computer vision through their ability to model long-range dependencies with self-attention \cite{Vaswani2017,Devlin2019,Dosovitskiy2021}. The integration of quantum computing and transformer-based models have made significant progress in recent years \cite{Cherrat2024,Ma2024,Zhang2025arxiv,Li2025quantumS}. Transformers are not only for long–range dependency modelling but also powerful \emph{learned compressors}: they convert a high-dimensional input stream into a compact set of information-rich feature vectors—exactly what a qubit-limited quantum back-end requires.

In this paper, we present \textbf{Genetic Transformer–Assisted Quantum Neural Networks (GTQNNs)},  
a hybrid architecture that couples a transformer encoder to a shallow variational
quantum circuit and optimizes the latter with the NSGA-II evolutionary algorithm.  Our new contributions in this paper include:

\begin{itemize}
\item \emph{Model design.}  
      The transformer compresses the high-dimensional input into an
      $n_{\text{qubits}}$-dimensional feature vector, which is then processed
      by a coherent QNN layer.

\item \emph{Multi-objective optimization.}  
      NSGA-II searches circuit space under two fitness objectives:  
      (i) maximise classification accuracy,  
      (ii) minimise the total number of primitive gates.  
      The second objective directly addresses NISQ hardware limitations.

\item \emph{Experimental validation.}  
      On four benchmarks—\textbf{Iris}, \textbf{Breast-Cancer Wisconsin},
      \textbf{MNIST}, and \textbf{Heart-Disease}—GTQNNs equal or surpass the
      best published QNN accuracies while requiring substantially fewer qubits
      and gates for most cases.

\item \emph{Fisher-spectrum analysis.}  
      A hybrid Fisher study shows the model operates far from a barren
      plateau; as $n_{\text{qubits}}$ increases, the leading eigen-directions
      concentrate in the QNN subspace, confirming that the transformer
      front-end has effectively reduced dimensionality.
\end{itemize}

These results demonstrate that GTQNNs deliver state-of-the-art performance
with a quantum-resource budget compatible with current NISQ devices.

The paper is organized as follows.  Section \ref{sec1} (Introduction) motivates the need for resource-efficient quantum machine-learning models and outlines our contributions.  Section \ref{sec2} details the \emph{Transformer-assisted Quantum Neural Network (TQNN)} architecture.  Section \ref{sec3} extends this idea into \emph{Genetic TQNNs (GTQNNs)}, coupling the model with an NSGA-II multi-objective genetic algorithm that co-optimizes classification accuracy and gate count.  Section \ref{exprement} (Experimental Results) reports performance on four benchmarks, highlighting accuracy gains and gate-depth savings versus state-of-the-art QNN baselines.  Section \ref{sec_inf_geom} (Fisher-Spectrum Analysis) examines the hybrid model’s trainability, showing via empirical Fisher eigen-spectra that GTQNNs avoid barren plateaus and shift curvature toward the quantum subspace as qubit budget grows.  Finally, Section \ref{discussion} (Conclusion and Discussion) summarises the findings, discusses current limitations and sketches future directions.

\section{Transformer-assisted quantum neural network (TQNN)}\label{sec2}

\subsection{Quantum neural networks}\label{sec:QNN_intro}

Quantum neural networks (QNNs) have become a focal point of current quantum-machine-learning (QML) research.  Although the field is still in its formative years, the prospect of genuine quantum advantage has attracted considerable attention~\cite{abbas2021power,Farhi2018,pointing2024simplicity,Sagingalieva2023,chen2024,maccormack2022branching}.  Much of that progress is fuelled by \emph{variational} techniques that couple a shallow quantum circuit to a classical optimizer, giving rise to a wide range of hybrid algorithms~\cite{Garcia2022}. A typical hybrid quantum neural network includes the three components: 
\begin{enumerate}[label=(\roman*)]
  \item \textbf{Data embedding.}  A \emph{feature–map} unitary
        \(U(\mathbf x)\) encodes a classical input
        \(\mathbf x\in\mathbb R^{N}\) into a quantum state,
        \(U(\mathbf x)\ket{0}^{\otimes n}\).
        Typical maps are tensor products of single–qubit phase
        rotations or collective entangling maps such as the
        ZZFeatureMap~\cite{Havlicek2019,Garcia2022}.
  \item \textbf{Variational processing.}  A depth–\(L\) parametrised
        circuit
        $$
          U_{\text{var}}(\boldsymbol\theta)
          =\prod_{\ell=1}^{L}\!
            \Bigl[
              \mathcal E_{\ell}\;
              \prod_{j=1}^{n} R_{y}^{(j)}(\theta_{\ell j})
            \Bigr]
        $$
        alternates trainable single–qubit rotations
        \(R_{y}(\theta)=e^{-i\theta\sigma_{y}/2}\)
        with fixed entangling layers \(\mathcal E_{\ell}\)
        (e.g.\ CNOT or \(\textsf{CZ}\) gates)~\cite{Sim2019,Zhang2023}.
        The weights \(\boldsymbol\theta\) are trainable parameters.
  \item \textbf{Measurement and loss.}
        Measuring an observable \(M\) (often
        \(Z^{\otimes m}\) or a shallow POVM) produces an expectation
        value \(f(\mathbf x;\boldsymbol\theta)=
        \bra{0}U^{\dagger}(\mathbf x,\boldsymbol\theta)\,M\,U(\mathbf x,\boldsymbol\theta)\ket{0}\).
        A classical cost—cross-entropy for classification or
        mean-squared error for regression—is formed from
        \(f\) and back-propagated to update \(\boldsymbol\theta\)
        with a classical optimiser such as \textsc{cobyla} or Adam.
\end{enumerate}

A number of quantum-neural-networks have been developed recently.  For example, \cite{alam2022deepqmlp} develops DeepQMLP, a scalable quantum-classical hybrid architecture patterned after conventional deep feed-forward networks. In this design, a sequence of shallow quantum-neural-network (QNN) blocks takes the place of the hidden layers in a multi-layer perceptron. Each QNN transforms its input into a fresh, trainable representation that is passed to the next block, building progressively richer features. Because every quantum block is shallow, the overall model suffers less from decoherence and gate errors, making it markedly more robust to the noise of current-generation quantum hardware.

Recent deep-learning models often rely on millions of trainable weights, making parameter-efficiency a central concern. A QNN with EfficientSU2 was introduced in \cite{liu2024training} for a training strategy that off-loads part of this burden to the exponentially large Hilbert space of a quantum processor. A classical network with $M$ parameters is re-encoded as a quantum neural network whose circuit contains only 
$O(polylog(M))$ adjustable rotation angles. These few angles are tuned on the quantum device and then mapped back to update the weights of the original classical network.

\cite{Singh2024} introduces a CFFQNN model that uses a quantum classical hybrid approach to process data. FFQNN is a QNN architecture that mirrors the flexibility of an a classical feed-forward neural network (FFNN): arbitrary hidden-layer widths, no intermediate measurements, and fully coherent operation throughout. The design cuts both circuit depth and CNOT gate count by more than 50 \% relative to leading QNN baselines, while keeping the qubit requirement independent of the number of input features.

These developments illustrate the rapid evolution of QNN
architectures: from variational ansätze that inherit the geometry
of kernel methods~\cite{Schuld2021} to explicitly neuron-like
coherent networks.  Each design balances expressivity,
trainability, and hardware constraints, continuing improvements for quantum computation.

\subsection{Transformer-based models}
Transformer architectures have emerged as a breakthrough in machine learning, especially in the context of natural language processing. The self-attention mechanism inherent in transformers enables these models to capture global relationships within the data, resulting in highly contextualized feature representations \cite{Vaswani2017}. The original transformer design, which was introduced for machine translation, has since been adapted to various domains including computer vision and reinforcement learning ~\cite{Devlin2019,Dosovitskiy2021,Baevski2020}.

In this paper, we present an integrated approach that uses the transformer encoder’s ability to reduce dimensionality and extract features may provide a more effective solution, addressing the challenge of mapping high-dimensional classical data onto the constrained input space of quantum circuits. Starting from a sequence  
\(\mathbf X=[x_{1},\dots,x_{T}]\subset\mathbb R^{d_{\text{in}}}\),  
the encoder linearly projects each token (after additional position encoding) into a query, key and value,

\begin{equation}\label{ea1}
\mathbf q_i = \mathbf W_Q x_i,\quad 
\mathbf k_i = \mathbf W_K x_i,\quad 
\mathbf v_i = \mathbf W_V x_i,
\end{equation}
here $W_Q, W_K, W_V$ are the query, key and value matrices.
\begin{equation}\label{ea12}
\mathbf z_i
=
\sum_{j=1}^{T}
\operatorname{softmax}\!\Bigl(
    \tfrac{\mathbf q_i^{\!\top}\mathbf k_j}{\sqrt{d_k}}
\Bigr)_{j}\,\mathbf v_j,
\qquad i=1,\dots,T,
\end{equation}
where \(d_k\) is the key dimension.   Stacking \(L\) attention layers—optionally separated by position-wise feed-forward blocks—yields a hierarchy  

\begin{equation}\label{ea3}
\mathbf H^{(L)}=\operatorname{Transformer}(\mathbf X)\in\mathbb R^{T\times d_{\text{model}}},
\end{equation}
whose rows act as \emph{learned principal components}: each captures a different, task-relevant pattern spread across the original \(T\) tokens.  
Because self-attention re-weights tokens globally, the model can discard redundant directions and concentrate information, accomplishing an adaptive dimensionality–reduction step within its \(\mathcal O(T^{2})\) compute budget. For hybrid quantum–classical pipelines this property is crucial—the encoder reduces the original, often hundreds-dimensional feature space down to a handful of dense channels \((\le n_{\text{qubits}})\), making it feasible to inject the data into a shallow variational quantum circuit.

\subsection{Transformer-assisted quantum neural network}
The proposed Transformer Quantum Neural Networks (TQNNs) model represents an innovative approach to hybrid quantum-classical computation. In this architecture, classical input data, such as images, are first flattened and passed through an embedding layer that maps each input element into a higher-dimensional space. This embedding mimics the word-to-vector transformation used in natural language processing, except that here each pixel or image patch is embedded into a vector space suitable for further processing. A transformer encoder, equipped with positional encodings to retain spatial relationships, then processes these embedded tokens. The self-attention mechanism inherent in the transformer allows the model to capture global interactions among the tokens, effectively condensing the information from a high-dimensional input into a more manageable form.

After the transformer encoder has processed the sequence of embedded tokens, an aggregation operation is performed to yield a single fixed-size feature vector. This step is critical because it transforms the output from a sequence of tokens into a compact representation that matches the input dimensionality requirements of the quantum neural network. Given the limited number of qubits available in current quantum hardware, reducing the input dimension to a number suitable for the qubit count is essential. This reduced representation is then mapped through classical linear layers to ensure compatibility with the QNN, which is implemented using parameterized quantum circuits and integrated into the overall model via quantum-classical interfaces as in Figure \ref{fig:hybrid_model}.

\tikzstyle{block} = [rectangle, draw, fill=blue!20, 
    minimum width=0.8cm, minimum height=2cm, text centered, rounded corners, font=\footnotesize]
\tikzstyle{smallblock} = [rectangle, draw, fill=green!30, 
    minimum width=0.2cm, minimum height=1.5cm, text centered, rounded corners, font=\footnotesize]
\tikzstyle{mainarrow} = [->, thick, >=Stealth, blue!70]
\tikzstyle{resarrow} = [->, thick, >=Stealth, dashed, orange!70]

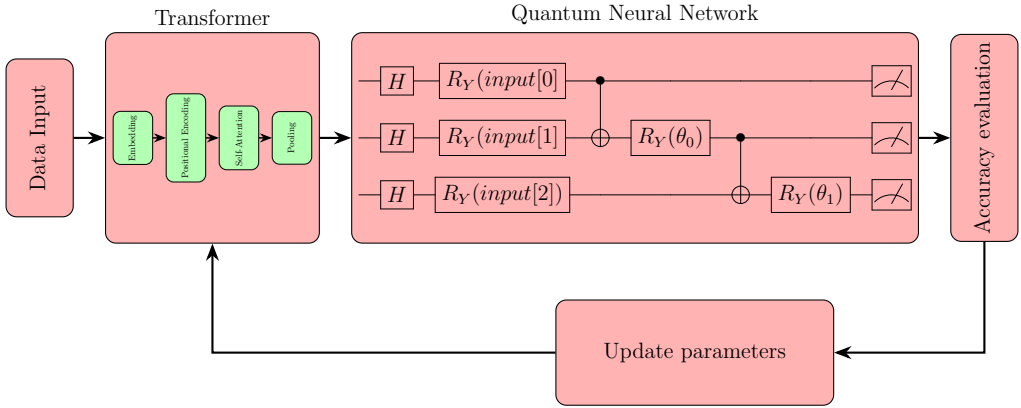
\begin{figure}[ht]
\centering
\begin{tikzpicture}[scale=0.7, transform shape, node distance=0.1cm, auto]
    \tikzstyle{startstop} = [rectangle, rounded corners, 
      minimum width=3cm, 
      minimum height=5cm,
      text centered, 
      draw=black, 
      fill=red!30]
      
     \tikzstyle{startstop1} = [rectangle, rounded corners, 
      minimum width=0.5cm, 
      minimum height=3cm,
      text centered, 
      draw=black, 
      fill=red!30] 
     
     \tikzstyle{startstop2} = [rectangle, rounded corners, 
      minimum width=5cm, 
      minimum height=2cm,
      text centered, 
      draw=black, 
      fill=red!30] 
    \tikzstyle{arrow} = [thick,->,>=Stealth]
    
    \node (data) [startstop1, text width=1cm] {\rotatebox{90}{Data Input}};
    
    \node (trans) [startstop, right=of data, xshift=0.5cm, minimum width=4cm, minimum height=4cm, label=above:Transformer] {
      \begin{tikzpicture}[scale=0.5, transform shape, node distance=0.01cm]
        \tikzstyle{block} = [rectangle, draw, fill=blue!20, 
            minimum width=0.8cm, minimum height=2cm, text centered, rounded corners, font=\footnotesize]
        \tikzstyle{smallblock} = [rectangle, draw, fill=green!30, 
            minimum width=1.5cm, minimum height=2cm, text centered, rounded corners, font=\footnotesize]
        \tikzstyle{arrow} = [->, thick,>=Stealth]
        
        \node (embed) [smallblock] {\rotatebox{90}{Embedding}};
        \node (posenc) [smallblock, anchor=west] at ($(embed.east)+(0.5cm,0)$) {\rotatebox{90}{Positional Encoding}};
        \node (attn) [smallblock, anchor=west] at ($(posenc.east)+(0.5cm,0)$) {\rotatebox{90}{Self-Attention}};
        \node (pool) [smallblock, anchor=west] at ($(attn.east)+(0.5cm,0)$) {\rotatebox{90}{Pooling}};
        
        \draw [arrow] (embed) -- (posenc);
        \draw [arrow] (posenc) -- (attn);
        \draw [arrow] (attn) -- (pool);
      \end{tikzpicture}
    };
    
    \node (qnn) [startstop, right=of trans, xshift=0.5cm, minimum width=6cm, minimum height=4cm,label=above:Quantum Neural Network] {
      \Qcircuit @C=1em @R=1em {
  & \gate{H} & \gate{R_Y(input[0]} &\ctrl{1} &  \qw                     &\qw &  \qw                  & \meter{} \\
  & \gate{H} & \gate{R_Y(input[1]} &\targ    & \gate{R_Y(\theta_0)}     &  \ctrl{1}  &        \qw               & \meter{} \\
  & \gate{H} & \gate{R_Y(input[2])} &\qw      &  \qw                    & \targ       & \gate{R_Y(\theta_1)} & \meter{}
}
    };
    
    \node (output) [startstop1, text width=1cm, right=of qnn, xshift=0.5cm]{\rotatebox{90}{Accuracy evaluation}};
     
    \node (update) [startstop2, text width=5cm, below = of output, yshift=-1cm,xshift=-5.5cm]{Update parameters}; 
        
    \draw [arrow] (data) -- (trans);
    \draw [arrow] (trans) -- (qnn);
    \draw [arrow] (qnn) -- (output);
    \draw [arrow] (output) |- (update); 
    \draw [arrow] (update) -| (trans);

\end{tikzpicture}
\caption{Transformer-assisted quantum neural network (TQNN)}
\label{fig:hybrid_model}
\end{figure}

For hybrid quantum–classical schemes the transformer encoder offers two key advantages:

\begin{enumerate}[label=(\alph*),leftmargin=1.8em]
  \item \textbf{Dimensionality reduction.}  
        After the transformer encoder has projected the data from the original
feature space $\mathbb{R}^{N}$ onto the subspace
$\mathbf H^{(L)}\!\in\!\mathbb{R}^{T\times d_{\text{model}}}$, we
\emph{average-pool} across the sequence dimension and obtain a single
$d_{\text{model}}$-dimensional embedding.  A final linear layer then
selects
\[
\mathbf h\in\mathbb{R}^{n},\qquad n=n_{\text{qubits}}\le d_{\text{model}},
\]
so that the length of $\mathbf h$ exactly matches the number of qubits
available for the quantum circuit.

  \item \textbf{Global feature extraction.}
      The row-normalised weights
      $$\alpha_{ij}=
        \operatorname{softmax}\!\bigl(
            \mathbf q_i^{\!\top}\mathbf k_j/\sqrt{d_k}
        \bigr)$$
      form a content-adaptive kernel: they let the classical front-end
      aggregate long-range correlations that would otherwise have to be
      captured by deeper entangling circuits on quantum hardware.

\end{enumerate}

Thus, coupling a transformer encoder to a variational quantum circuit
creates a \emph{division of labour}:  
the classical transformer compresses and mixes the raw
\(N\)-dimensional input into an \(n\)-dimensional, information-dense
feature vector,
while the quantum layer leverages superposition and entanglement
to process those \(n\) channels
in a Hilbert space of dimension \(2^{\,n}\).
Such an integrated architecture promises a more faithful exploitation
of each paradigm's strengths than earlier hybrids that relied on
convolutions or ad-hoc quantum embeddings alone.

\section{Genetic transformer-assisted quantum neural networks (GTQNNs)} \label{sec3}

\subsection{ Multi-Objective Genetic Algorithm}
Genetic algorithms solve optimization problems by emulating natural evolution.  They maintain a population of candidate solutions and, generation after generation, apply selection, crossover and mutation to create new offspring.  Individuals that score higher on the objective (or objectives) are preferentially chosen to propagate and gradually guide the population toward areas of greater fitness within the search space.  After a prescribed number of iterations—or once improvement stalls—the algorithm returns the best‐performing individuals as approximate optima of the fitness function over the enormous configuration landscape \cite{Miettinen1999,Deb2002}.

Genetic algorithms repeatedly pick “parent” solutions from the current pool and applying genetic operators to them.  For each offspring, two parents are selected, combined through crossover, and then modified by mutation, producing a child that carries a mix of their genetic material.  New parent pairs are drawn for every child until stopping conditions are met.  The success of a genetic algorithm hinges on these operators: 1) selection chooses which individuals win the right to reproduce, biasing sampling toward high-fitness solutions while still preserving diversity; 2) mutation introduces random perturbations to a child’s genome, enabling the search to jump to remote regions of the landscape and escape local optima; 3) crossover swaps segments of genetic code between two parents, creating larger, coordinated changes and promoting the discovery of novel combinations of useful genes. The precise design of these operators determines both the efficiency and the scope of the evolutionary search.

Genetic algorithms employ explicit stopping criteria to ensure both efficiency and reliable convergence. Typical criteria include: (i) detecting fitness convergence or saturation, when successive generations show negligible improvement; (ii) enforcing a required performance threshold, so evolution halts once a solution meets the target accuracy; and (iii) imposing a hard cap on the number of generations to bound runtime. Equipped with such safeguards, genetic algorithms can tackle large, complex optimisation problems while avoiding needless computation. In our experiment, we only impose the number of generation as the stop criteria. 

For numerous real-world tasks, it is advantageous to frame the search as an \emph{evolutionary multi-objective optimization} (EMO) problem, in which several, often competing, objective functions must be minimized or maximized simultaneously \cite{ADB14,Miettinen1999,Deb2002}. As with single-objective cases, multi-objective formulations can include constraints that restrict the set of admissible solutions; the algorithm must therefore find Pareto-optimal individuals that satisfy all such feasibility requirements.

\tikzstyle{startstop} = [rectangle, rounded corners, 
minimum width=3cm, 
minimum height=1cm,
text centered, 
draw=black, 
fill=red!30]

\tikzstyle{description} = [rectangle, 
minimum width=2cm, 
minimum height=1cm, 
text centered, 
text width=2cm, 
draw=black, 
fill=orange!30]

\tikzstyle{decision} = [diamond, 
minimum width=2cm, 
minimum height=1cm, 
text centered, 
text width=2cm, 
draw=black, 
fill=green!30]
\tikzstyle{arrow} = [thick,->,>=stealth]

\tikzstyle{process} = [rectangle, 
minimum width=3cm, 
minimum height=1cm, 
text centered, 
text width=4cm, 
draw=black, 
fill=blue!20]

\tikzstyle{processV} = [rectangle, 
minimum width=1cm, 
minimum height=2cm, 
text centered, 
text width=1cm, 
draw=black, 
fill=blue!20]

\tikzstyle{processY} = [rectangle, 
minimum width=3cm, 
minimum height=1cm, 
text centered, 
text width=4cm, 
draw=black, 
fill=yellow!10]

\tikzstyle{processB} = [rectangle, 
minimum width=3cm, 
minimum height=1cm, 
text centered, 
text width=6cm, 
draw=black, 
fill=yellow!20]

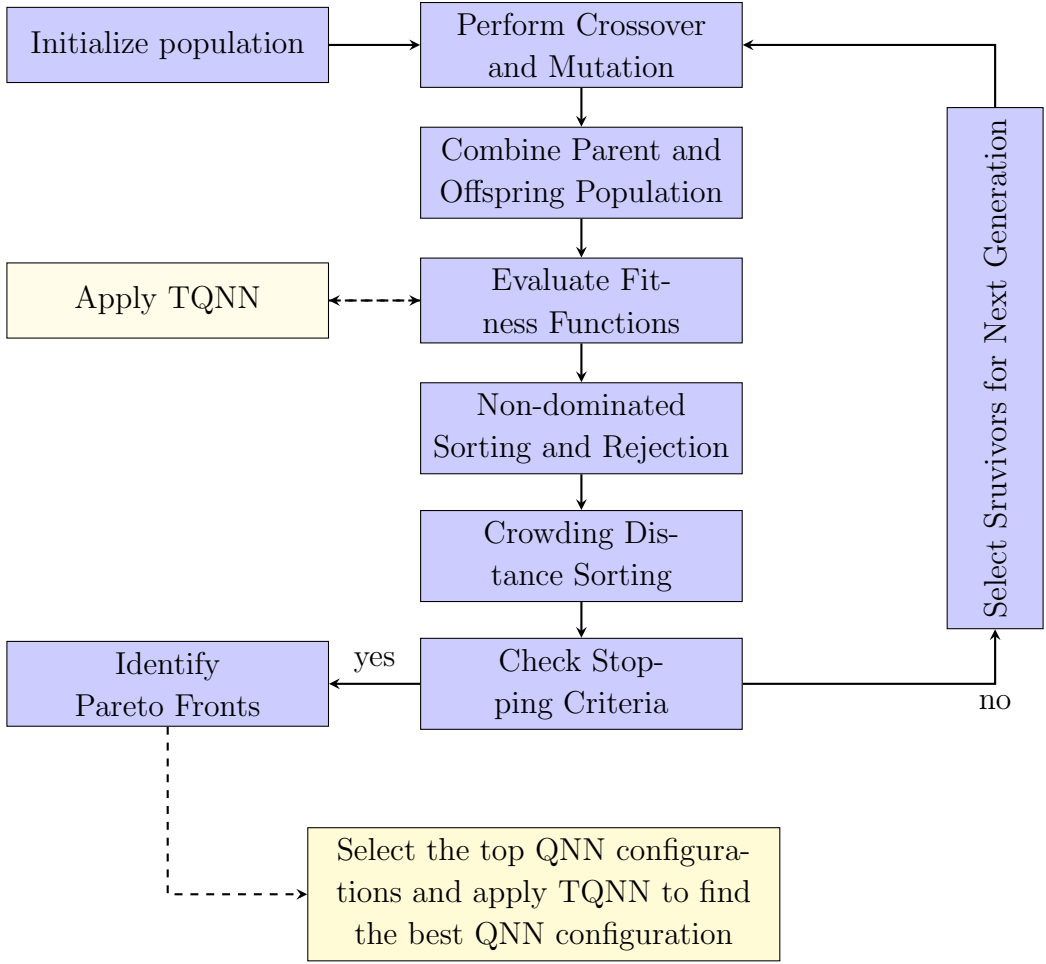
\begin{figure}
\centering

\begin{tikzpicture}[node distance=0.5cm]
\node (1) [process] {Initialize population};
\node (2) [process, right of=1, xshift=5cm]   {Perform Crossover and Mutation};
\node (4) [process, below of=2, yshift=-1.2cm] {Combine Parent and Offspring Population};
\node (5) [process, below of=4, yshift=-1.2cm] {Evaluate Fitness Functions};
\node (6) [process, below of=5, yshift=-1.2cm] {Non-dominated Sorting and Rejection};
\node (7) [process, below of=6, yshift=-1.2cm] {Crowding Distance Sorting};
\node (8) [process, below of=7, yshift=-1.2cm] {Check Stopping Criteria};
\node (9) [process, left of=8, xshift= -5cm] {Identify Pareto Fronts};
\node (10) [processV, right of=8, xshift=5cm, yshift=4.2cm] {\rotatebox{90}{Select Sruvivors for Next Generation}};
\node (11) [processY, left of=5, xshift=-5cm] {Apply TQNN};
\node (12) [processB, below of=9, xshift=5cm,yshift=-2.3cm] {Select the top QNN configurations and apply TQNN to find the best QNN configuration};


\draw [arrow] (1)--(2) ;
\draw [arrow] (2)--(4) ;
\draw [arrow](4) -- (5);
\draw [arrow] (5) -- (6);
\draw [arrow] (6) -- (7);
\draw [arrow] (7) -- (8);
\draw [arrow] (8) -- node[anchor=south] {yes} (9);
\draw [arrow] (8) -| node[anchor=north] {no} (10);
\draw [arrow] (10) |- (2);

\draw [arrow,dashed] (5) -- (11);
\draw [arrow,dashed] (11) -- (5);
\draw [arrow,dashed] (9) |- (12);
\end{tikzpicture}
  \caption{Genetic Transformer Quantum Neural Networks (GTQNNs)}
  \label{NSGA-II}
\end{figure}

Optimality in multi–objective optimisation is formalised via a partial
order called \emph{dominance}.  
Throughout this work we confine ourselves to \emph{unconstrained} problems,
i.e.\ no equality, inequality, or bound constraints limit the feasible set.
Let
\(\mathbf x^{(1)},\mathbf x^{(2)} \in \mathbb R^{d}\)
be two feasible decision vectors evaluated by
\(m\) objective functions
\(f_1,\dots,f_m : \mathbb R^{d}\!\to\!\mathbb R\).
We say that
\(\mathbf x^{(1)}\) \textbf{dominates}
\(\mathbf x^{(2)}\),
denoted
\(\mathbf x^{(1)} \prec \mathbf x^{(2)}\),
iff

\begin{enumerate}[label=(\roman*)]
\item (\textit{non-worseness}) 
      \[
        f_k\!\bigl(\mathbf x^{(1)}\bigr)
        \;\le\;
        f_k\!\bigl(\mathbf x^{(2)}\bigr)
        \quad\text{for all } k \in \{1,\dots,m\};
      \]
\item (\textit{strict improvement})  
      there exists at least one index \(k^\star\) such that
      \[
        f_{k^\star}\!\bigl(\mathbf x^{(1)}\bigr)
        \;<\;
        f_{k^\star}\!\bigl(\mathbf x^{(2)}\bigr).
      \]
\end{enumerate}

A solution is called \emph{non-dominated} when no other member of the current
population is strictly better in every objective.  Non-dominated points exhibit an intrinsic \emph{trade–off}: any gain in one
objective unavoidably incurs a loss in at least one other objective.
This property encourages the algorithm to preserve a diverse spectrum of
candidates rather than collapsing prematurely onto a single compromise.
The set of all mutually non-dominated solutions constitutes the
\emph{Pareto front}---a frontier along which every incremental improvement
in one objective demands a compensating sacrifice in another.

The Non-dominated Sorting Genetic Algorithm~II (NSGA-II) was proposed by
Deb~\cite{Deb2002} to remedy two shortcomings of earlier EMO schemes:
lack of elitism and poor diversity maintenance.  Today it is one of the
most popular algorithms in evolutionary multi-objective optimisation.
In our work we embed a TQNN inside the
NSGA-II loop to serve as the fitness evaluator
(Figure~\ref{NSGA-II}).
We will show that this
\emph{TQNN-assisted NSGA-II} improves prediction accuracies compared
with the classical version, yielding superior Pareto fronts for the benchmark data sets.

\subsection{Fitness function and genetic quantum feature map}\label{featureMap}
The fitness function serves as the objective (cost) measure for each trial
TQNN: it uses a variational quantum circuit and returns a scalar score
that reflects both predictive quality and hardware efficiency.
Drawing on the metrics in~\cite{ARG21,Wang2025,Wang2025-1},
we define two fitness criteria (see Eq.~\eqref{FitnessFunc})
with a twofold aim: \textit{Maximize classification accuracy}; and \textit{Minimize gate cost}, quantified as the total count of primitive gates required to realize the circuit on quantum hardware. 
\begin{equation}\label{FitnessFunc} 
\left\{
\begin{array}{lr}
\text{(maximize)} \text{Fitness 1} = \text{Classification accuracy}\\ 
\text{(minimize)} \text{Fitness 2 } =\text{Gate count} \\
\end{array}
\right. 
\end{equation}

Assume that the number of qubits is $N$. the variational quantum circuit starts with $N$ Hadamard gates, followed by $N$ rotation gates with respect to the $Y$ axis,   $RY$. The transformer module will generate an output of length $N$ to ensure that each $RY$ takes one output as the rotation angle of $RY$ gates. As a result, the encoding scheme for each individual population  in the NSGA-II algorithm has $\binom{N}{2}=\frac{N *(N-1)}{2} $ bits with $1$ indicting a CNOT gate and $RY$ gate, and $0$ otherwise. 

The evolutionary search begins with a randomly generated population, each individual encoding a candidate quantum circuit.  Every circuit is trained and scored on the training set with the multi-objective fitness function.  Individuals with higher fitness are preferentially chosen to reproduce: selection picks the parents, crossover recombines their circuit descriptions, and mutation introduces random edits.  The resulting offspring form the next generation of circuits.  This cycle of evaluation and variation repeats until the stopping criteria are satisfied.

Because the fitness objectives simultaneously maximize accuracy and minimize gate cost, the Pareto fronts returned by TQNN reveal how different circuits trade hardware economy for predictive power.  Examining these fronts helps us identify solutions that deliver near-maximal accuracy at a fraction of the gate budget, as well as those that push accuracy to its limit regardless of cost.  A detailed exploration of the full, multi-dimensional Pareto surface would further clarify how a specific circuit design influence accuracy and where the sharpest cost-benefit gains lie.

\section{Experimental Results }\label{exprement}

\subsection{Experimental Procedure}\label{procudure}
The research was trained and evaluated on the supercomputers at Arizona State University with NVIDIA’s CUDA acceleration framework \cite{ASUComputing2023}. The supercomputers provide ASU researchers access to a state-of-the-art system including NVIDIA A100. All experiments were performed on the IBM Qiskit \texttt{AerSimulator} (QASM mode, 1024 shots). 

With the training and test sets prepared, we use jMetalPy \cite{jMetalPy} to launch the evolutionary
search for high-performing variational circuits as outlined earlier.
The run begins with a population of 20 (Population size = 20, offspring population size= 20) random chromosomes,  each
encoded as a binary string that describes a candidate TQNN circuit.
At every generation we apply a crossover with probability \(p_c = 0.90\), bit-flip mutation with probability
\(p_m = \frac{1}{\text{length of string}}\) and other NSGA-II operations to produce the next population.

For each individual circuit we perform an \emph{inner} training loop of 50 epochs (
epochs- = 50) using a mini-batch size of 32, then
evaluate (i) classification accuracy on the validation set and
(ii) circuit cost, measured as the total number of single- and
two-qubit gates. The evolutionary process is allowed to run for 50 or 60 generations (Generations = 50, 60).

After the final generation we extract the \(k=10\) non-dominated
solutions with the highest accuracy.  
Each of these circuits is retrained from scratch for
100 or more outer epochs (epochs = 100 or more ), and the best-performing model is
retained as the final GTQNN.  
The entire workflow—population initialisation, genetic operators,
fitness evaluation, Pareto selection, and final retraining—is depicted
in Fig.~\ref{NSGA-II}, with each stage represented by a distinct
rectangular block.

This multi-stage procedure is designed to achieve optimal solutions for complex optimization tasks while leveraging quantum computing principles an integrated approach that combines classical and quantum techniques. Initially, the Transformer model was employed to select relevant features from the dataset, effectively narrowing down the input dimensionality. The selected features were then processed through a Quantum Neural Network, where quantum circuits encoded these features into quantum states suitable for classification tasks. The experiments on the following datasets demonstrate that GTQNNs deliver state-of-the-art performance with a quantum-resource budget compatible with current NISQ devices.

\subsection{Iris dataset}\label{sec5-4}

The Iris data set comprises 150 samples drawn from three species—
\textit{I.\ setosa}, \textit{I.\ versicolor}, and \textit{I.\ virginica}.
Each sample is described by four real-valued features
(sepal length, sepal width, petal length, petal width), forming a
$150\times4$ matrix whose rows are specimens and whose columns are
measurements.  We apply the optimization workflow of Section~\ref{procudure} to discover
an optimal variational-circuit architecture for GTQNN.  Because the problem involves only four input features, the
outer training loop is run for a modest
(epochs = 250), which is sufficient for full
convergence on this data set. 
\begin{table}[htbp]
\centering
\resizebox{0.85\textwidth}{!}{%
\begin{tabular}{>{\raggedright\arraybackslash}p{3.5cm} c c c c c c c c}
\multirow{2}{*}{Method} & \multicolumn{8}{c}{Number of Qubits} \\
\cline{2-9}
 & 3 & 4 & 5 & 6 & 7 & 8 & 9 & 10 \\
\hline \\[-1em]
\parbox{3.5cm}{\raggedright GTQNN\\[0.3em]}
 &  &  & &  &  & &  &  \\[0.5em]

\parbox{3.5cm}{\scriptsize Accuracy}
 & 1 & 0.90 & 0.8667 & 0.9667 & 0.90 & 0.9667 & 1 & 1 \\[0.5em]
 
\parbox{3.5cm}{\scriptsize Gate count}
 & 8 & 16 & 20 & 28 & 26 & 48 & 58 & 78 \\[1.5em]

\hline\hline
\parbox{3.5cm}{\raggedright QNN with EfficientSU2 \cite{liu2024training}}
 & \multicolumn{8}{>{\raggedright\arraybackslash}p{%
      \dimexpr\linewidth-3.5cm-8\tabcolsep\relax}}{%
      \scriptsize Reported accuracies range from 0.90 to 0.95 with one layer QNN (8 qubits, $ > $ 32 gates (32 parameters))}\\[1.5em]
\hline

\hline\hline
\parbox{3.5cm}{\raggedright DeepQMLP \cite{alam2022deepqmlp}}
 & \multicolumn{8}{>{\raggedright\arraybackslash}p{%
      \dimexpr\linewidth-3.5cm-8\tabcolsep\relax}}{%
      \scriptsize Reported accuracy close to 1 with 4 or more parametric layers ( 4 qubits, each layer has 16 gates).}\\[1.5em]
\hline
\end{tabular}}
\caption{Quantum gate count and accuracy comparison for the Iris dataset.}
\label{table_compareAccuracy_Iris}
\end{table}

\begin{figure}[h!]
    \centering
    \begin{subfigure}[b]{0.22\textwidth}
        \centering
        \includegraphics[width=\textwidth]{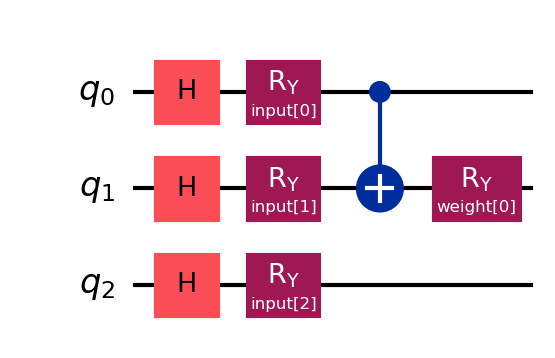}
        \caption{3 qubits}
    \end{subfigure}
    \hfill
    \begin{subfigure}[b]{0.22\textwidth}
        \centering
        \includegraphics[width=\textwidth]{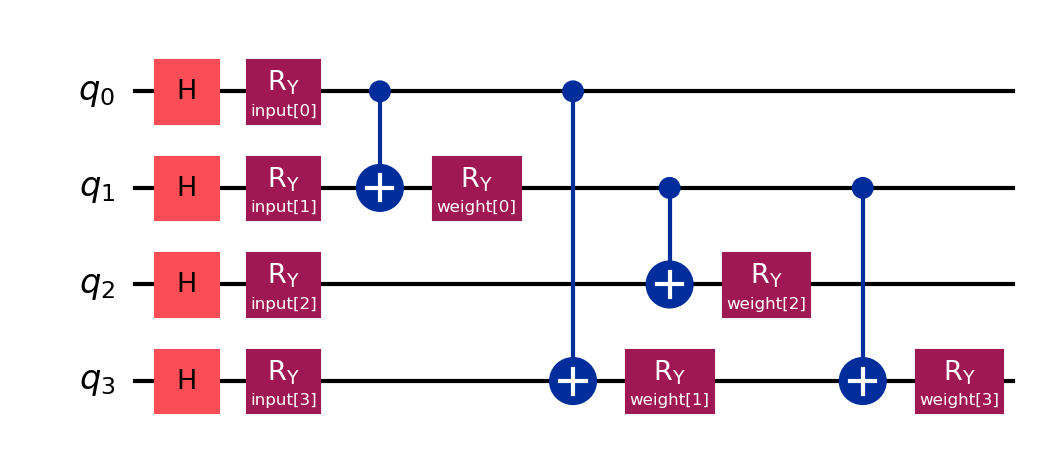}
        \caption{4 qubits}
    \end{subfigure}
    \hfill
    \begin{subfigure}[b]{0.22\textwidth}
        \centering
        \includegraphics[width=\textwidth]{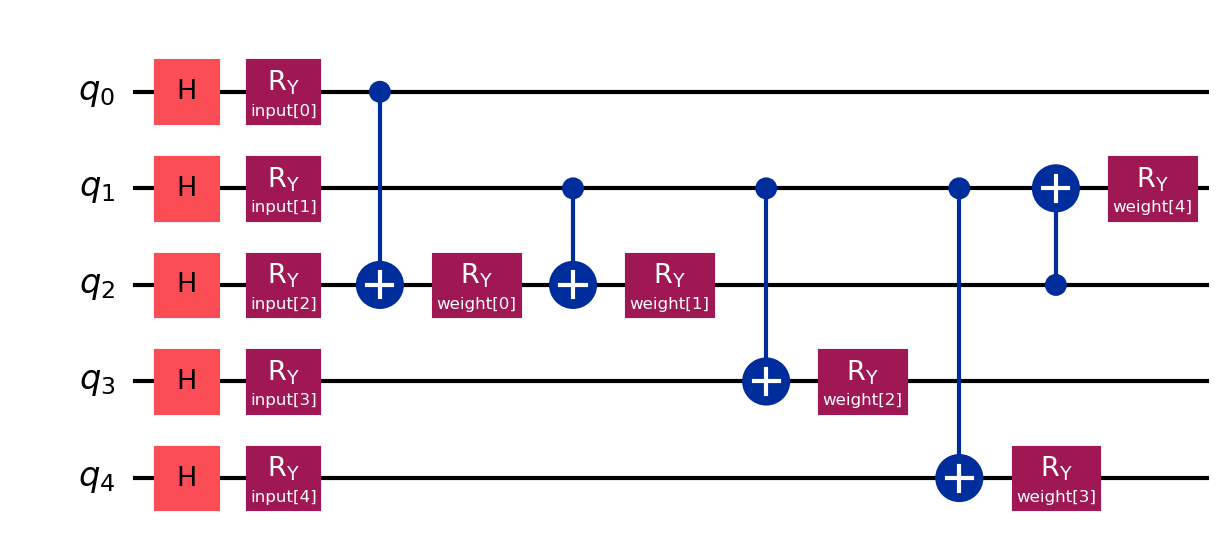}
        \caption{5 qubits}     
    \end{subfigure}

    \vskip\baselineskip

    \begin{subfigure}[b]{0.22\textwidth}
        \centering
        \includegraphics[width=\textwidth]{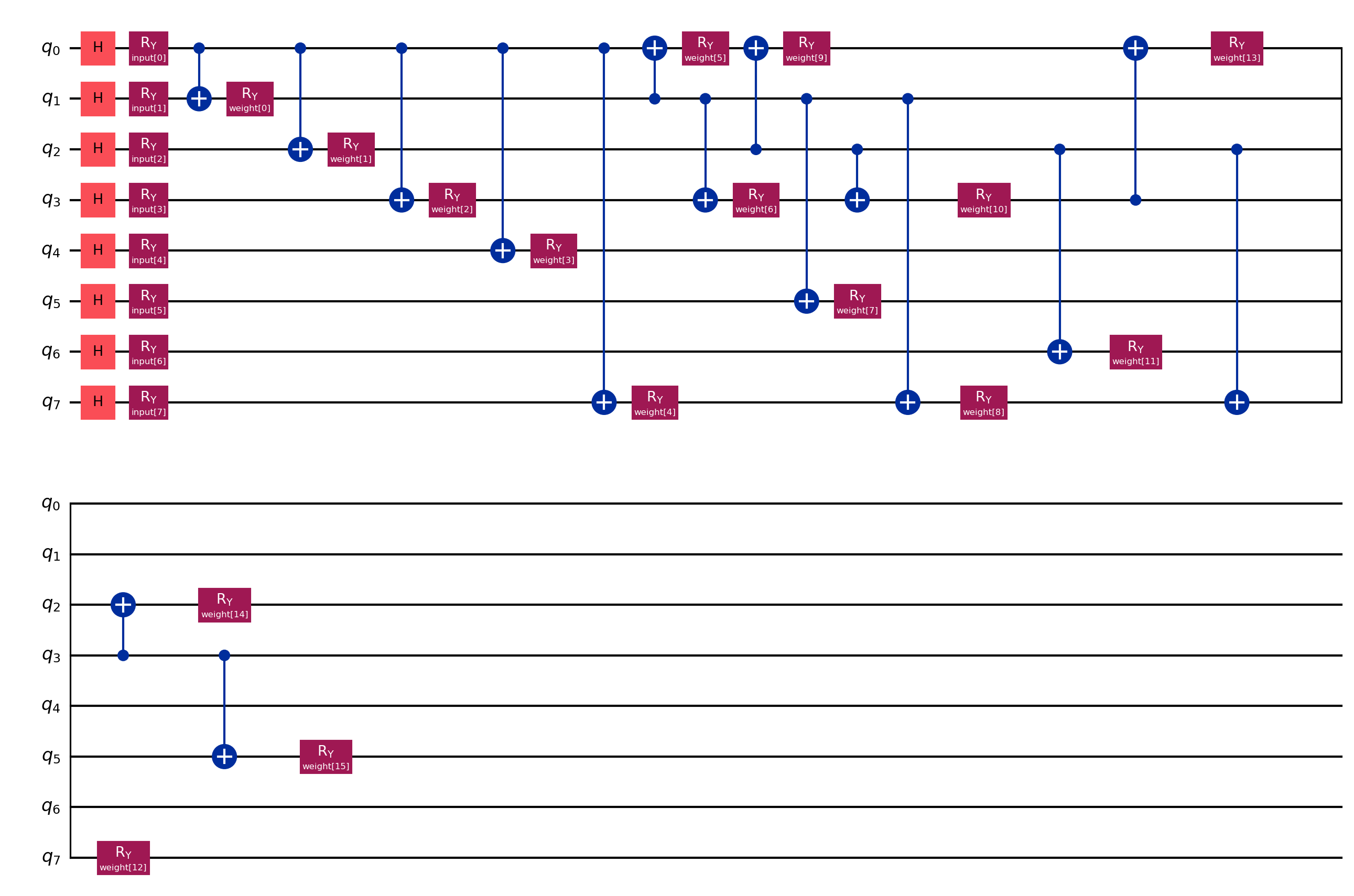}
        \caption{8 qubits}
    \end{subfigure}
    \hfill
    \begin{subfigure}[b]{0.22\textwidth}
        \centering
        \includegraphics[width=\textwidth]{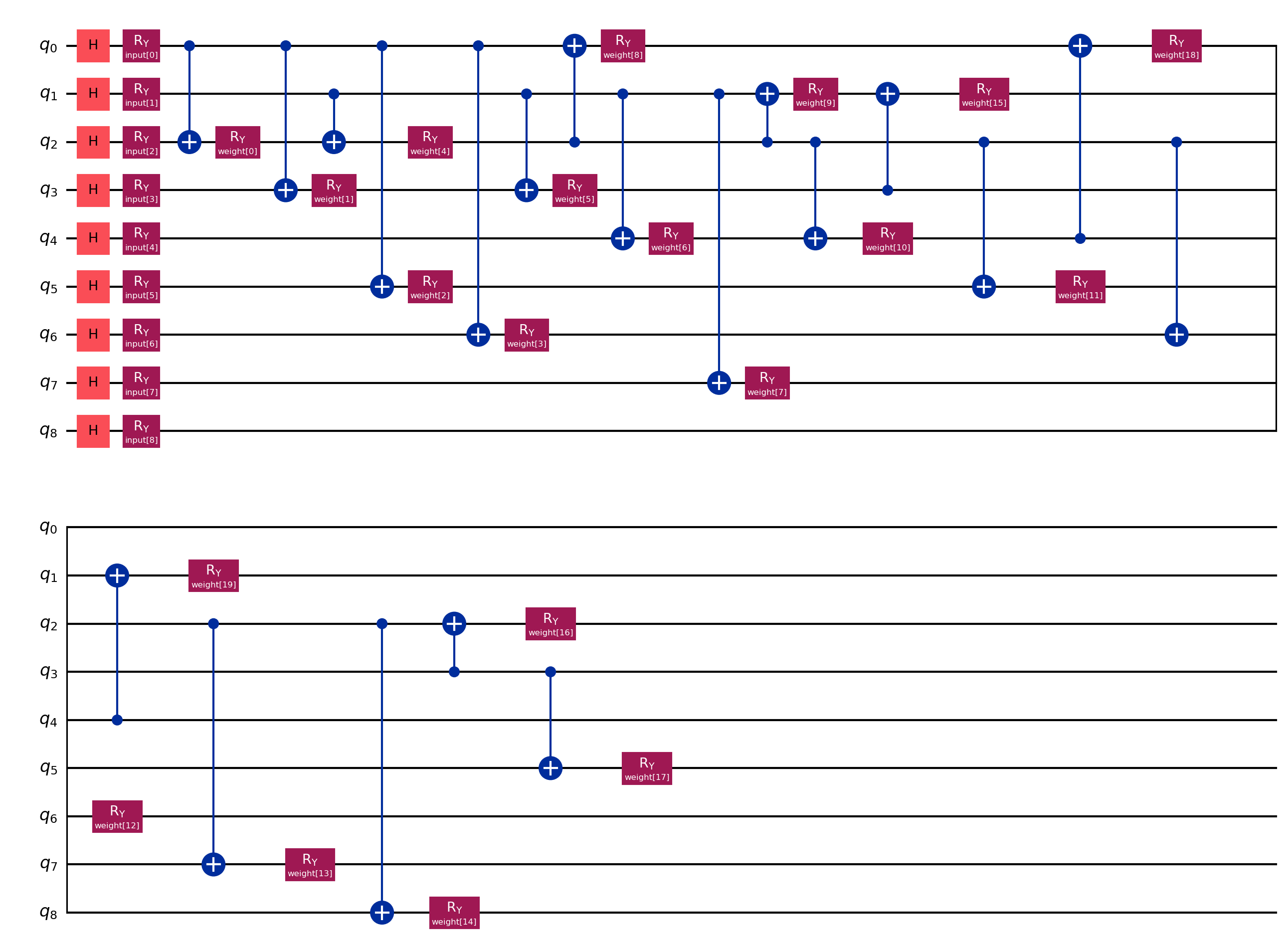}
        \caption{9 qubits}
    \end{subfigure}
    \hfill
    \begin{subfigure}[b]{0.22\textwidth}
        \centering
        \includegraphics[width=\textwidth]{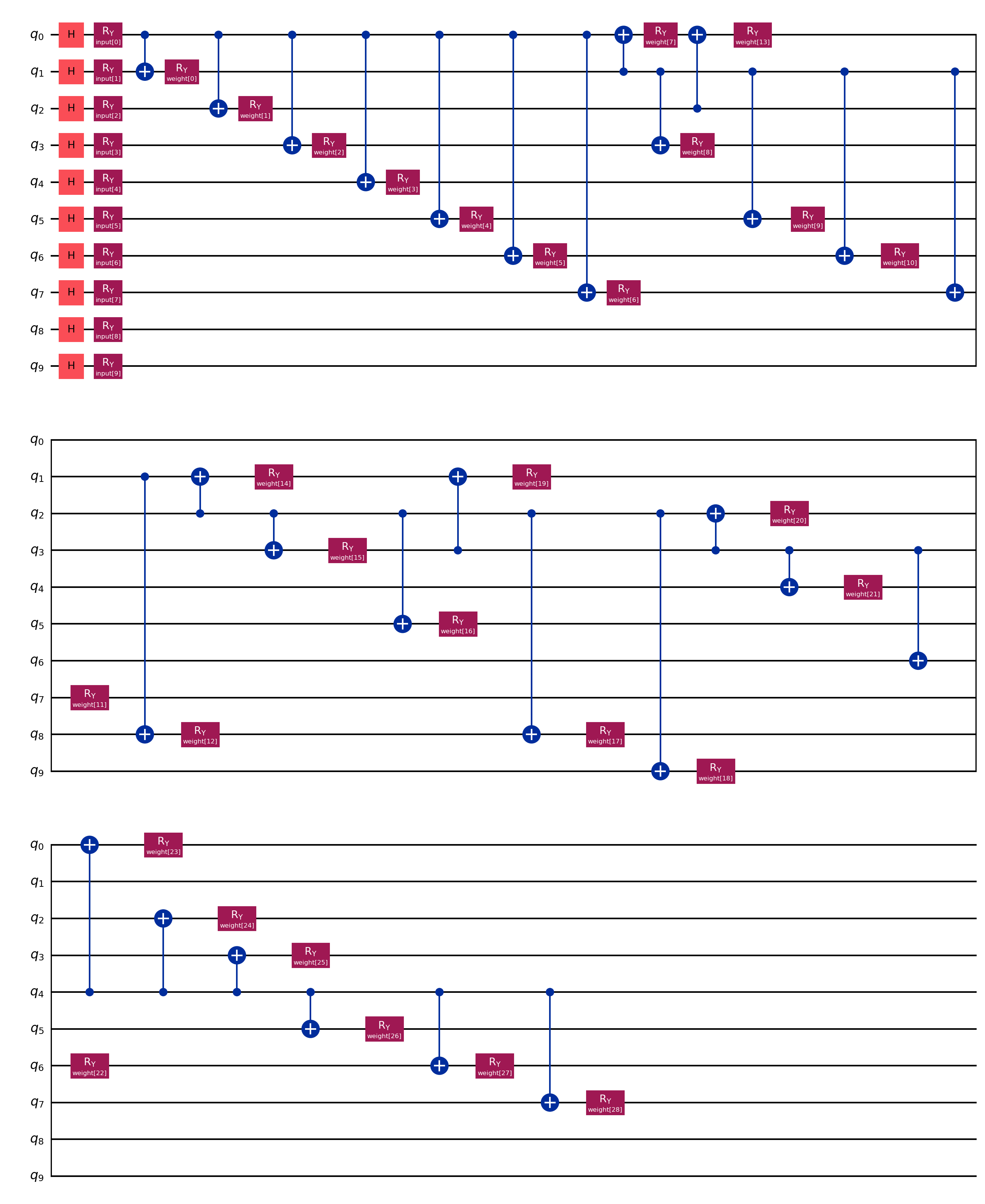}
        \caption{10 qubits}
    \end{subfigure}

    \caption{Optimal Variational Quantum Circuits for Iris dataset}
    \label{fig:circuitIris}
\end{figure}

Table \ref{table_compareAccuracy_Iris} contrasts our genetic transformer-assisted quantum neural network (GTQNN) with two recent baselines—QNN with EfficientSU2 \cite{liu2024training} and DeepQMLP \cite{alam2022deepqmlp}—on the Iris-flower classification task.
For each qubit budget we report the best accuracy obtained by the search as well as the corresponding circuit size (row “Gate count”).  With only 3 qubits the GTQNN already reaches perfect accuracy; performance remains around 0.90 up to 8 qubits and returns to 100 \% at 9–10 qubits, while gate counts grow from 8 to 78.  QNN with EfficientSU2 \cite{liu2024training}, evaluated at a single layer QNN (8 qubits, more than 32 gates), yields an accuracy band of 0.90–0.95—competitive but with greater gate cost for most qubits than GTQNN.  DeepQMLP \cite{alam2022deepqmlp} attains unit accuracy with more gates after 50 training epochs, matching the GTQNN’s score at a similar qubit count but with more  gate depth used by our 4-qubit configuration. Table \ref{fig:circuitIris} shows some corresponding optimal variational quantum circuits from GTQNN. Overall, the tables show that the GTQNN achieves state-of-the-art accuracy across a wide range of qubit counts while remaining substantially shallower than the published benchmarks, underscoring its suitability for near-term quantum hardware.

\subsection{Breast cancer dataset}
 
The Breast-Cancer-Wisconsin (diagnostic) data set from \texttt{scikit-learn} contains 569 observations, each described by 30 positive real-valued features extracted from medical-image analysis of breast tissue (e.g., mean radius, texture, perimeter, area, compactness, and related statistics). The target is binary: 212 tumours are labelled \emph{malignant} and 357 \emph{benign}.  We apply the optimization workflow of Section~\ref{procudure} to discover
an optimal variational-circuit architecture for GTQNN. The outer training loop converges in only 100 epochs, already delivering competitive classification accuracy, so we adopt outer epochs as 100 for this data set.

\begin{table}[htbp]
\centering
\resizebox{0.85\textwidth}{!}{
\begin{tabular}{>{\raggedright\arraybackslash}p{3.5cm} c c c c c c c c}
\multirow{2}{*}{Method} & \multicolumn{8}{c}{Number of Qubits} \\
\cline{2-9}
 & 3 & 4 & 5 & 6 & 7 & 8 & 9 & 10 \\
\hline \\[-1em]
\parbox{3.5cm}{\raggedright GTQNN\\[0.5em]}
 &  &  &  &  &  &  &  & \\[2em]

\parbox{3.5cm}{\scriptsize  Accuracy }
 & 0.9737 & 0.9737 & 0.9737 & 0.9737 & 0.9737 & 0.9737 & 0.9737 & 0.9825  \\[0.5em]

\parbox{3.5cm}{\scriptsize  Gate count }
 & 12 & 10& 20 & 28 & 28 & 40 & 48 & 50 \\[1.5em]
\hline

\hline\hline
\parbox{3.5cm}{\raggedright CFFQNN \cite{Singh2024}}
 & \multicolumn{8}{>{\raggedright\arraybackslash}p{%
      \dimexpr\linewidth-3.5cm-8\tabcolsep\relax}}{%
      \scriptsize Reported accuracies: about 0.85 with a layer structure
of [3,2,1], $>$ 35 gates (parameters). Use PCA to reduce its dimension to 7}\\[1.5em]
\hline

\end{tabular}}
\caption{Quantum gate count and accuracy comparison for Breast cancer dataset}
\label{table_compareAccuracy_cancerdata}
\end{table}

\begin{figure}[h!]
    \centering
    \begin{subfigure}[b]{0.22\textwidth}
        \centering
        \includegraphics[width=\textwidth]{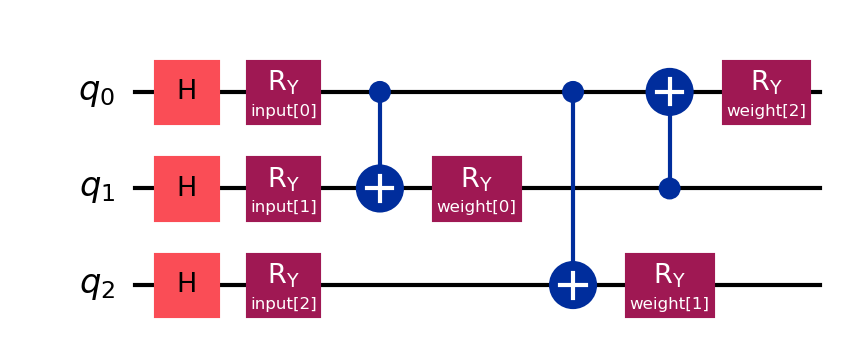}
        \caption{3 qubits}
    \end{subfigure}
    \hfill
    \begin{subfigure}[b]{0.22\textwidth}
        \centering
        \includegraphics[width=\textwidth]{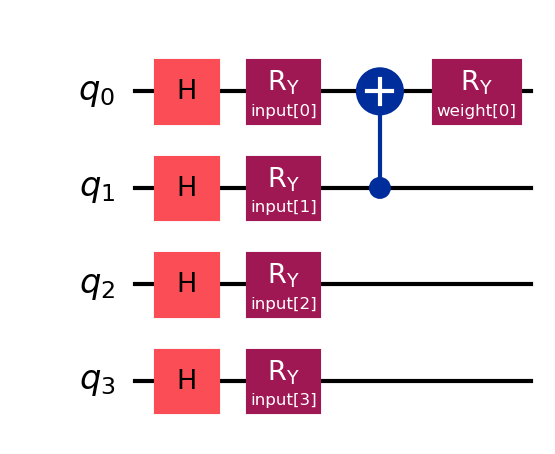}
        \caption{4 qubits}
    \end{subfigure}
    \hfill
    \begin{subfigure}[b]{0.22\textwidth}
        \centering
        \includegraphics[width=\textwidth]{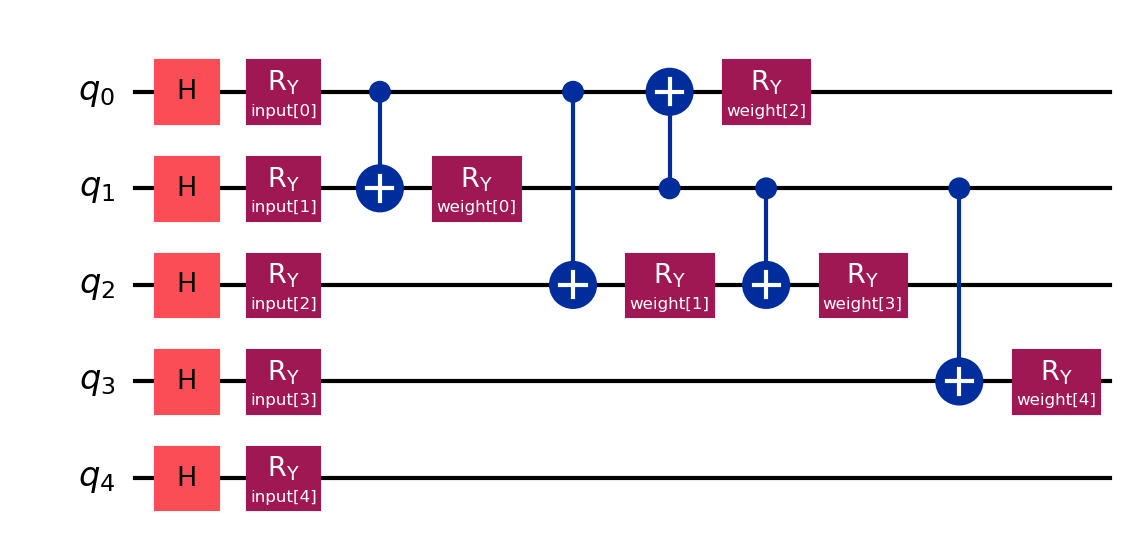}
        \caption{5 qubits}     
    \end{subfigure}
    \hfill
    \begin{subfigure}[b]{0.22\textwidth}
        \centering
        \includegraphics[width=\textwidth]{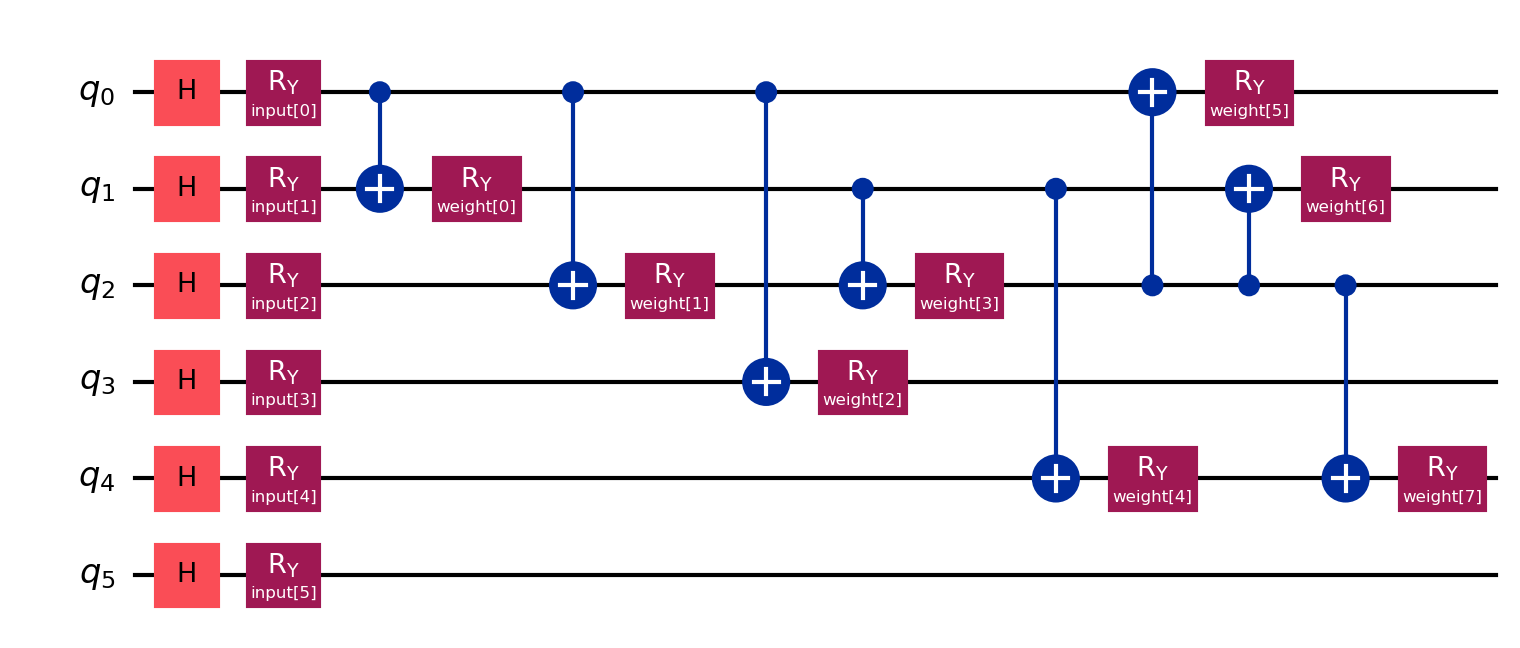}
        \caption{6 qubits} 
    \end{subfigure}

    \vskip\baselineskip

    \begin{subfigure}[b]{0.22\textwidth}
        \centering
        \includegraphics[width=\textwidth]{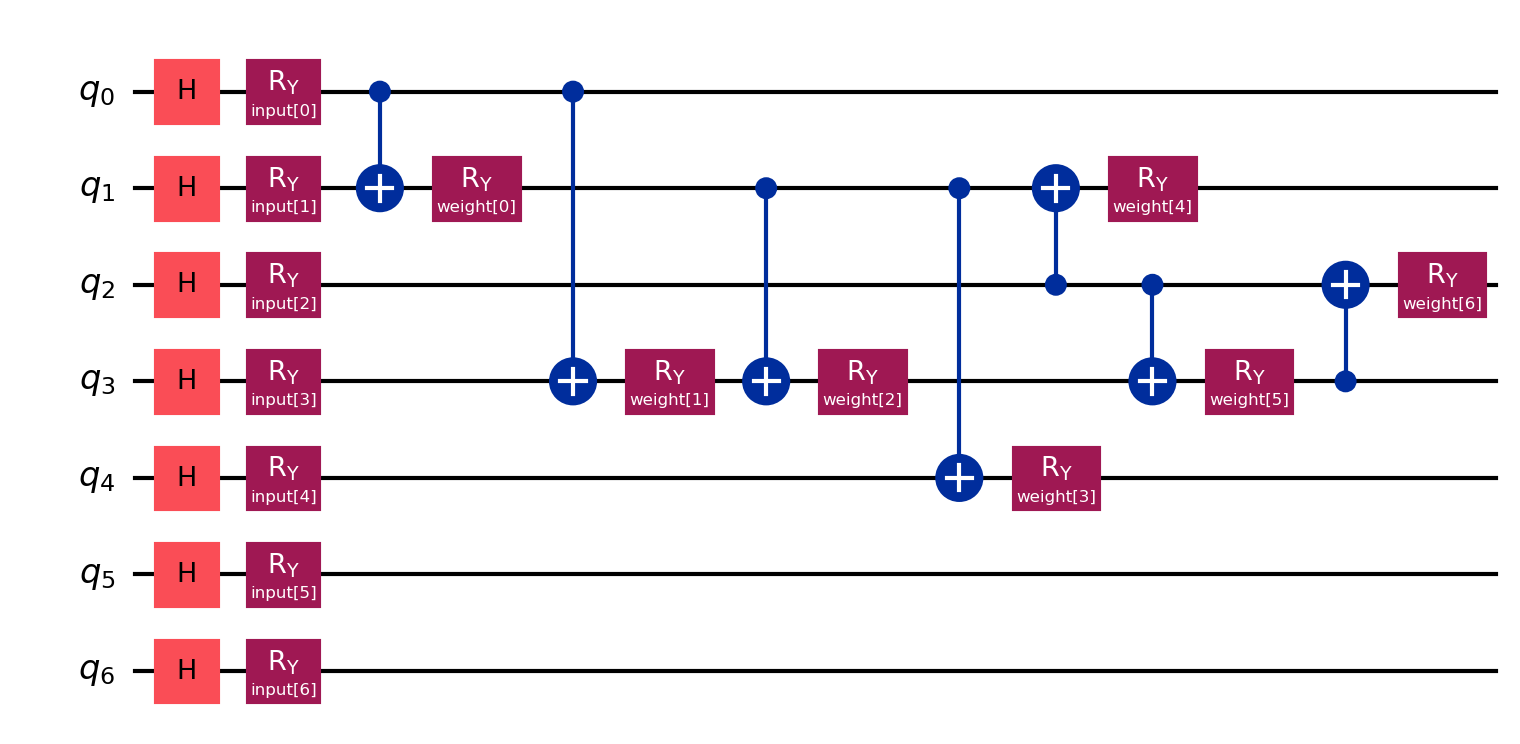}
        \caption{7 qubits} 
    \end{subfigure}
    \hfill
    \begin{subfigure}[b]{0.22\textwidth}
        \centering
        \includegraphics[width=\textwidth]{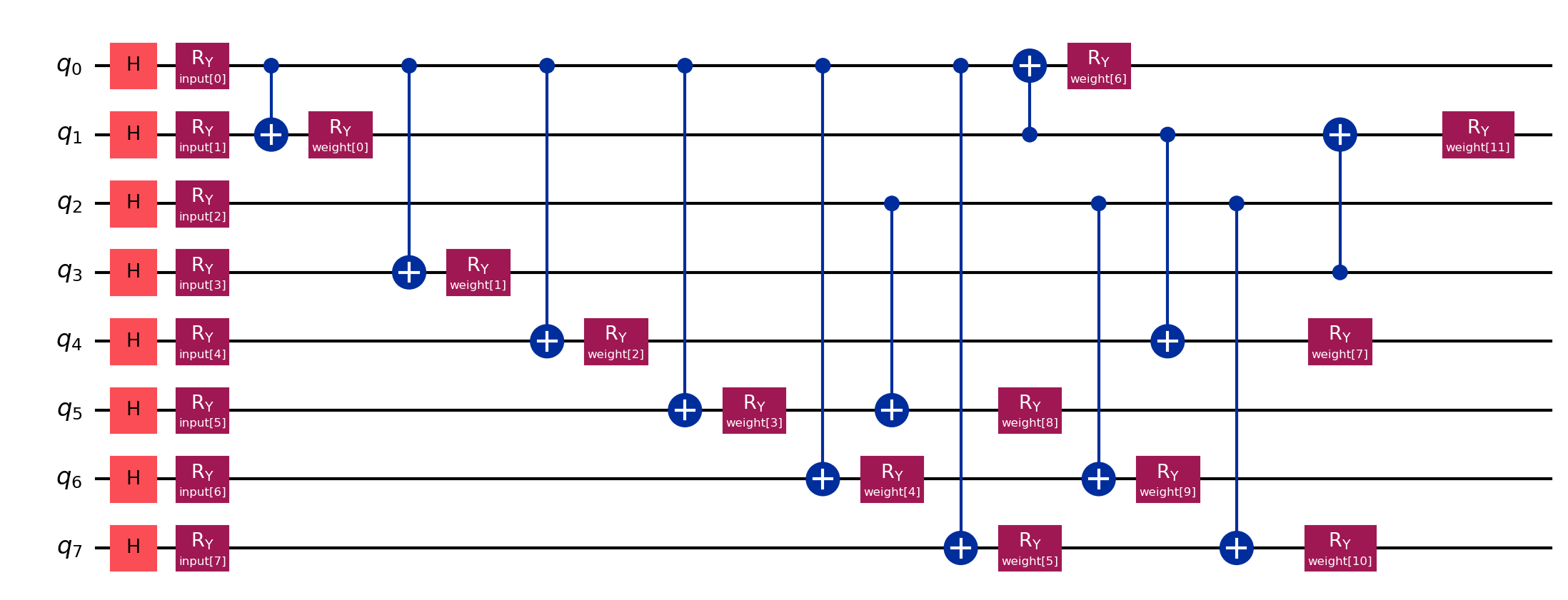}
        \caption{8 qubits}
    \end{subfigure}
    \hfill
    \begin{subfigure}[b]{0.22\textwidth}
        \centering
        \includegraphics[width=\textwidth]{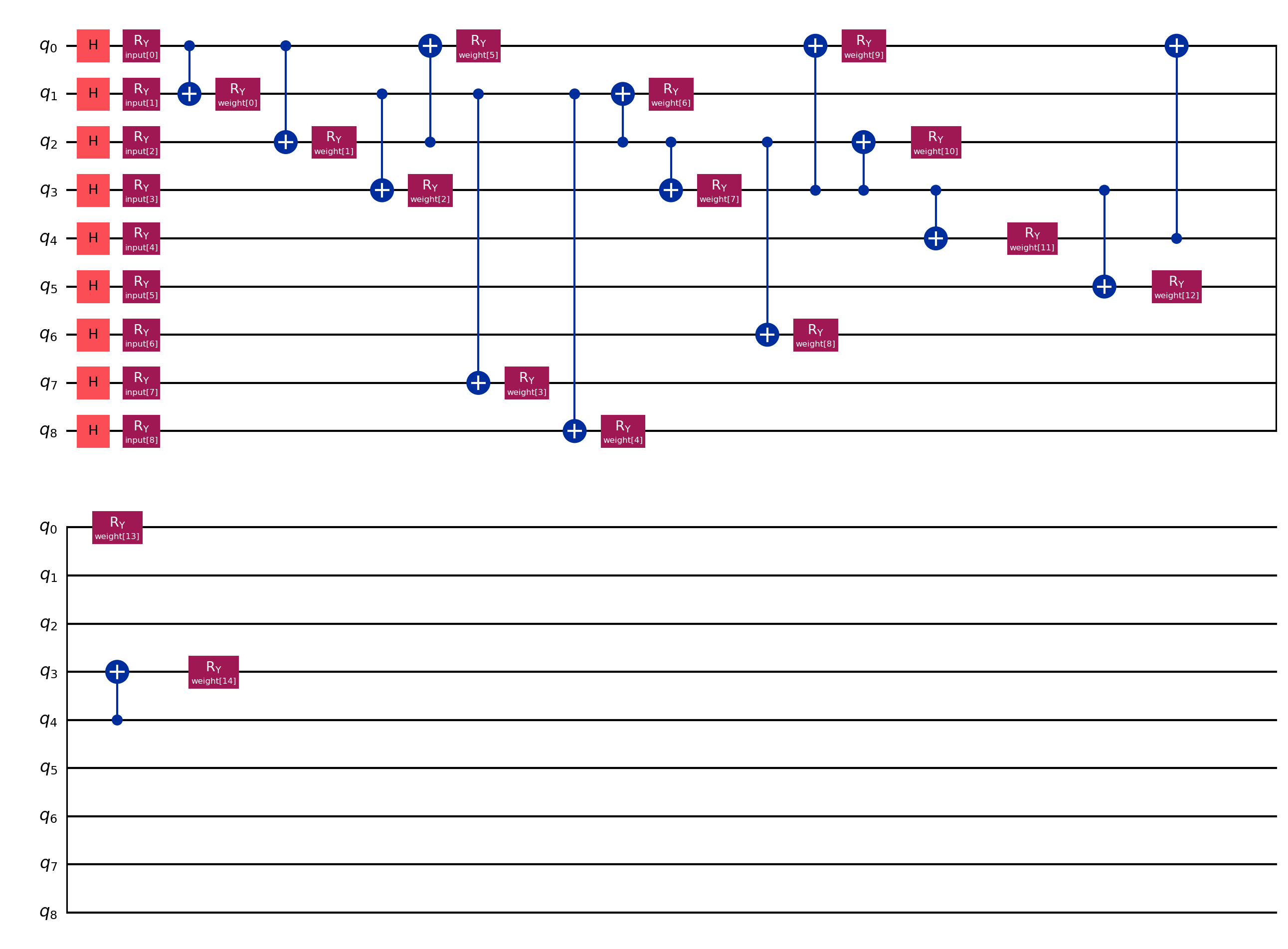}
        \caption{9 qubits}
    \end{subfigure}
    \hfill
    \begin{subfigure}[b]{0.22\textwidth}
        \centering
        \includegraphics[width=\textwidth]{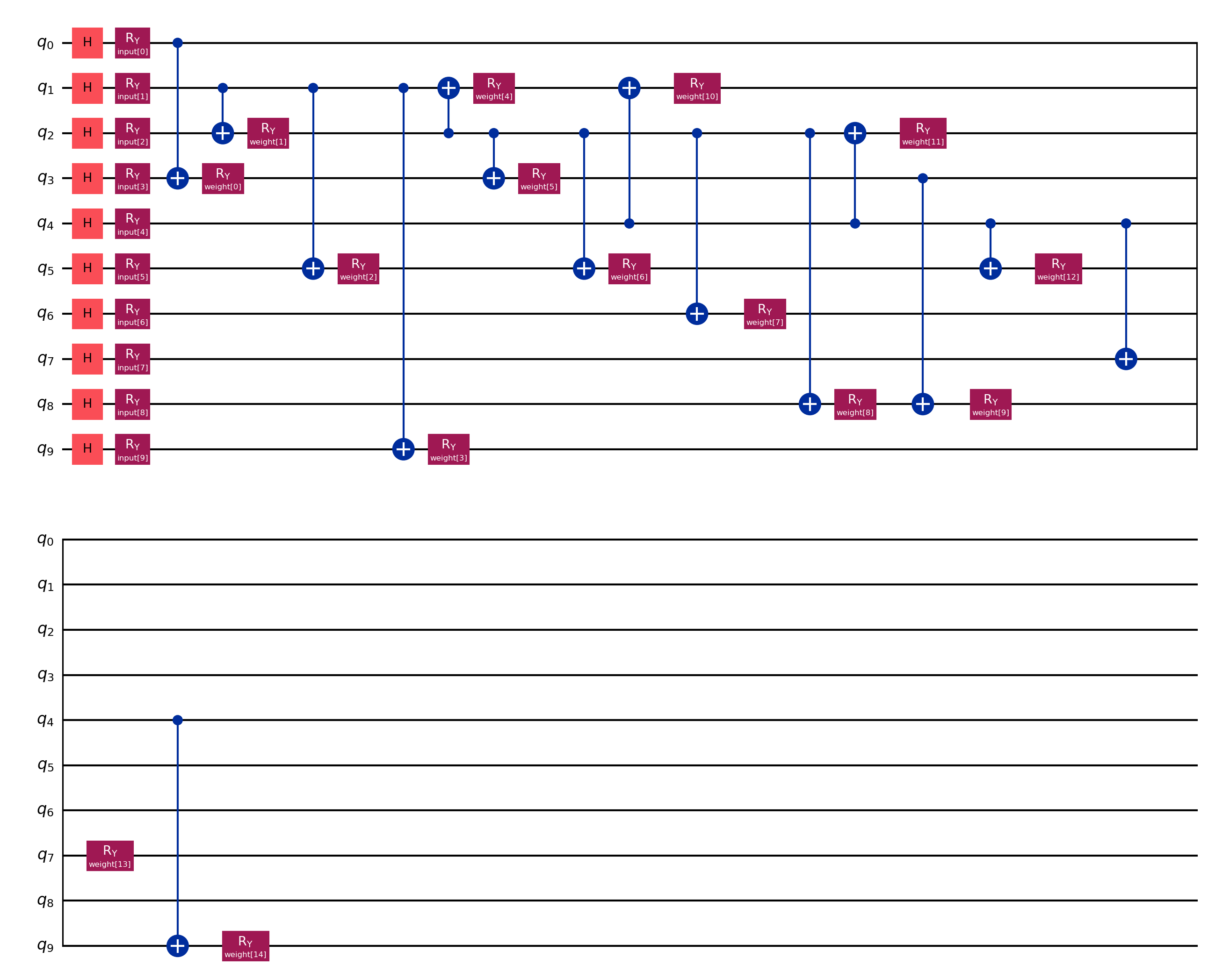}
        \caption{10 qubits}
    \end{subfigure}

    \caption{Optimal Variational Quantum Circuits for Breast cancer dataset}
    \label{fig:main1-breastCancer}
\end{figure}

Table \ref{table_compareAccuracy_cancerdata} benchmarks our genetic transformer-assisted QNN (GTQNN) against the coherent feed-forward QNN (CFFQNN) of \cite{Singh2024} on the Breast-Cancer-Wisconsin diagnostic data.  For every qubit budget between 3 and 10, the evolutionary search yields GTQNN circuits whose accuracies cluster tightly around $0.974$, rising to $0.983$ when ten qubits are available.  Gate counts remain modest—only 10 to 12 primitive gates for the 3- and 4-qubit models and 50 gates at the 10-qubit setting—demonstrating that deeper circuits are not required to sustain high performance.   Table \ref{fig:main1-breastCancer} shows some corresponding optimal variational quantum circuits from GTQNN.

By contrast, the CFFQNN baseline, evaluated at a single architecture $[3,2,1]$ with more than 35 gates after PCA compression to seven features, reports an accuracy of roughly $0.85$.  Thus GTQNN improves classification accuracy by more than ten percentage points while using fewer gates at comparable or smaller qubit counts.  These results highlight the advantage of coupling the transformer front-end and genetic search with shallow quantum layers: the hybrid system achieves state-of-the-art accuracy yet keeps circuit depth low enough for near-term hardware execution.

\subsection{MNIST dataset}\label{sec5-3}

\begin{table}[htbp]
\centering
\resizebox{0.85\textwidth}{!}{
\begin{tabular}{>{\raggedright\arraybackslash}p{3.5cm} c c c c c c c c}
\multirow{2}{*}{Method} & \multicolumn{8}{c}{Number of Qubits} \\
\cline{2-9}
 & 3 & 4 & 5 & 6 & 7 & 8 & 9 & 10 \\
\hline \\[-1em]
\parbox{3.5cm}{\raggedright GTQNN\\[0.5em]}
 &  &  &  &  &  &  &  &  \\[2em]

\parbox{3.5cm}{\scriptsize  Accuracy}
 & 0.973 & 0.98 & 0.9867 & 0.9667 & 0.98 & 0.9933 & 0.9533 & 0.9733\\[0.5em]

\parbox{3.5cm}{\scriptsize  Gate count}
 & 8 & 14& 16 & 26 & 36 & 42 & 50 & 64 \\[1.5em]

\hline\hline
\parbox{3.5cm}{\raggedright QNN with EfficientSU2 \cite{liu2024training}}
 & \multicolumn{8}{>{\raggedright\arraybackslash}p{%
      \dimexpr\linewidth-3.5cm-8\tabcolsep\relax}}{%
      \scriptsize Reported accuracy around 0.85 with 26 layers, 13 qubits, $>$ 728 gates (parameters).}\\[1.5em]
\hline
\end{tabular}}
\caption{Quantum gate count and accuracy comparison for MNIST dataset.}
\label{table_compareAccuracy_MNIST}
\end{table}

\begin{figure}[h!]
    \centering
    \begin{subfigure}[b]{0.22\textwidth}
        \centering
        \includegraphics[width=\textwidth]{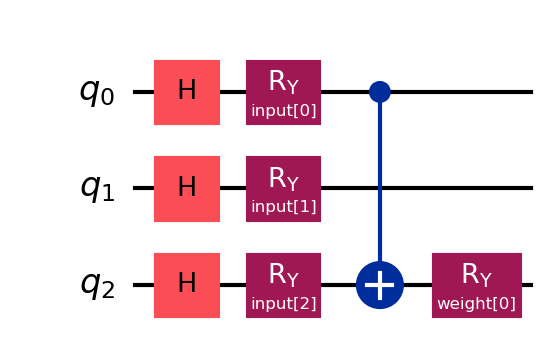}
        \caption{3 qubits}
    \end{subfigure}
    \hfill
    \begin{subfigure}[b]{0.22\textwidth}
        \centering
        \includegraphics[width=\textwidth]{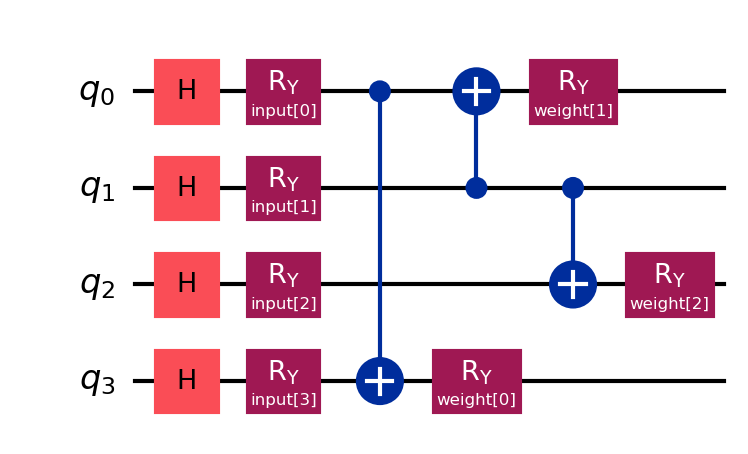}
        \caption{4 qubits}
    \end{subfigure}
    \hfill
    \begin{subfigure}[b]{0.22\textwidth}
        \centering
        \includegraphics[width=\textwidth]{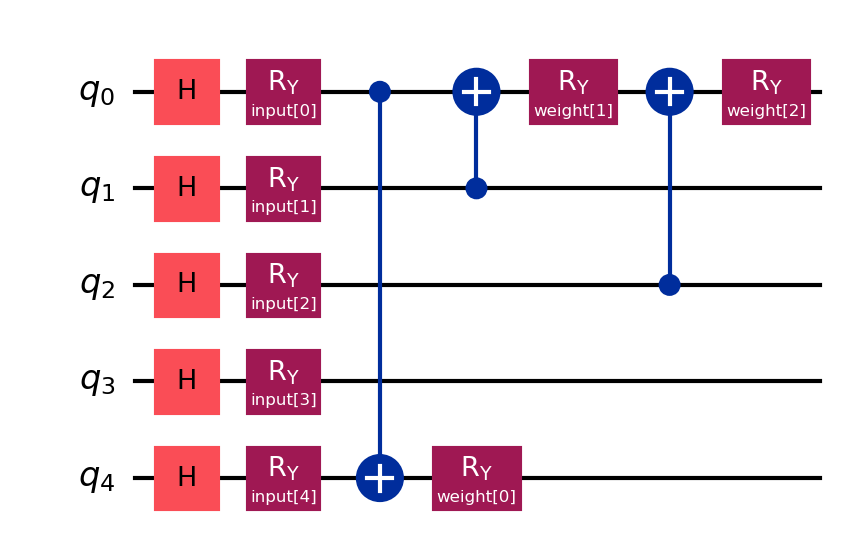}
        \caption{5 qubits}     
    \end{subfigure}
    \hfill
    \begin{subfigure}[b]{0.22\textwidth}
        \centering
        \includegraphics[width=\textwidth]{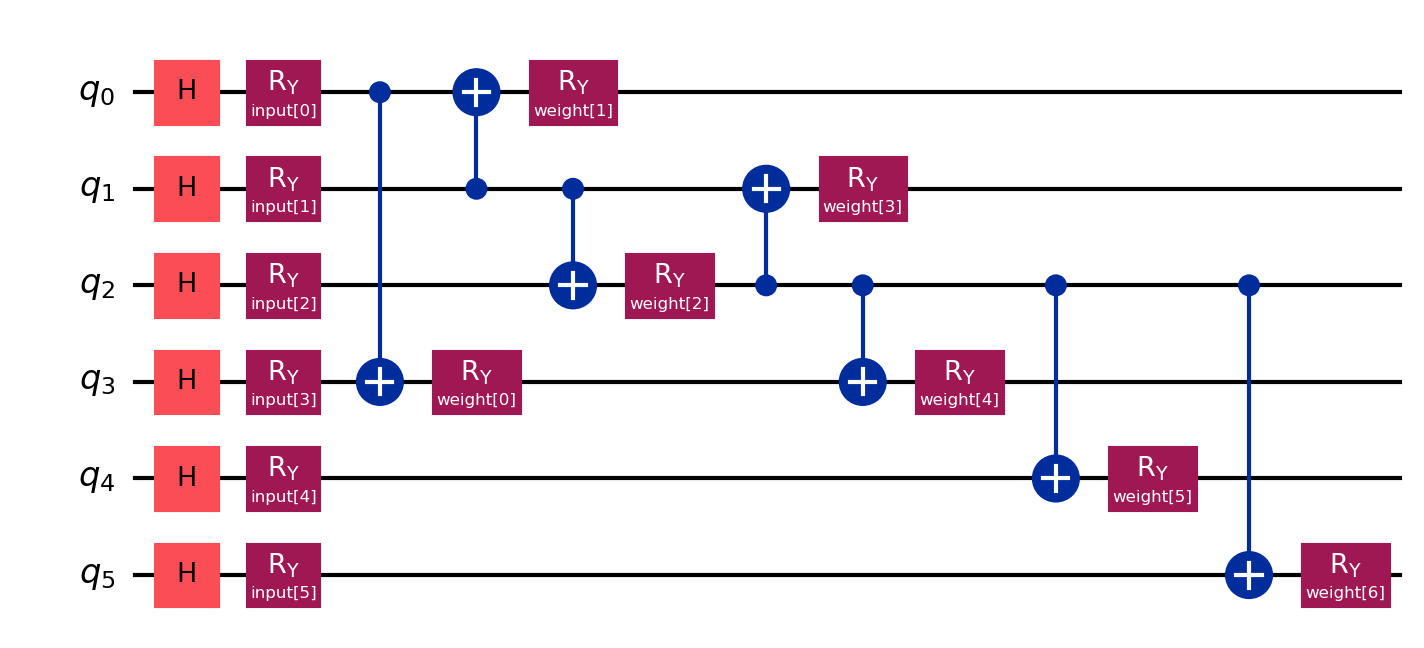}
        \caption{6 qubits} 
    \end{subfigure}

    \vskip\baselineskip

    \begin{subfigure}[b]{0.22\textwidth}
        \centering
        \includegraphics[width=\textwidth]{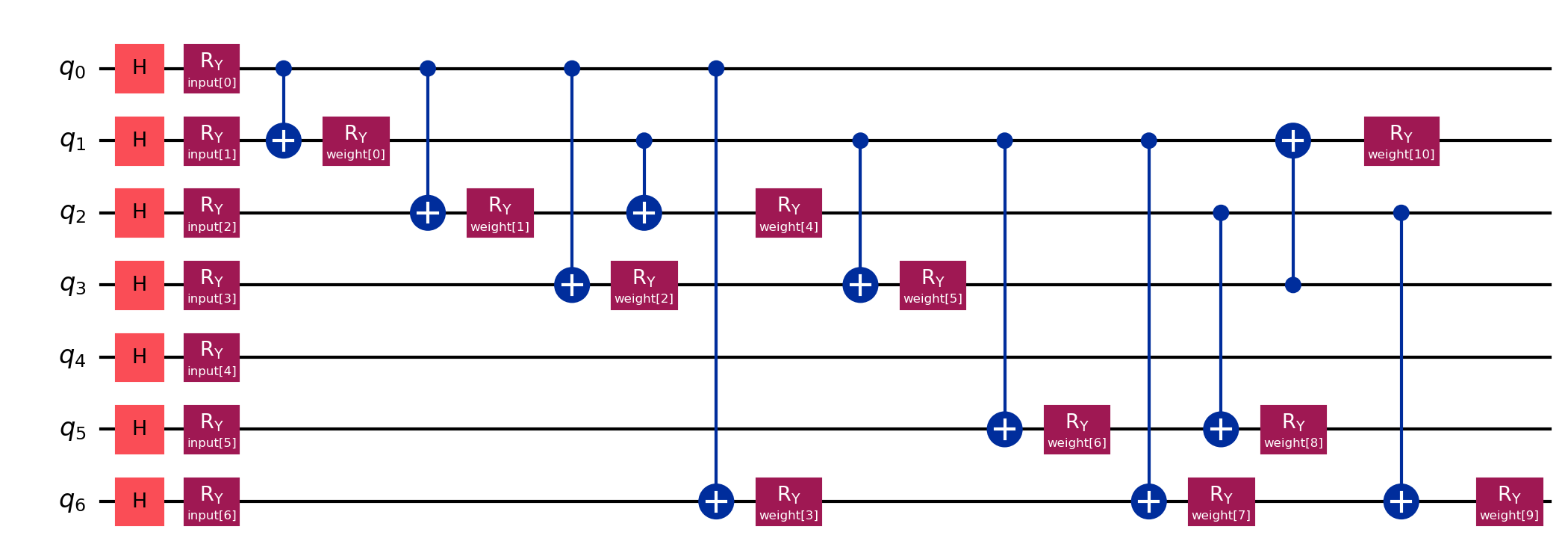}
        \caption{7 qubits} 
    \end{subfigure}
    \hfill
    \begin{subfigure}[b]{0.22\textwidth}
        \centering
        \includegraphics[width=\textwidth]{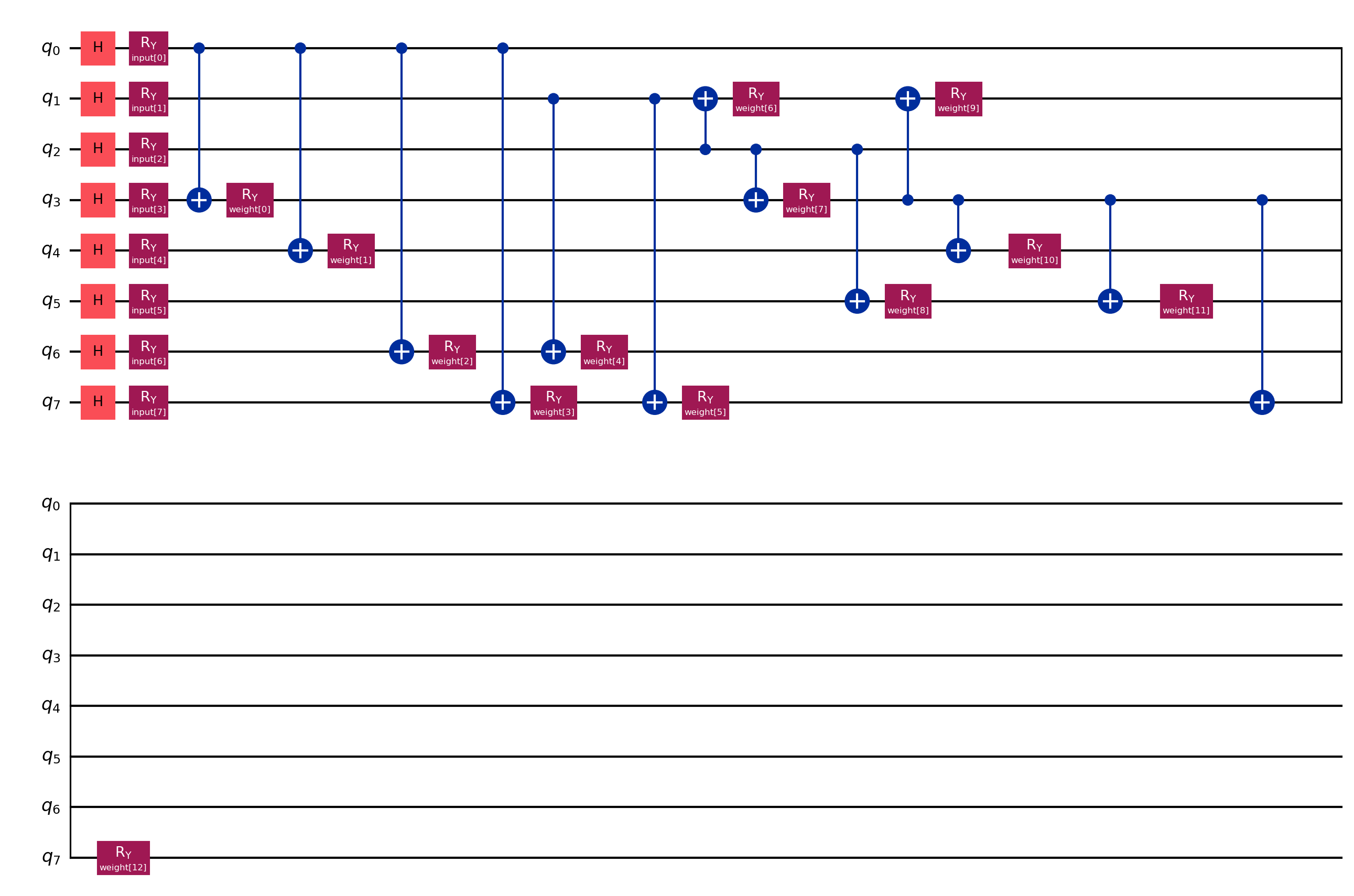}
        \caption{8 qubits}
    \end{subfigure}
    \hfill
    \begin{subfigure}[b]{0.22\textwidth}
        \centering
        \includegraphics[width=\textwidth]{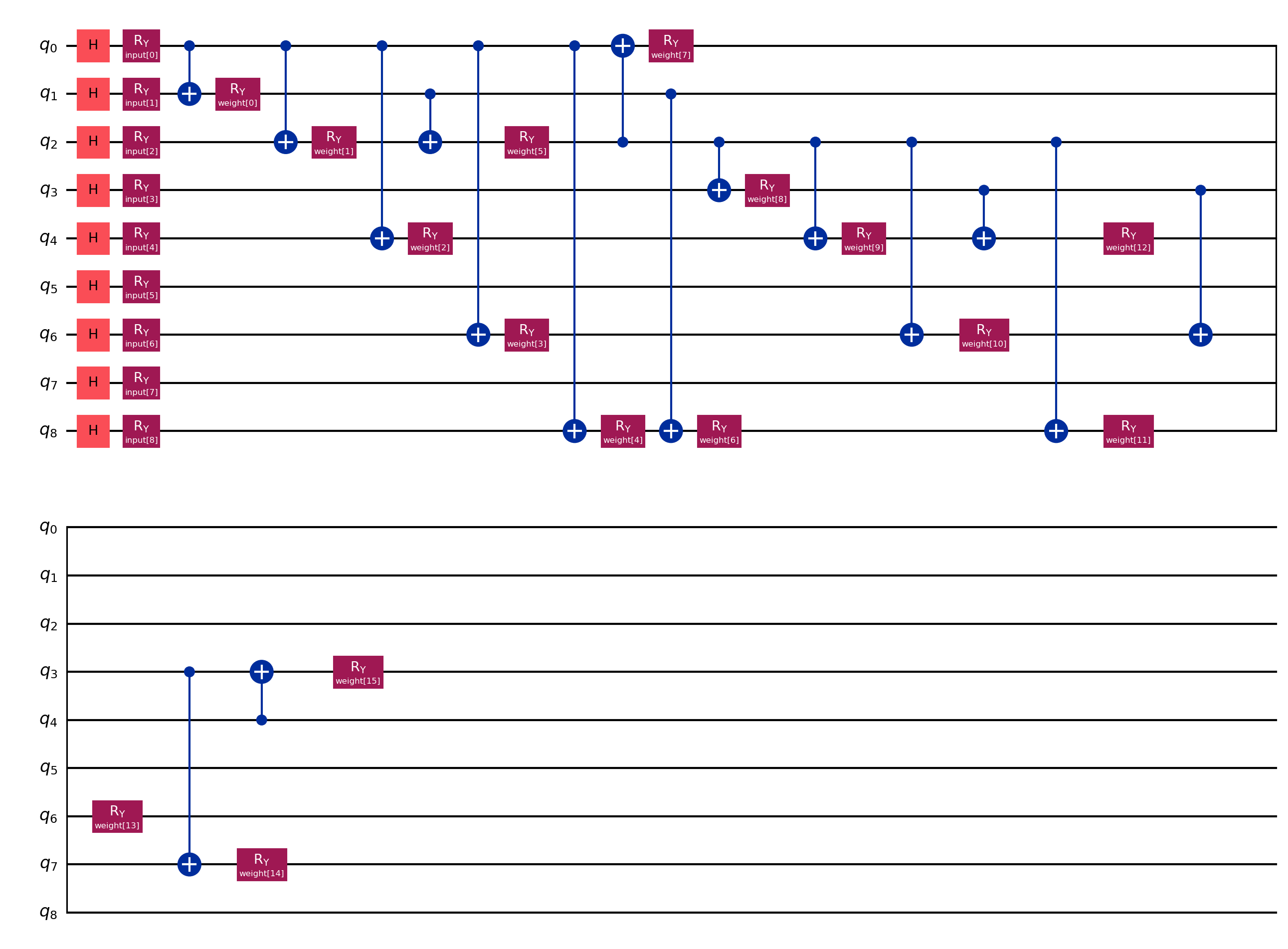}
        \caption{9 qubits}
    \end{subfigure}
    \hfill
    \begin{subfigure}[b]{0.22\textwidth}
        \centering
        \includegraphics[width=\textwidth]{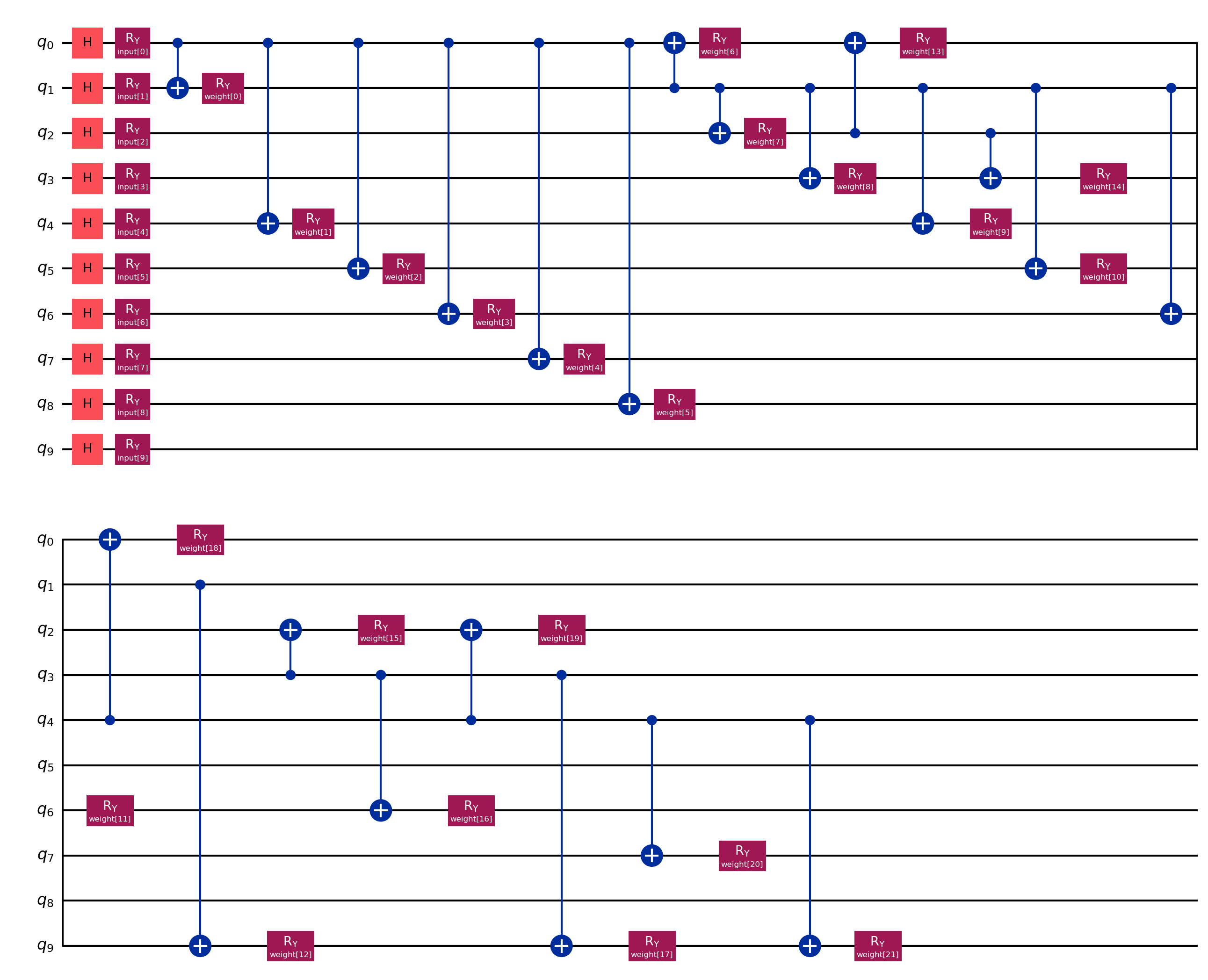}
        \caption{10 qubits}
    \end{subfigure}

    \caption{Optimal Variational Quantum Circuits for MNIST dataset}
    \label{fig:main1-MNIST}
\end{figure}

The MNIST data set contains $60\,000$ $28\times28$ grey-scale images of handwritten digits for training and $10\,000$ images for testing, each labelled $0\!-\!9$. We only limit our classification to the three digits 1,2 and 3. We flatten the pixels to a 784-dimensional vector, pass them through the transformer front-end described in Section \ref{procudure}, and feed the resulting feature vector to the genetic transformer-assisted QNN (GTQNN).  The outer training loop is run for $500$ epochs.
The eight optimal variational circuits are depicted in Fig. \ref{fig:main1-MNIST}; gate counts for each appear in the second row of Table \ref{table_compareAccuracy_MNIST}.

The GTQNN achieves accuracies between $0.973$ (3 qubits, 8 gates) and $0.993$ (8 qubits, 42 gates), staying above $0.95$ for every qubit setting.
Increasing the qubit budget from 3 to 10 roughly quadruples the gate count (8 → 64) yet produces only marginal accuracy gains beyond 8 qubits, indicating that the transformer front-end already extracts a compact, informative representation.
By comparison, QNN with EfficientSU2 \cite{liu2024training} reports about $0.85$ accuracy while using 13 qubits and over $ 700$ gates—an order of magnitude more hardware than any of our GTQNN configurations. Overall, the table shows that the evolutionary GTQNN generates shallower circuits than QNN with EfficientSU2 and attains near-state-of-the-art accuracy across all qubit budgets, demonstrating its suitability for resource-constrained, near-term quantum processors.

\subsection{Heart Disease dataset}\label{sec5-4}
\begin{table}[htbp]
\centering
\resizebox{0.85\textwidth}{!}{
\begin{tabular}{>{\raggedright\arraybackslash}p{3.5cm} c c c c c c c c}
\multirow{2}{*}{Method} & \multicolumn{8}{c}{Number of Qubits} \\
\cline{2-9}
 & 3 & 4 & 5 & 6 & 7 & 8 & 9 & 10 \\
\hline \\[-1em]
\parbox{3.5cm}{\raggedright GTQNN\\[0.5em]}
 &  & &  & &  &  &  &  \\[2em]

\parbox{3.5cm}{\scriptsize  Accuracy}
 & 0.9854 & 0.9854 & 0.9756 & 0.9854 & 0.9854 & 0.9805 & 0.8829 & 0.8488 \\[0.5em]
 
\parbox{3.5cm}{\scriptsize  Gate count}
 & 10 & 18& 24 & 24 & 20 & 40 & 56 & 62 \\[1.5em]
\hline

\hline\hline
\parbox{3.5cm}{\raggedright CFFQNN \cite{Singh2024}}
 & \multicolumn{8}{>{\raggedright\arraybackslash}p{%
      \dimexpr\linewidth-3.5cm-8\tabcolsep\relax}}{%
      \scriptsize Reported accuracies: about 0.78 with a layer structure
of [2,3,1], corresponding to 26 gates (18 CNOT gates). Use PCA to reduce its dimension to 7}\\[1.5em]
\hline

\end{tabular}}
\caption{Quantum gate count and accuracy comparison for Heart Disease dataset}
\label{table_compareAccuracy_heart}
\end{table}

\begin{figure}[h!]
    \centering
    \begin{subfigure}[b]{0.22\textwidth}
        \centering
        \includegraphics[width=\textwidth]{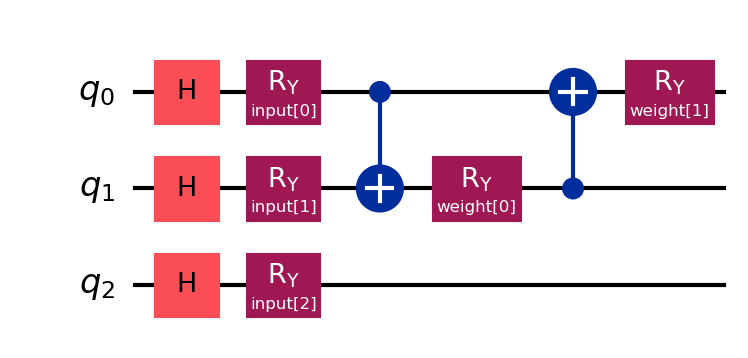}
        \caption{3 qubits}
    \end{subfigure}
    \hfill
    \begin{subfigure}[b]{0.22\textwidth}
        \centering
        \includegraphics[width=\textwidth]{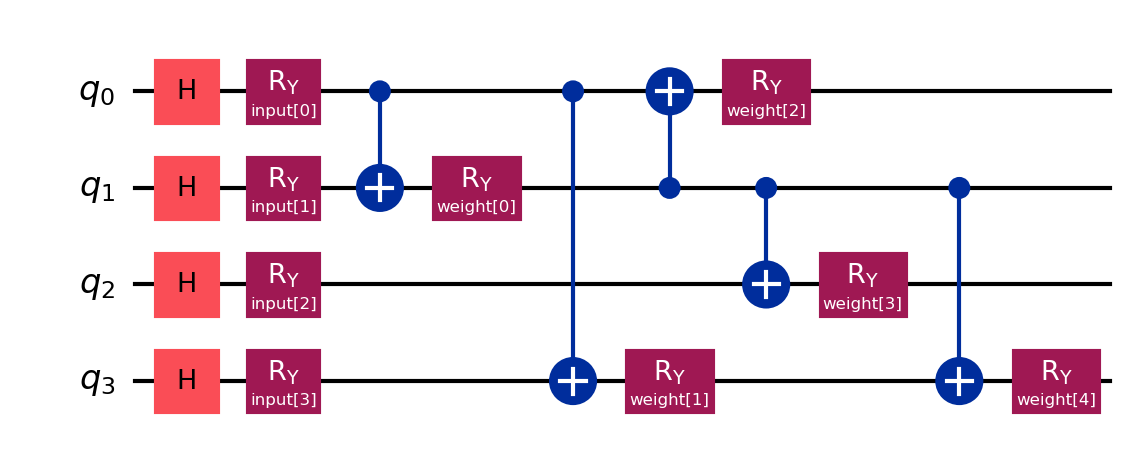}
        \caption{4 qubits}
    \end{subfigure}
    \hfill
    \begin{subfigure}[b]{0.22\textwidth}
        \centering
        \includegraphics[width=\textwidth]{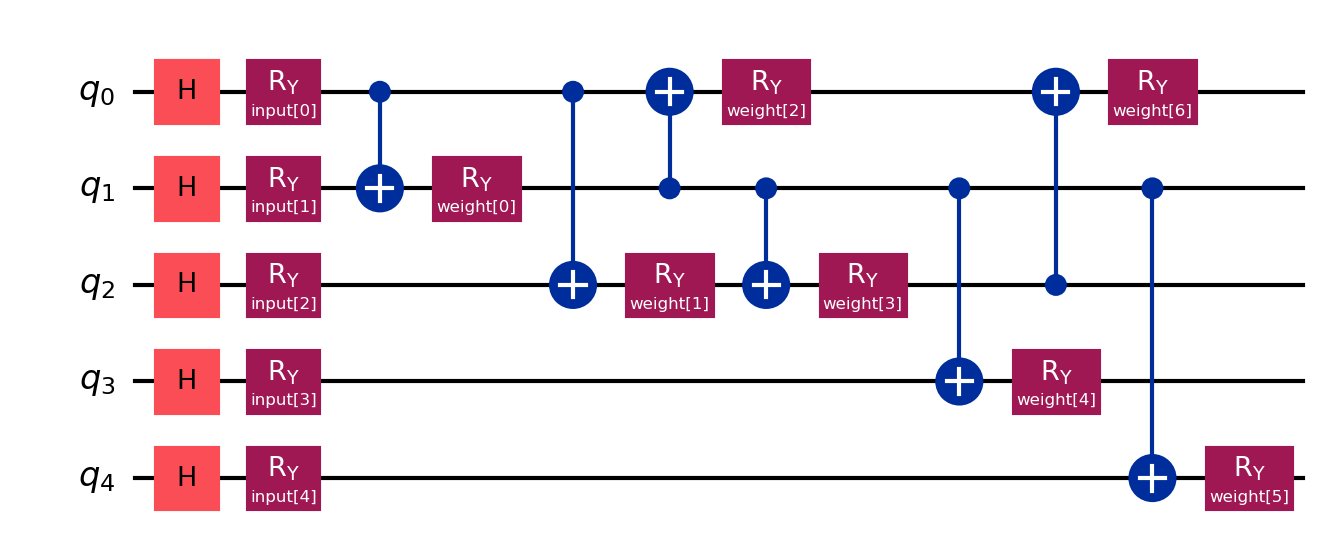}
        \caption{5 qubits}     
    \end{subfigure}
    \hfill
    \begin{subfigure}[b]{0.22\textwidth}
        \centering
        \includegraphics[width=\textwidth]{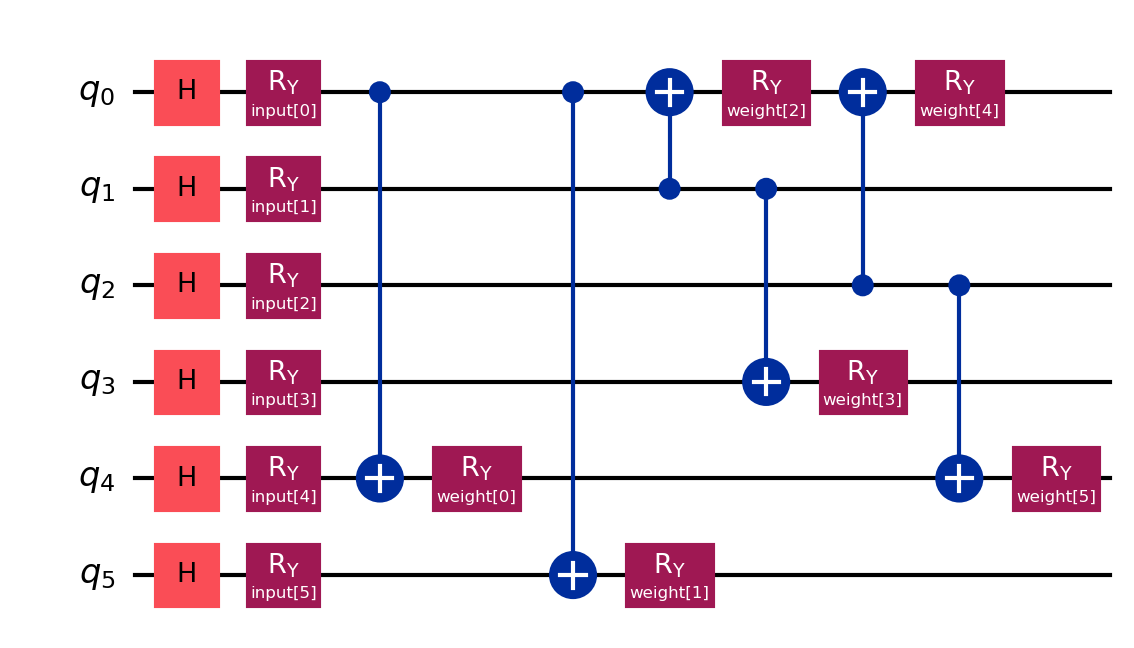}
        \caption{6 qubits} 
    \end{subfigure}

    \vskip\baselineskip

    \begin{subfigure}[b]{0.22\textwidth}
        \centering
        \includegraphics[width=\textwidth]{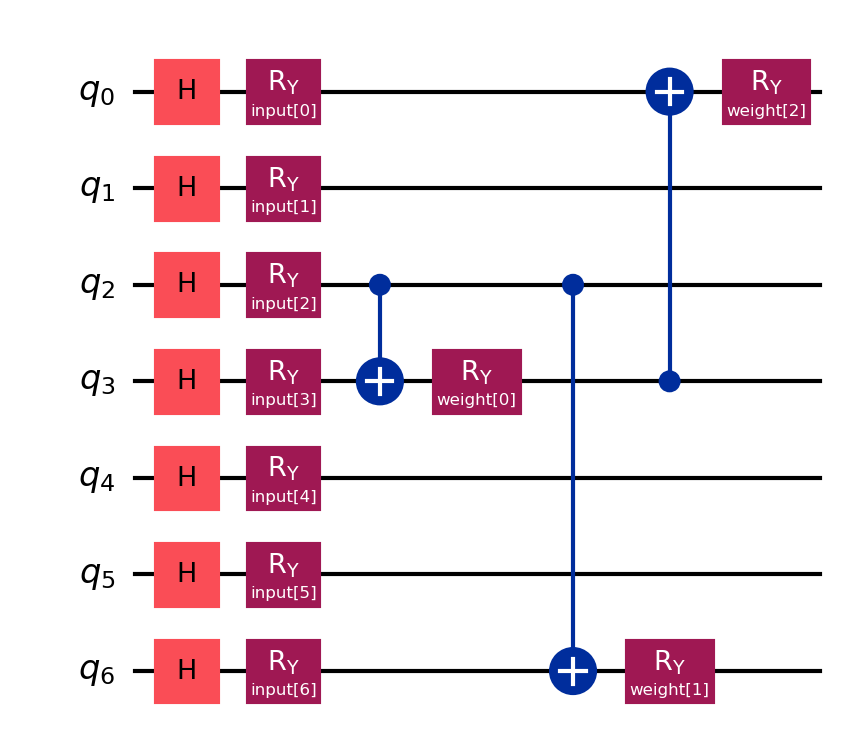}
        \caption{7 qubits} 
    \end{subfigure}
    \hfill
    \begin{subfigure}[b]{0.22\textwidth}
        \centering
        \includegraphics[width=\textwidth]{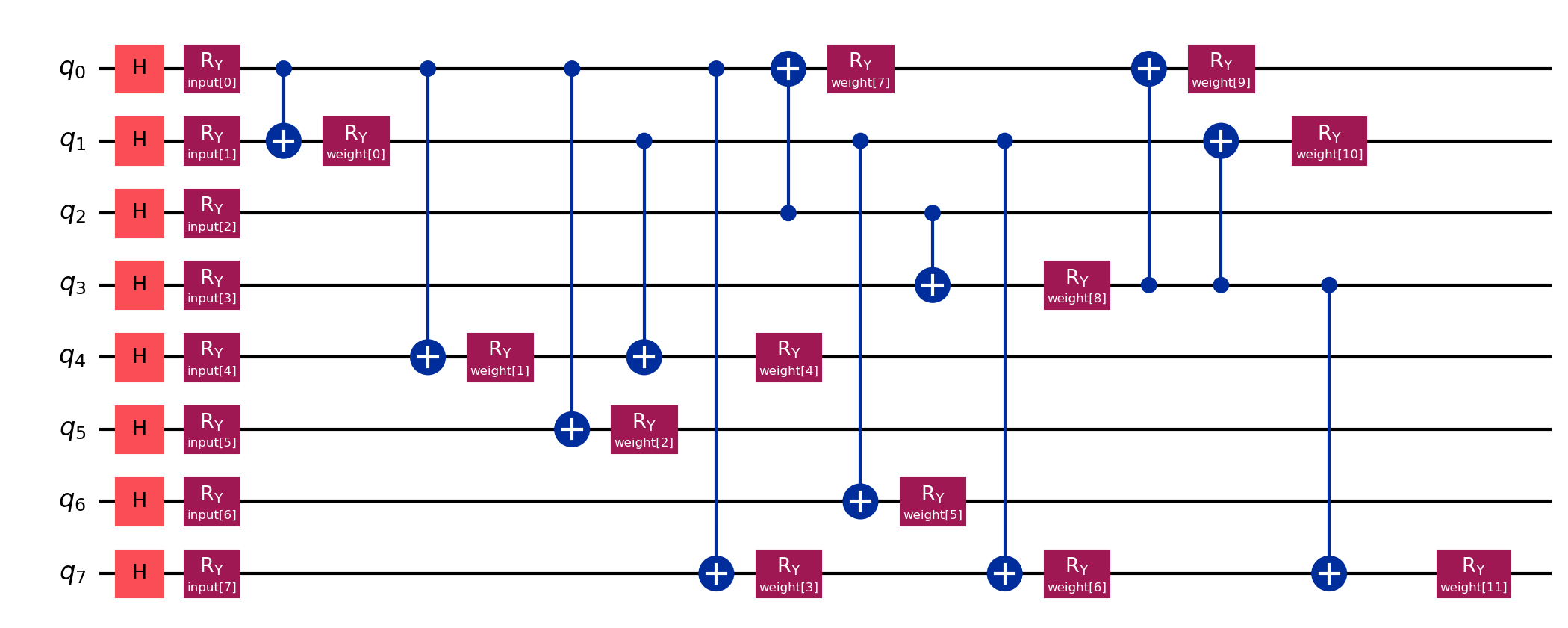}
        \caption{8 qubits}
    \end{subfigure}
    \hfill
    \begin{subfigure}[b]{0.22\textwidth}
        \centering
        \includegraphics[width=\textwidth]{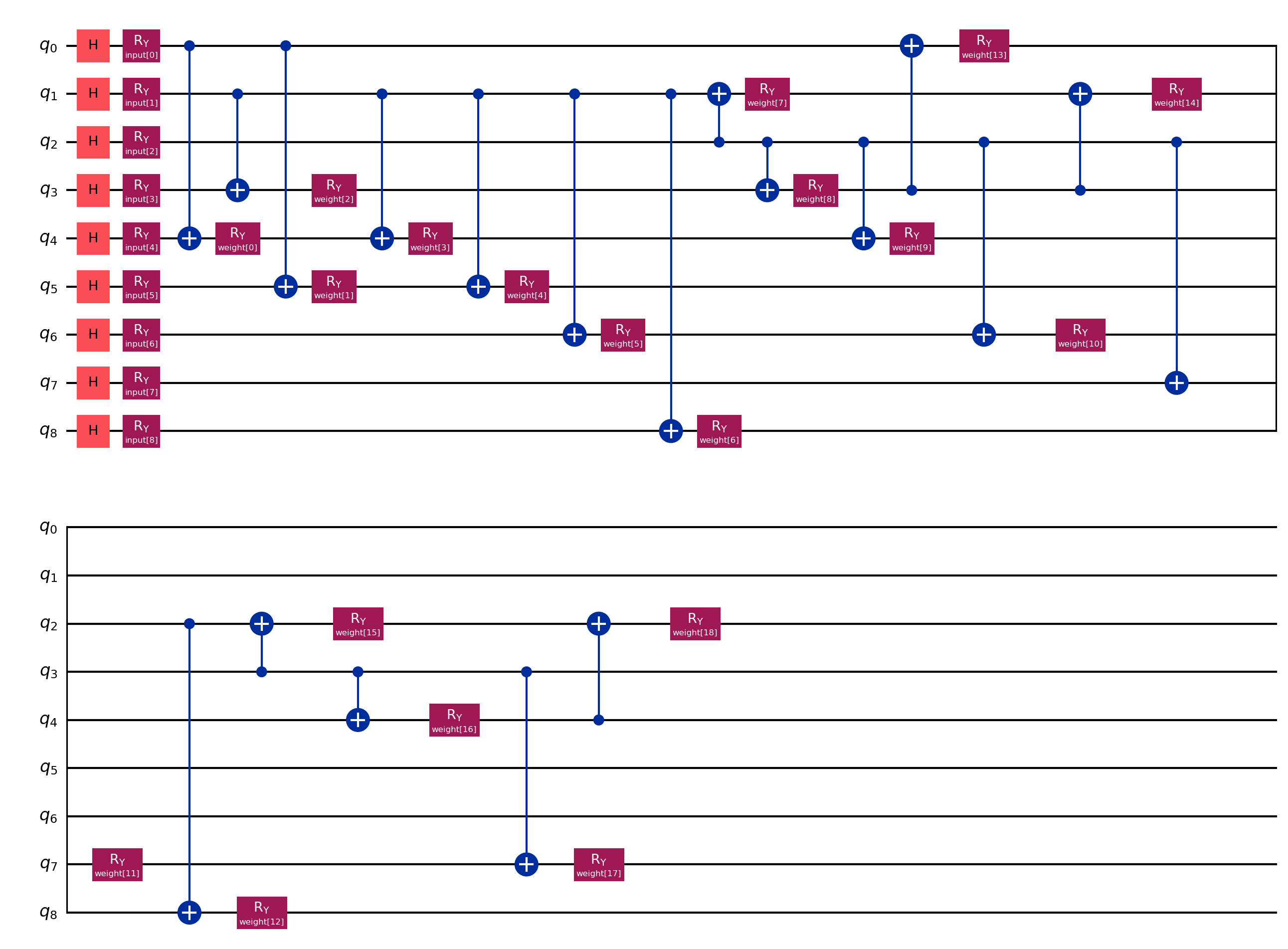}
        \caption{9 qubits}
    \end{subfigure}
    \hfill
    \begin{subfigure}[b]{0.22\textwidth}
        \centering
        \includegraphics[width=\textwidth]{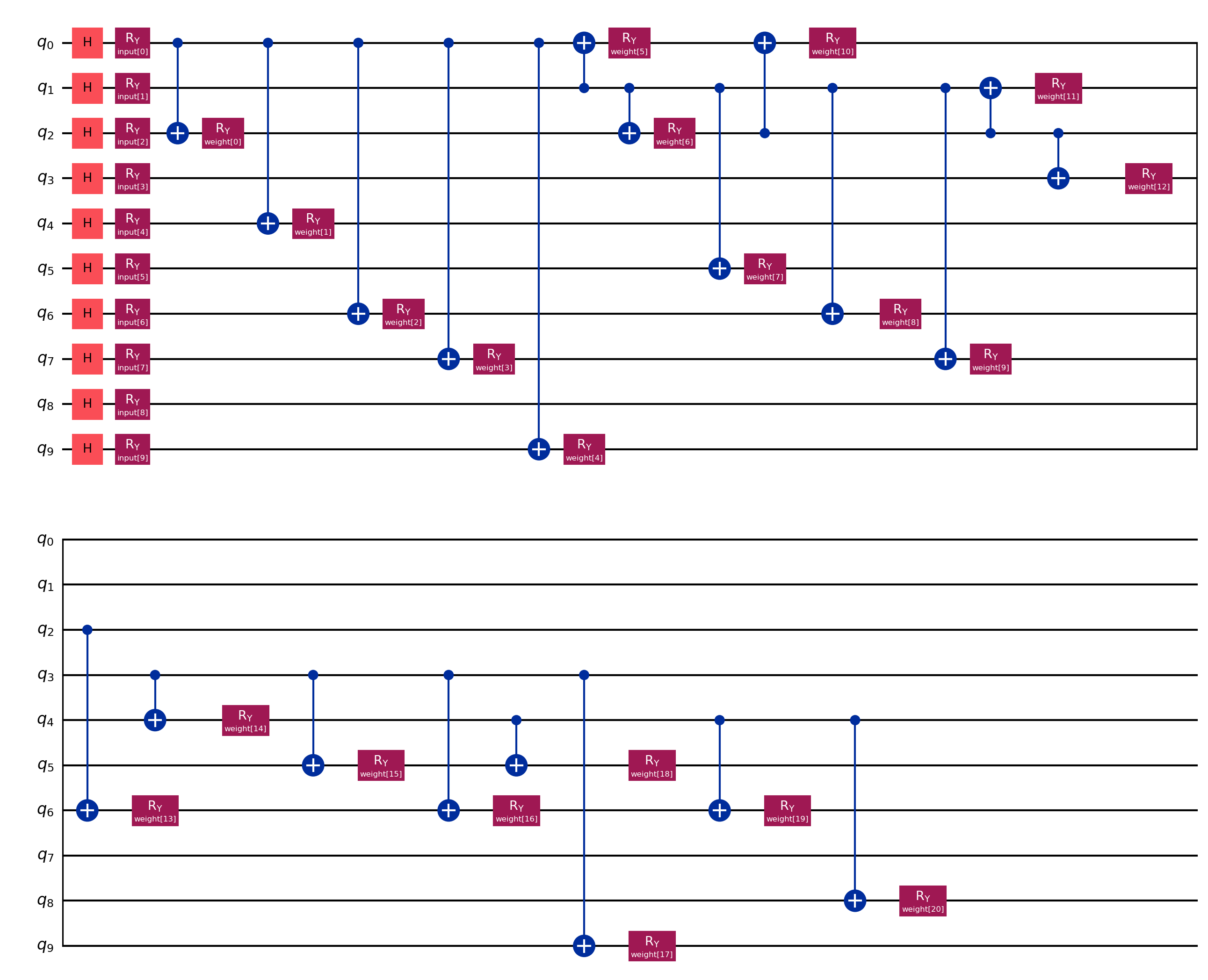}
        \caption{10 qubits}
    \end{subfigure}

    \caption{Optimal Variational Quantum Circuits for Heart Diseast dataset}
    \label{fig:main1-Heart}
\end{figure}

The Heart-Disease data set (Kaggle, LAPP 2024) comprises 918 patient records, each described by 13 clinical attributes such as age, resting-blood pressure, serum cholesterol, and exercise-induced angina.  The binary target indicates the presence or absence of heart disease.  Following the workflow of Section \ref{procudure}, the transformer front-end encodes the 13 features and feeds them to the genetic transformer-assisted QNN (GTQNN). The best circuit per generation is then retrained for 400 outer epochs, and its test accuracy is reported.  Gate counts for the selected circuits appear in the second row of Table \ref{table_compareAccuracy_heart}.  The eight optimal variational circuits are depicted in Fig.\ref{fig:main1-Heart};

With as few as 3 qubits (10 gates) the GTQNN already reaches an accuracy of 0.985, matching its best score at 4, 6, and 7 qubits while using no more than 24 gates.  Accuracy remains above 0.98 until the 8-qubit setting, after which it drops as the circuit grows deeper (56–62 gates).  In comparison, the coherent feed-forward QNN (CFFQNN) of \cite{Singh2024}, evaluated at a fixed $[2,3,1]$ architecture with 42 gates (18 CNOTs) after PCA reduction to seven features, reports only 0.78 accuracy more than 20\% points below every GTQNN configuration up to 8 qubits.  The table therefore shows that the evolutionary GTQNN achieves near-perfect classification with significantly shallower circuits, underscoring its advantage for resource-constrained quantum hardware.


\section{Fisher spectrum analysis}\label{sec_inf_geom}
In this section, we approach the notion of the contribution of contribution (``energy split") between transformer and QNN from an information geometry perspective. We first define measures that apply to both classical and quantum models, and subsequently use them to study the contributions of Transformer and QNN in GTQNN.

\subsection{The Fisher information}\label{sec_fisher_info}
The Fisher information is a central notion in a wide range of disciplines, from statistical physics to computational neuroscience \cite{frieden_2004,rissanen1996fisher}.  When we cast a neural network as a parametric statistical model, Fisher information quantifies how much insight a particular parameter vector \(\theta\) provides.  Writing the joint density of data pairs as, for $\theta\in\Theta\subset[-1,1]^d$ 
\begin{equation}\label{eq_emp_fisher1}
p(x,y;\theta)=p(y\mid x;\theta)\,p(x),
\qquad
x\in\mathcal X\subset\mathbb R^{s_{\mathrm{in}}},\;
y\in\mathcal Y\subset\mathbb R^{s_{\mathrm{out}}},
\end{equation}
we obtain probabilities by a post-processing step that depends on the network type:
soft-max for classical models, parity mapping for quantum models.  
Here \(p(x)\) is a fixed prior, whereas \(p(y\mid x;\theta)\) captures the model’s input–output relation for a given \(\theta\). The full parameter manifold \(\Theta\) becomes Riemannian under the Fisher metric.  
The corresponding matrix  
\begin{equation}\label{eq_emp_fisher2}
F(\theta)=
\operatorname E_{(x,y)\sim p}\!
\Bigl[
  \nabla_\theta\log p(x,y;\theta)\,
  \nabla_\theta\log p(x,y;\theta)^{\!\top}
\Bigr]
\in\mathbb R^{d\times d}
\end{equation}
is positive semidefinite, so all of its eigenvalues are real and non-negative.  
In practice we estimate it with the \emph{empirical} Fisher

\begin{equation}\label{eq_emp_fisher}
\tilde F_k(\theta)=
\frac1k\sum_{j=1}^{k}
  \nabla_\theta\log p(x_j,y_j;\theta)\,
  \nabla_\theta\log p(x_j,y_j;\theta)^{\!\top},
\end{equation}
where \(\{(x_j,y_j)\}_{j=1}^{k}\) are i.i.d. samples from the same joint
distribution \(p(x,y;\theta)\) \cite{abbas2021power,kunstner2019limitations}.  
Because these samples are drawn from the true model, the approximation is
consistent: \(\displaystyle\lim_{k\to\infty}\tilde F_k(\theta)=F(\theta)\)
\cite{kunstner2019limitations}. This is ensured in our numerical analysis by design. By the above definition, it follow that the Fisher information matrix is positive semidefinite and therefore, has non-negative, real numbers as its eigenvalues. 

The Fisher information conveniently measures how sensitive a network’s outputs are to movements in parameter space, making it the natural metric for natural-gradient optimization—which updates along directions that most efficiently lower the loss \cite{amari1998natural,abbas2021power}. In the numerical experiment we identified the full empirical Fisher parameter in the GTQNN as two disjoint blocks, \(\mathcal T\) (Transformer) and \(\mathcal Q\) (QNN). We diagonalize \(\hat F\), so the
eigen-pairs \((\lambda_k,u_k)\) already contain the effect of those
off-diagonal blocks.  The Fisher eigen-pair satisfies
\(F\,u=\lambda u\).
Write \(u=(u_{\mathcal T},u_{\mathcal Q})^{\!\top}\) and partition
\(F\) accordingly.  Neglecting small off-diagonal blocks,

\begin{equation}\label{eq_emp_fisher4}
  \lambda
  \; \approx\;
  u_{\mathcal T}^{\!\top}F_{\mathcal TT}u_{\mathcal T}
  \;+\;
  u_{\mathcal Q}^{\!\top}F_{\mathcal QQ}u_{\mathcal Q},
\end{equation}
where $F_{\mathcal TT}$ and $F_{\mathcal QQ}$ can be viewed as the Fisher within transformer and QNN. The squared components of \(u\) indeed quantify how much the
transformer versus the QNN moves when one steps along the
eigen-direction.  Given a normalized Fisher eigen-vector
\(u\in\mathbb R^{D}\), the quantity
\(\displaystyle 
    \sum_{j\in\mathcal B} u_j^{2}
\)
is the fraction of the vector’s \emph{energy} that lives in a block
\(\mathcal B\subset\{1,\dots,D\}\).
With two disjoint blocks, \(\mathcal T\) (Transformer) and \(\mathcal Q\) (QNN),

\begin{equation}\label{eq_emp_fisher5}
 1
 \;=\;
 \|u\|_2^{2}
 \;\approx\;
 \underbrace{\sum_{j\in\mathcal T} u_j^{2}}_{\text{Transformer share}}
 +\;
 \underbrace{\sum_{j\in\mathcal Q} u_j^{2}}_{\text{QNN share}},
\end{equation}
so those two sums \emph{directly} give the percentages printed in the
diagnostic figures \ref{fig:main1FisherIris} and \ref{fig:main1FisherMNIST}. 

\subsection{Experiments of Fisher matrices}

\begin{figure}[h!]
    \centering
    \begin{subfigure}[b]{0.22\textwidth}
        \centering
        \includegraphics[width=\textwidth]{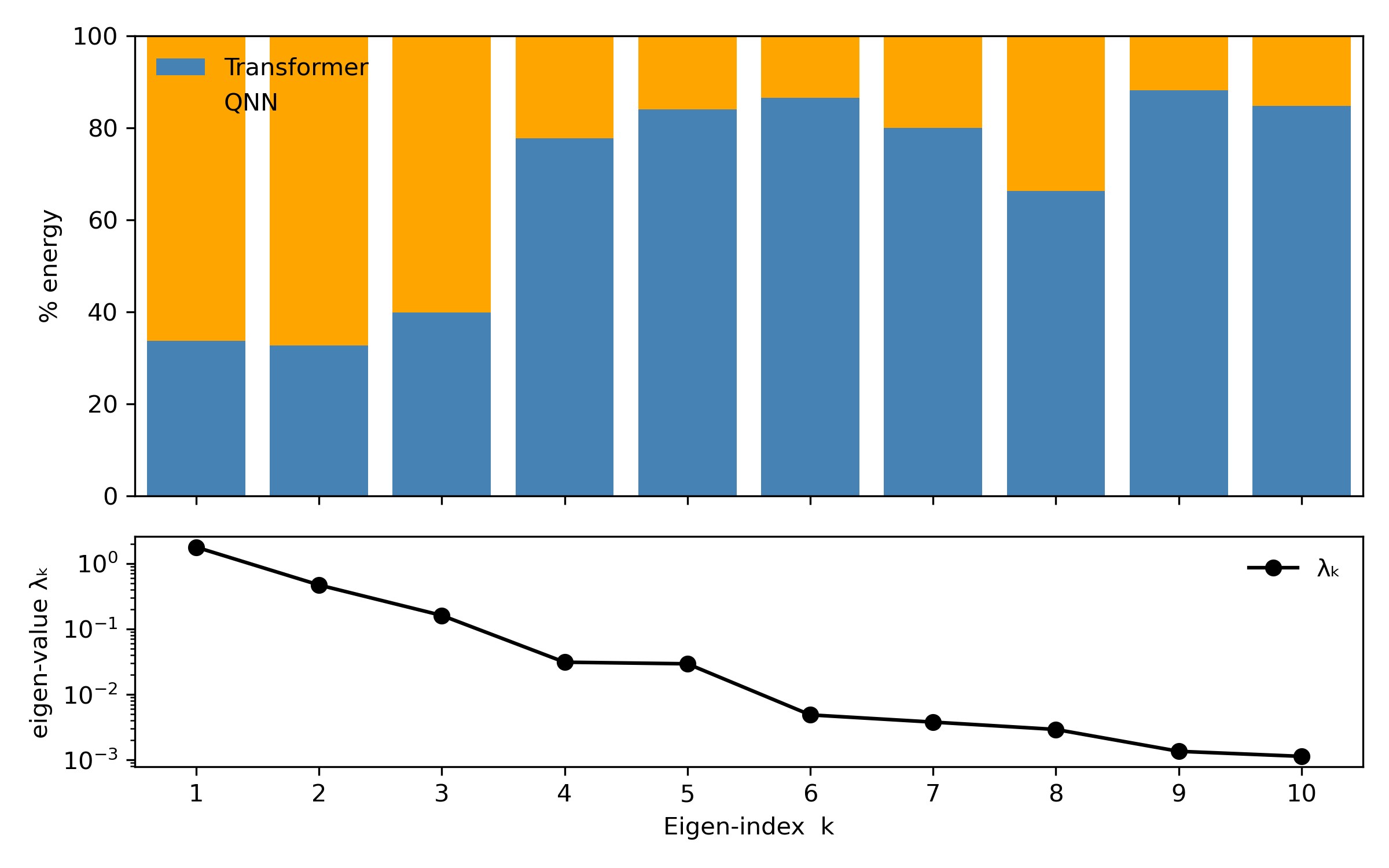}
        \caption{3 qubits}
    \end{subfigure}
    \hfill
    \begin{subfigure}[b]{0.22\textwidth}
        \centering
        \includegraphics[width=\textwidth]{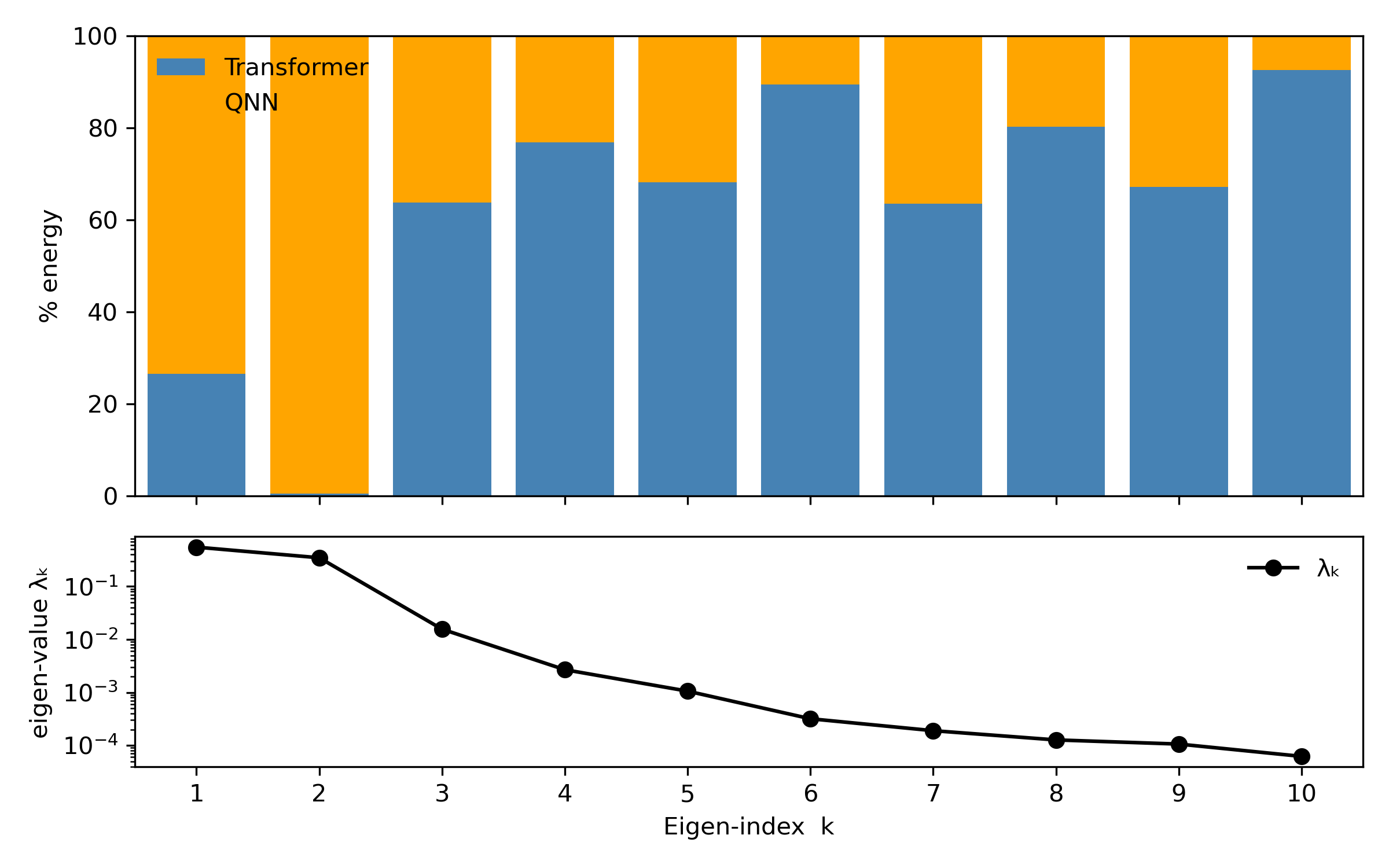}
        \caption{4 qubits}
    \end{subfigure}
    \hfill
    \begin{subfigure}[b]{0.22\textwidth}
        \centering
        \includegraphics[width=\textwidth]{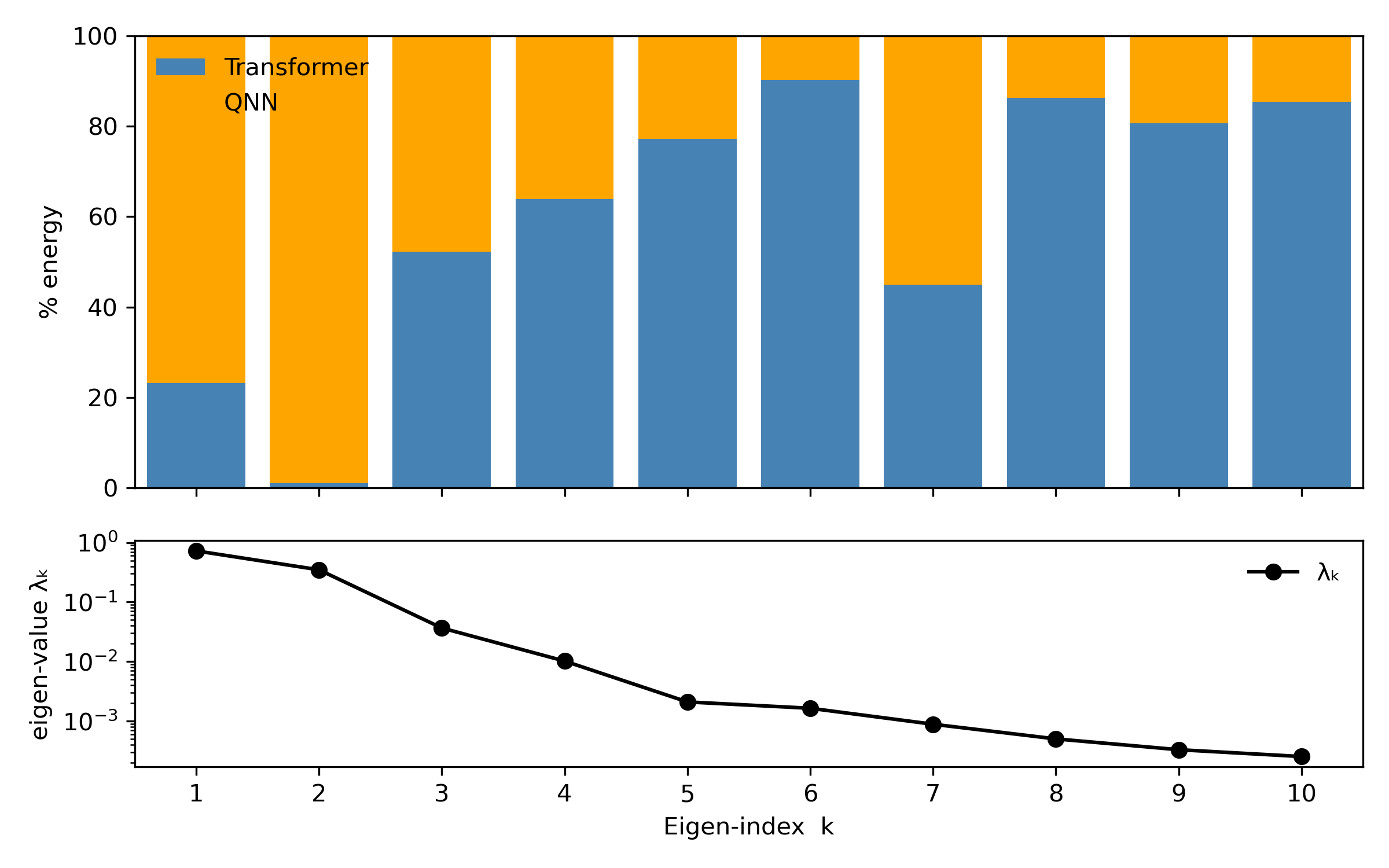}
        \caption{5 qubits}     
    \end{subfigure}
    \hfill
    \begin{subfigure}[b]{0.22\textwidth}
        \centering
        \includegraphics[width=\textwidth]{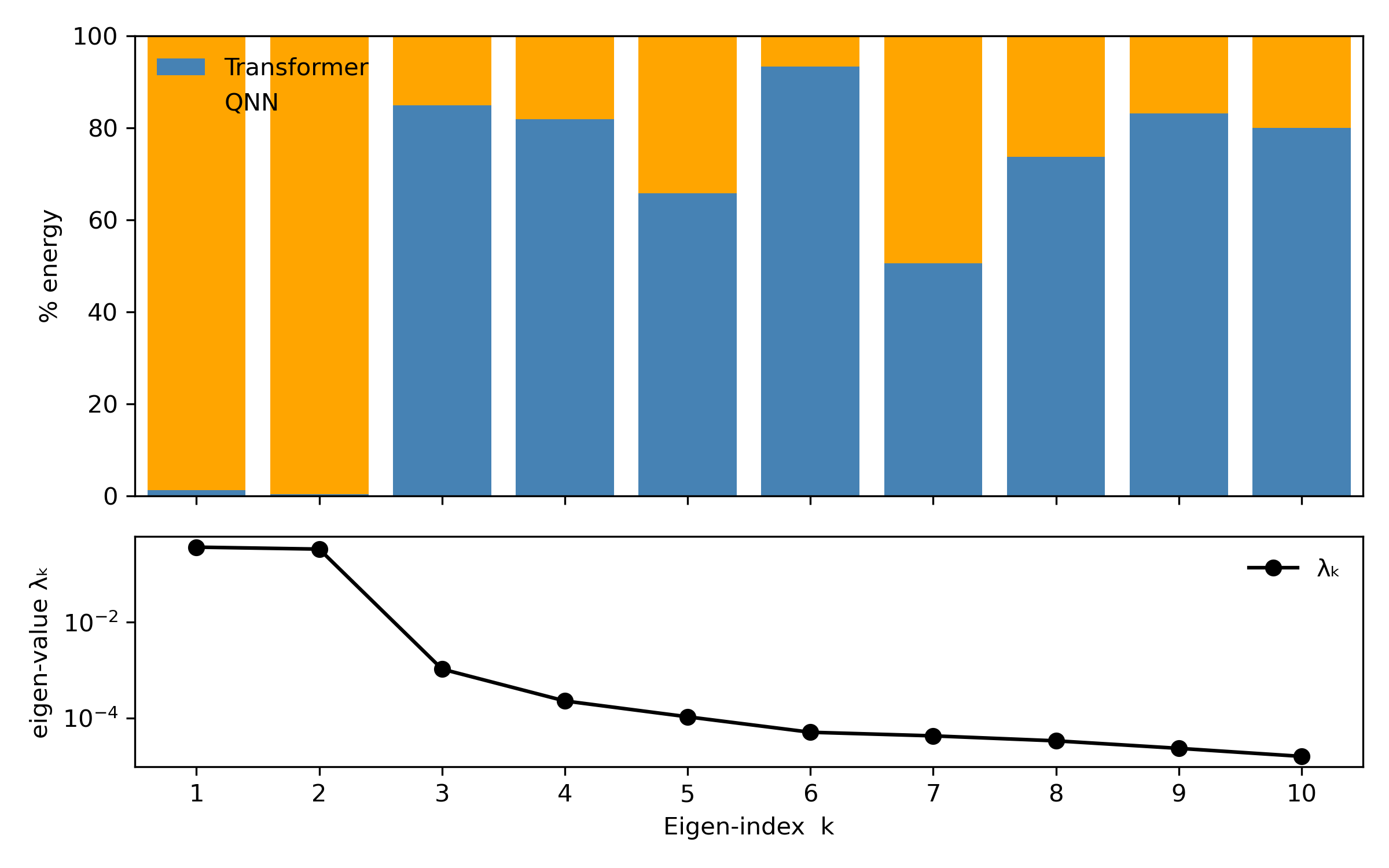}
        \caption{6 qubits} 
    \end{subfigure}

    \vskip\baselineskip

    \begin{subfigure}[b]{0.22\textwidth}
        \centering
        \includegraphics[width=\textwidth]{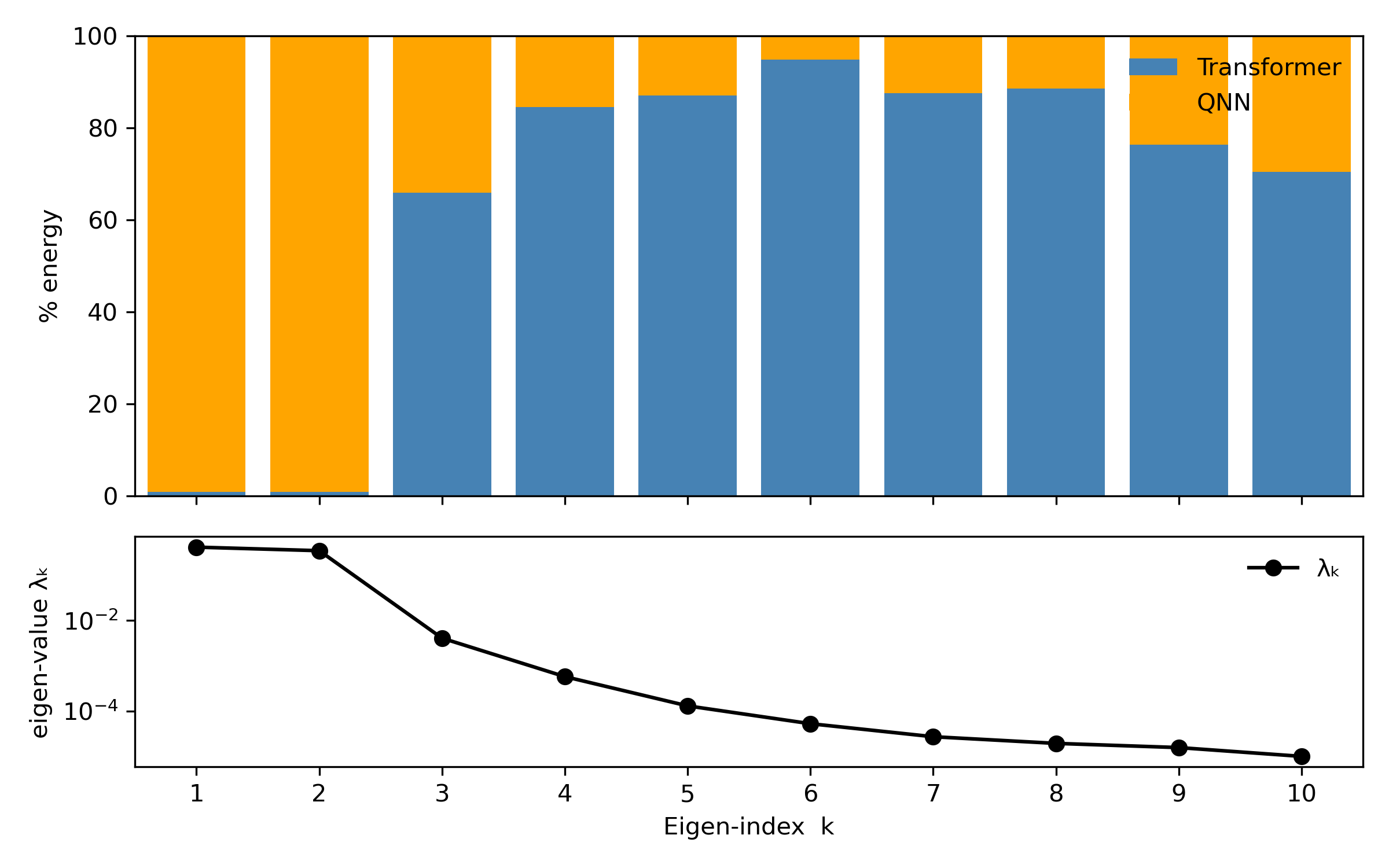}
        \caption{7 qubits} 
    \end{subfigure}
    \hfill
    \begin{subfigure}[b]{0.22\textwidth}
        \centering
        \includegraphics[width=\textwidth]{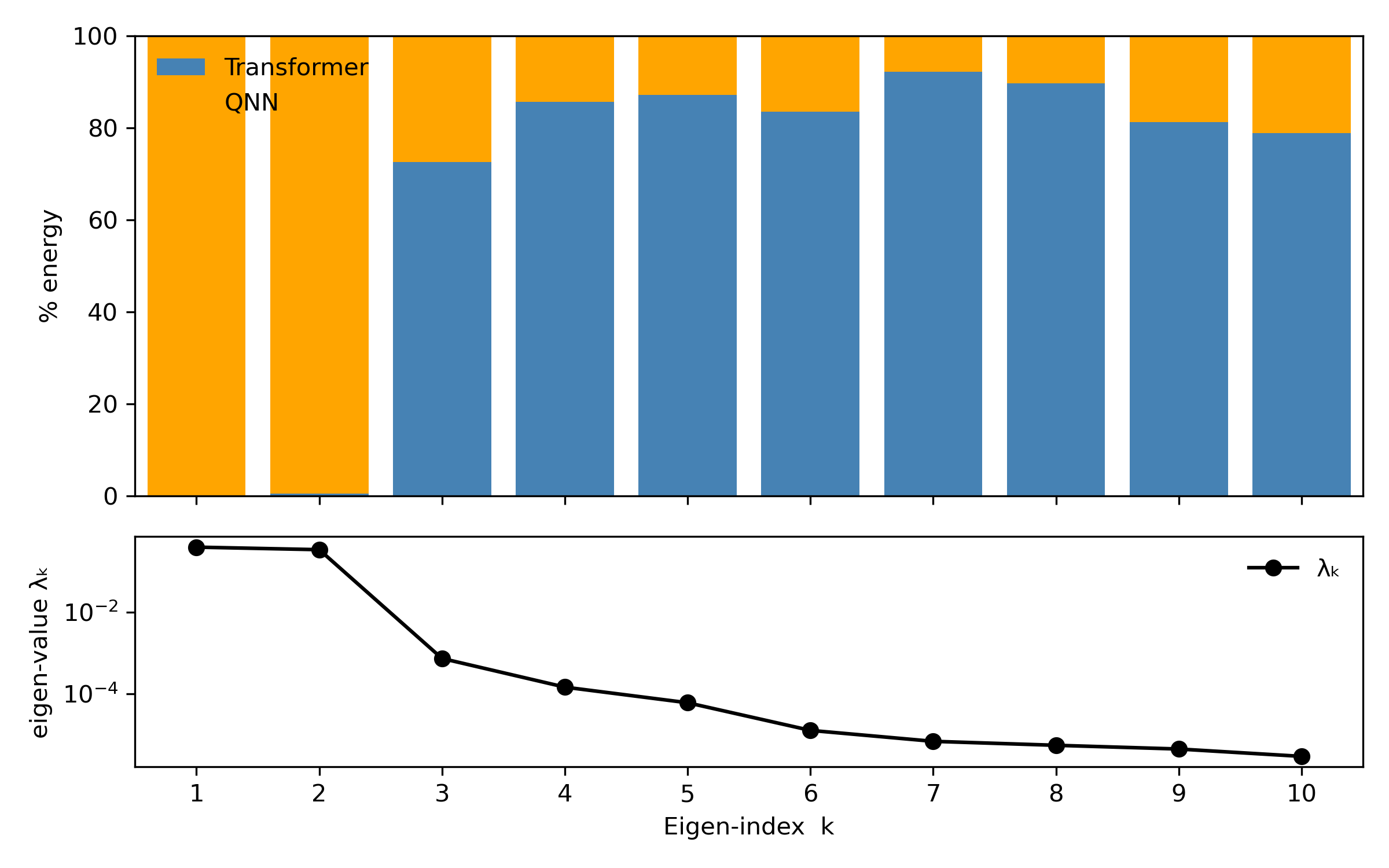}
        \caption{8 qubits}
    \end{subfigure}
    \hfill
    \begin{subfigure}[b]{0.22\textwidth}
        \centering
        \includegraphics[width=\textwidth]{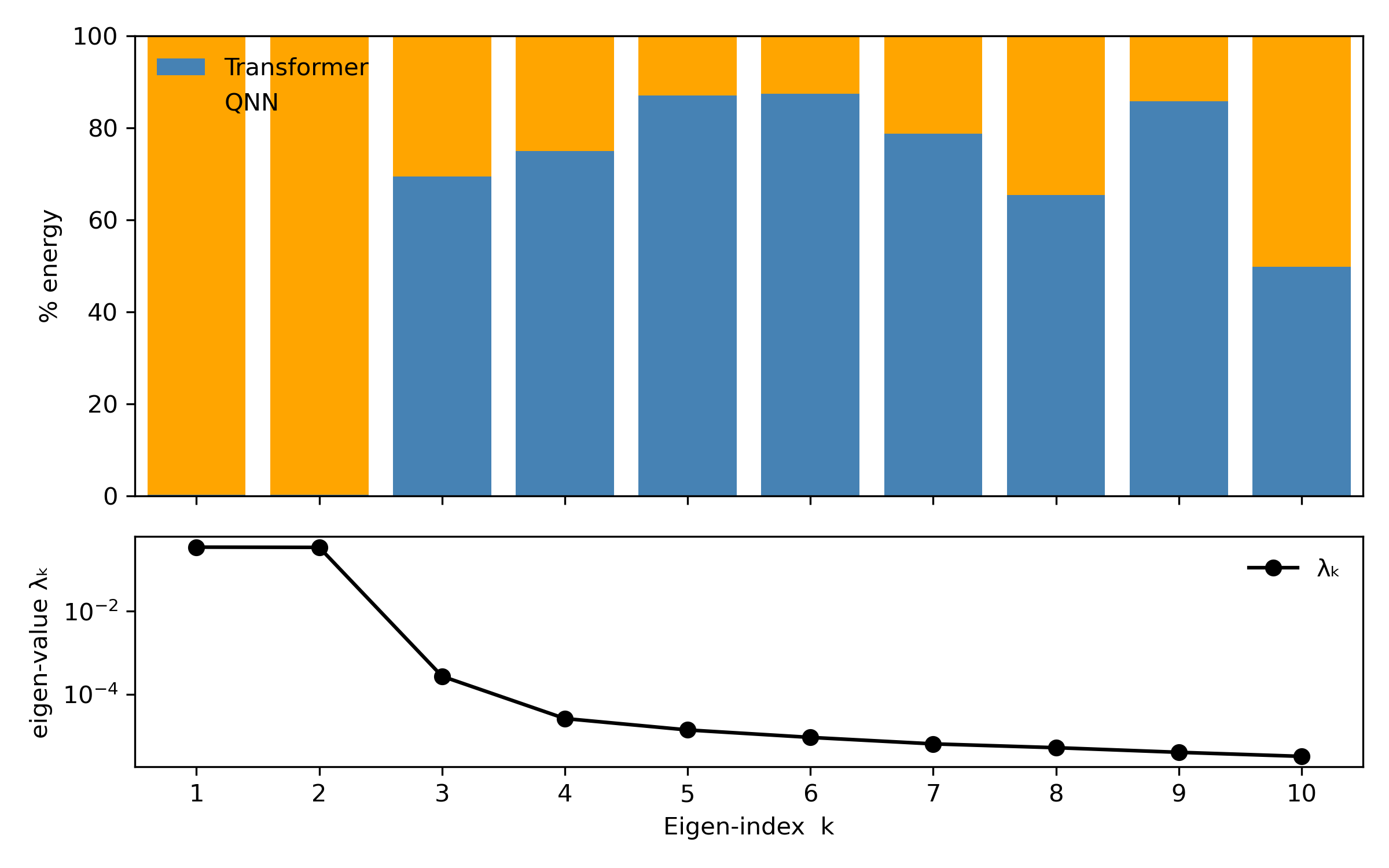}
        \caption{9 qubits}
    \end{subfigure}
    \hfill
    \begin{subfigure}[b]{0.22\textwidth}
        \centering
        \includegraphics[width=\textwidth]{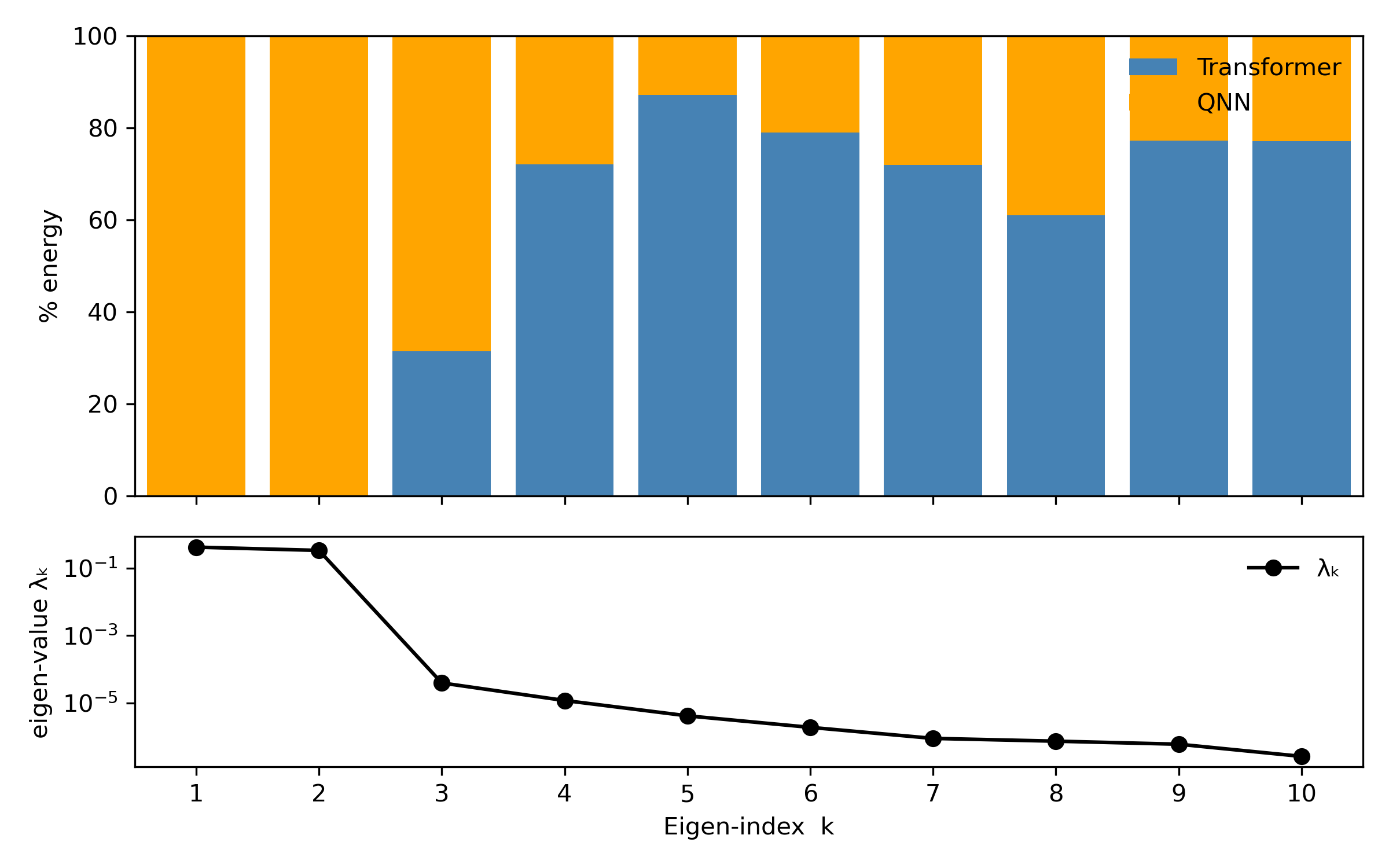}
        \caption{10 qubits}
    \end{subfigure}

    \caption{Above: contribution (energy) of Transformer (blue for Transformer and orange for QNN) for Iris dataset; Below: the ten largest Fisher eigenvalues}
    \label{fig:main1FisherIris}
\end{figure}

\begin{figure}[h!]
    \centering
    \begin{subfigure}[b]{0.22\textwidth}
        \centering
        \includegraphics[width=\textwidth]{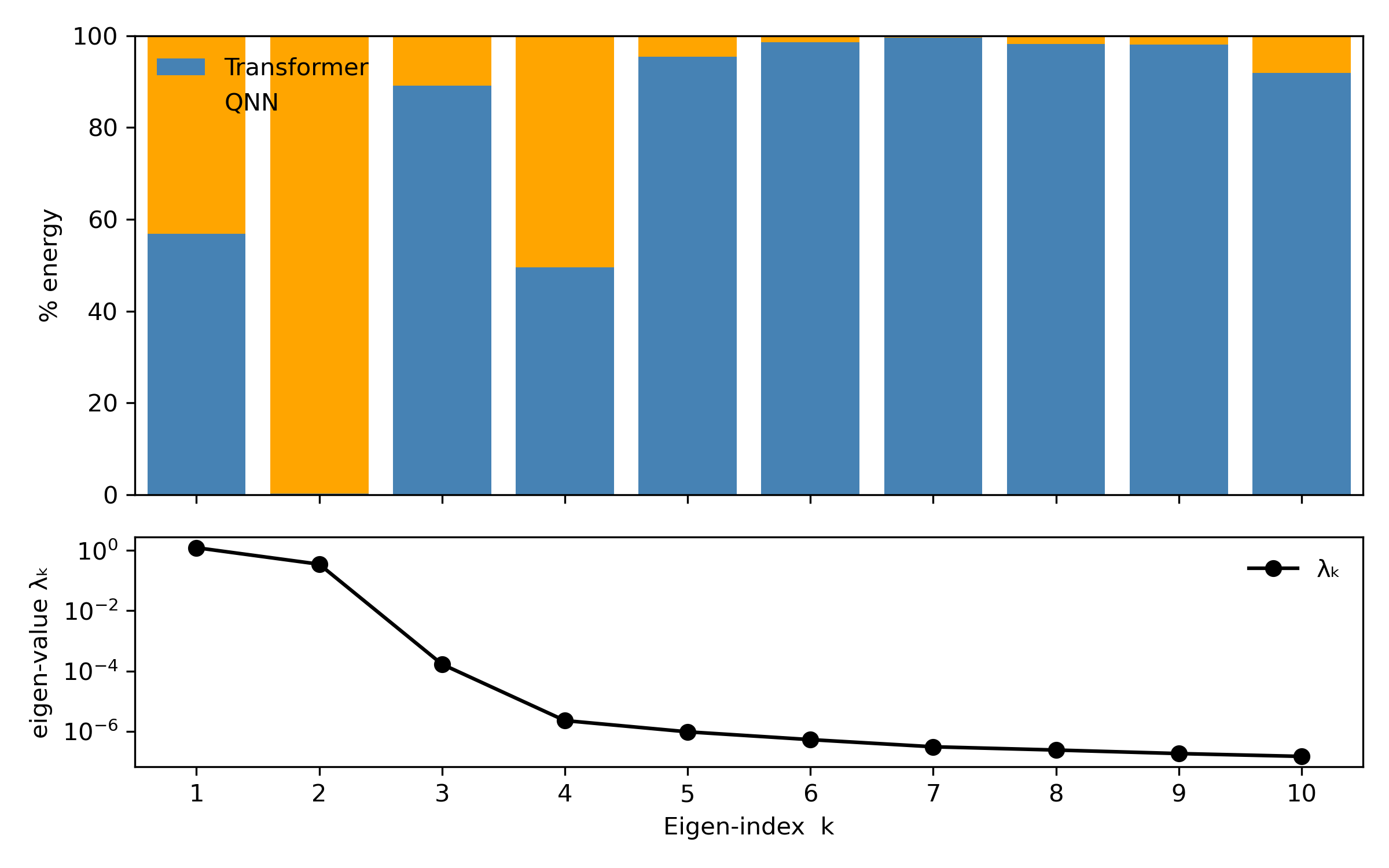}
        \caption{3 qubits}
    \end{subfigure}
    \hfill
    \begin{subfigure}[b]{0.22\textwidth}
        \centering
        \includegraphics[width=\textwidth]{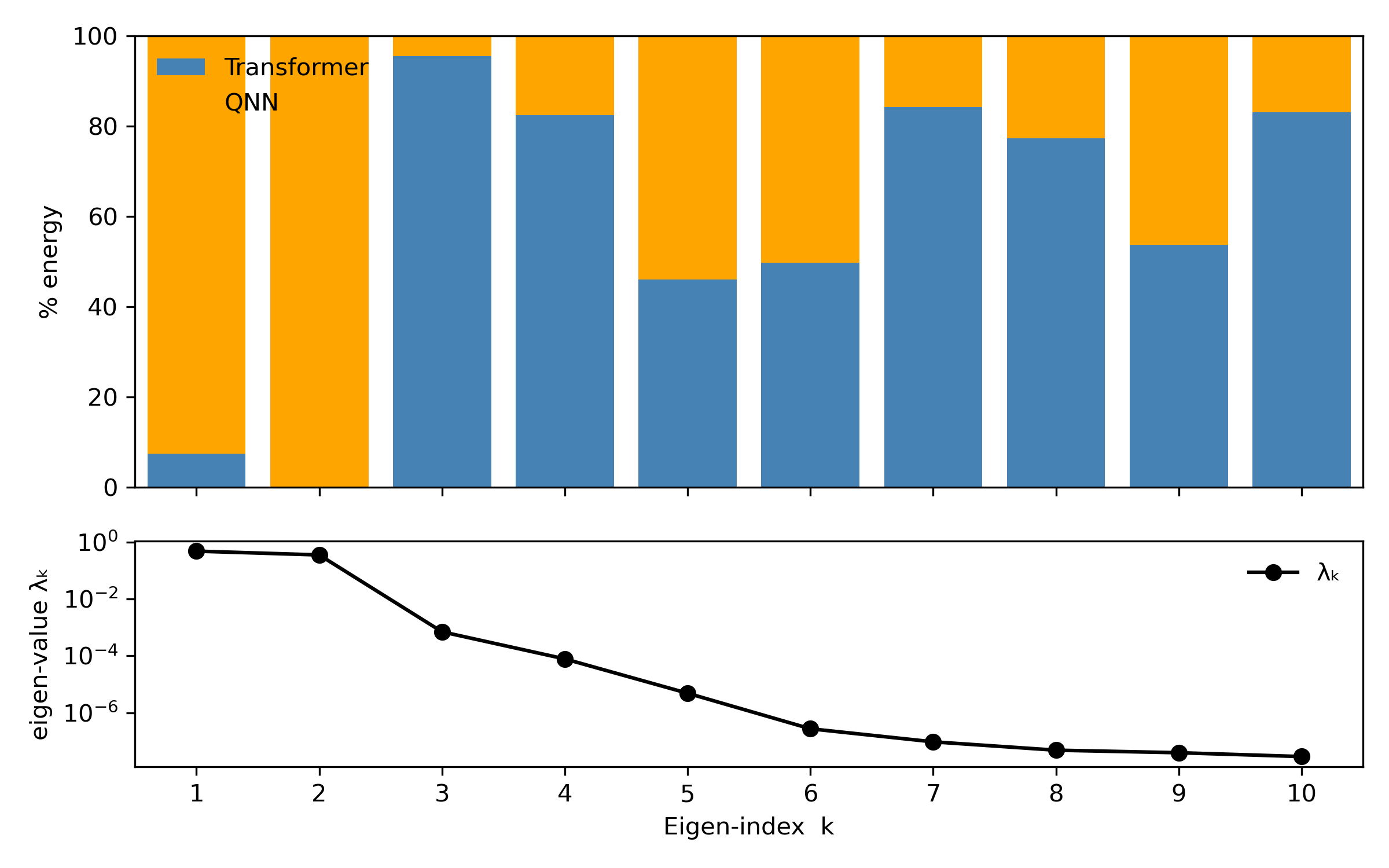}
        \caption{4 qubits}
    \end{subfigure}
    \hfill
    \begin{subfigure}[b]{0.22\textwidth}
        \centering
        \includegraphics[width=\textwidth]{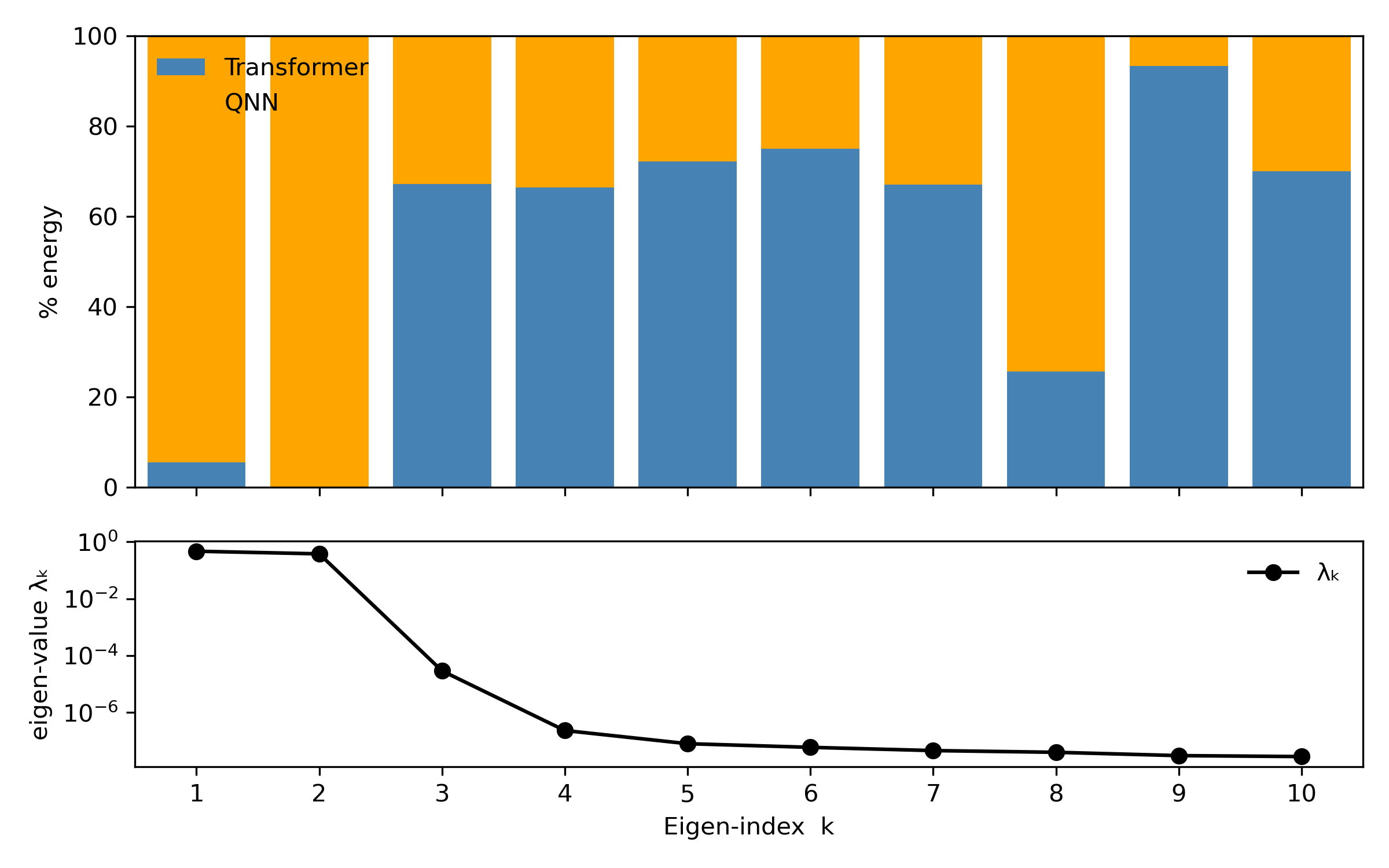}
        \caption{5 qubits}     
    \end{subfigure}
    \hfill
    \begin{subfigure}[b]{0.22\textwidth}
        \centering
        \includegraphics[width=\textwidth]{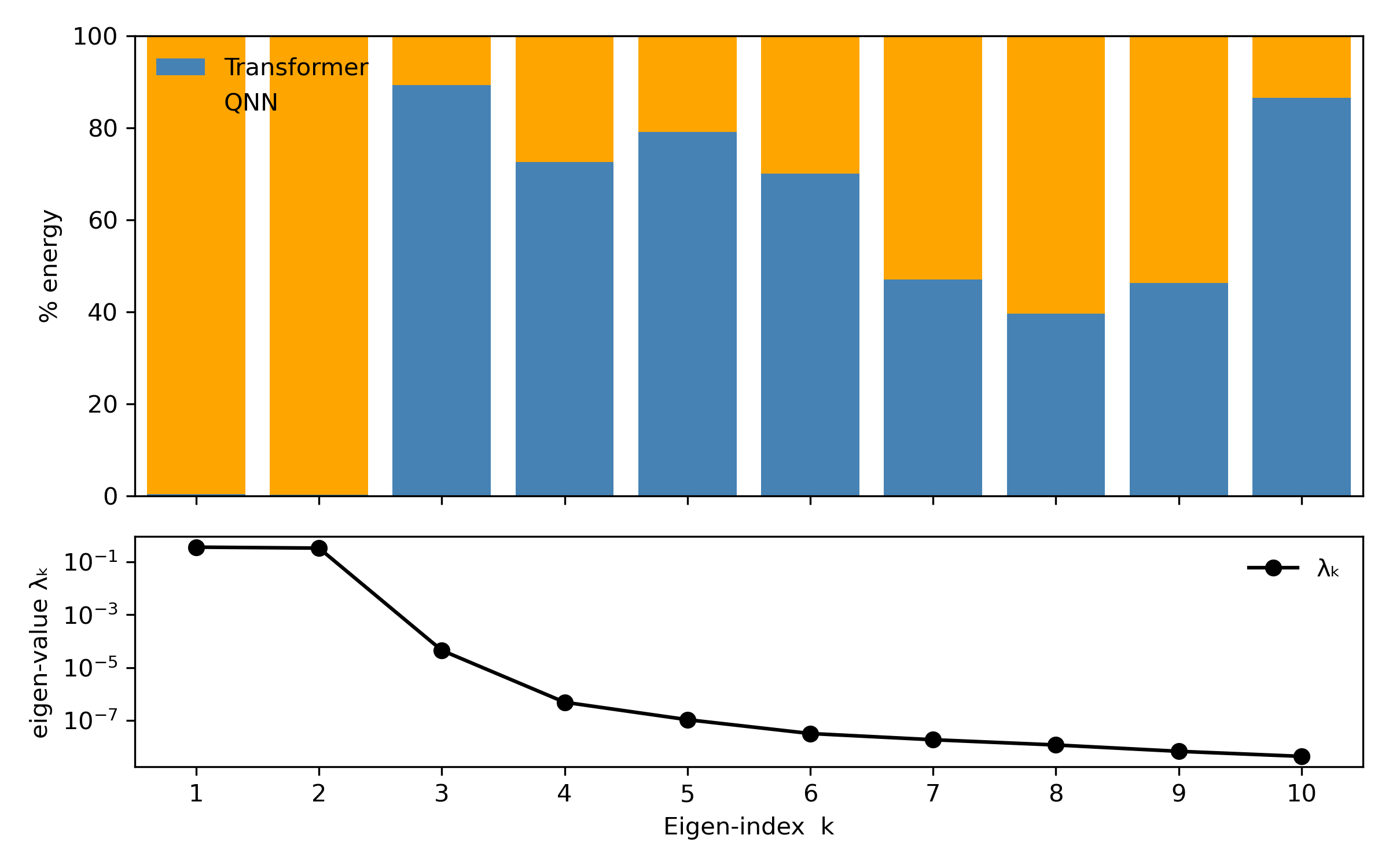}
        \caption{6 qubits} 
    \end{subfigure}

    \vskip\baselineskip

    \begin{subfigure}[b]{0.22\textwidth}
        \centering
        \includegraphics[width=\textwidth]{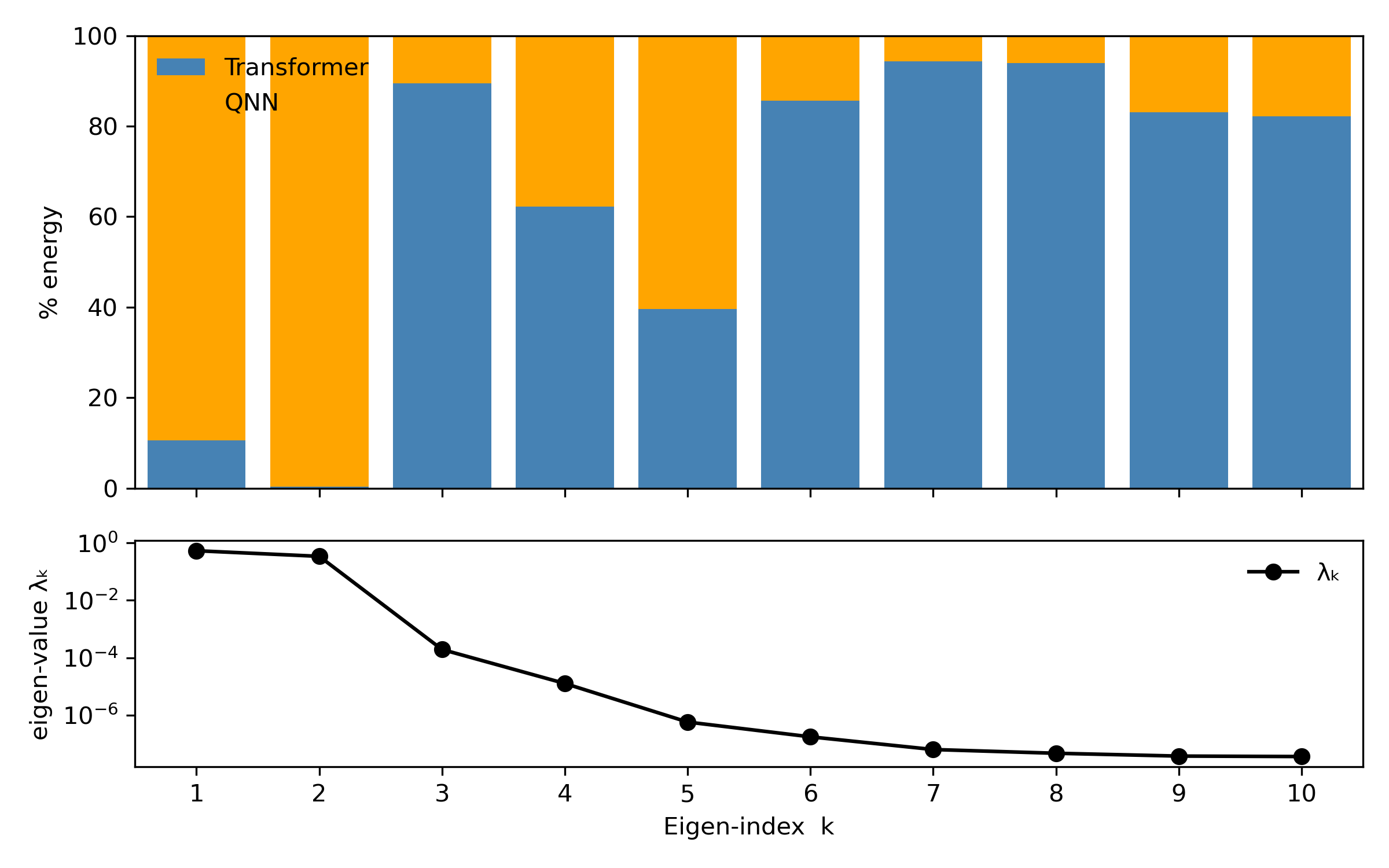}
        \caption{q=7} 
    \end{subfigure}
    \hfill
    \begin{subfigure}[b]{0.22\textwidth}
        \centering
        \includegraphics[width=\textwidth]{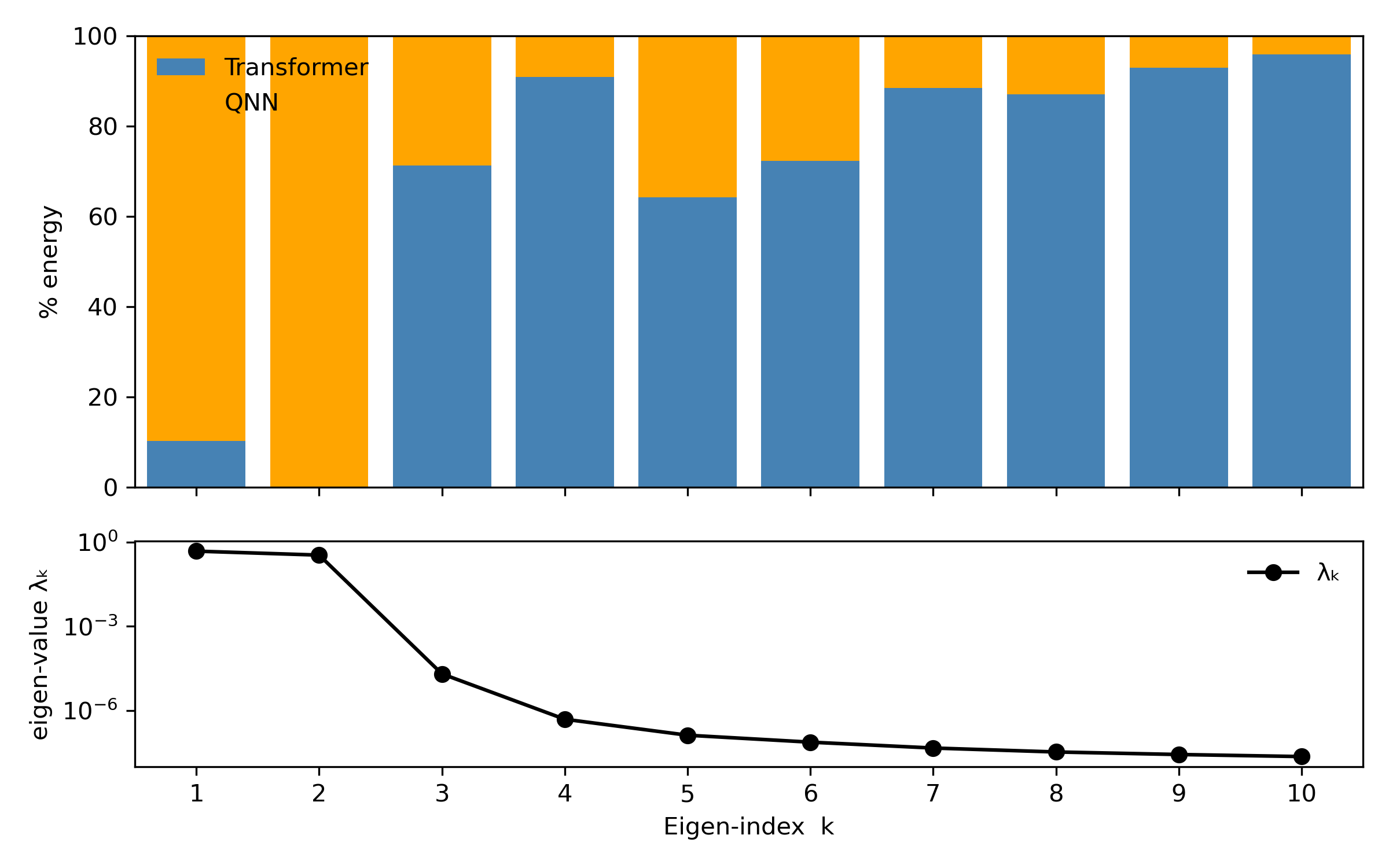}
        \caption{8 qubits}
    \end{subfigure}
    \hfill
    \begin{subfigure}[b]{0.22\textwidth}
        \centering
        \includegraphics[width=\textwidth]{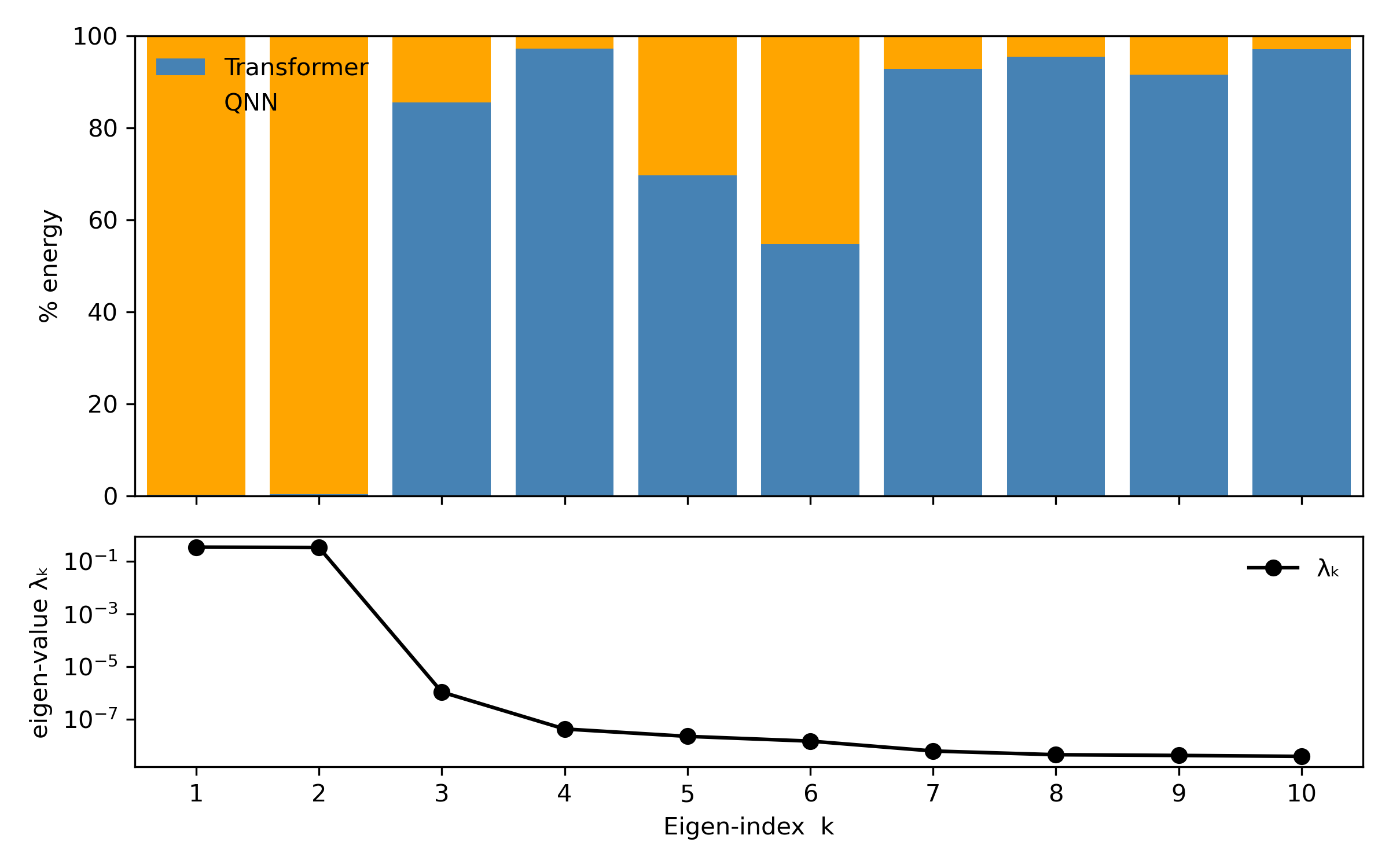}
        \caption{9 qubits}
    \end{subfigure}
    \hfill
    \begin{subfigure}[b]{0.22\textwidth}
        \centering
        \includegraphics[width=\textwidth]{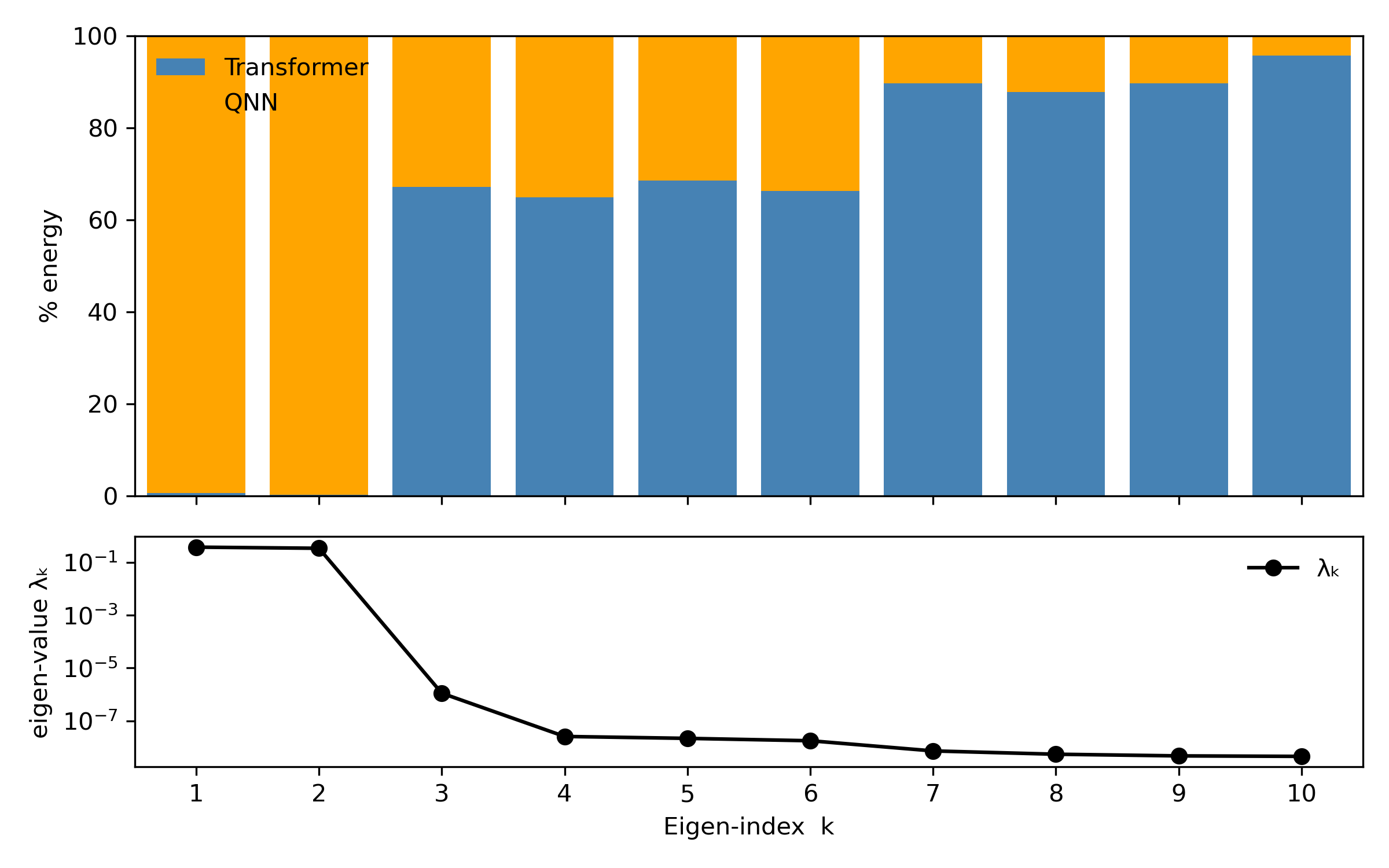}
        \caption{10 qubits}
    \end{subfigure}
    \caption{Above: contribution (energy) of Transformer for MNIST dataset (blue for Transformer and orange for QNN); Below: the 10 largest Fisher eigenvalues.}
    \label{fig:main1FisherMNIST}
\end{figure}

Figures \ref{fig:main1FisherIris} and \ref{fig:main1FisherMNIST} show the eigen-value split for Iris and MNIST datasets. Fisher information for other datasets are similar. We use the maximum sample size as 256 and the ten largest eigenvalues and their parameter-space eigenvectors are returned. The simulations show that the two largest Fisher eigen-values are $\mathcal O(1)$ and all subsequent eigen-values drop by $\sim 2$–$3$ orders of
magnitude. Recall that a barren-plateau regime would force
\(\operatorname{tr}\!\bigl(\E_\theta[F(\theta)]\bigr)\)
to vanish exponentially in the number of qubits~\cite{abbas2021power}.
Our empirical spectrum shows
\[
  \lambda_0 = \mathcal O(1), \qquad
  \sum_{k\le 10}\lambda_k \gg 0,
\]
so TQNN does sit in a barren plateau. As the qubit number increases, the QNN does contribute most of the strongest mode, confirming that it injects meaningful curvature rather than collapsing to a flat region. The hybrid Fisher spectrum indicates that the model is far from a barren plateau, its leading curvature is concentrated in
more QNN direction as the qubit number increases, confirming that additional qubits enlarge the expressivity and that the transformer front-end has already compressed the input features into a lower-dimensional representation.

\section{Conclusion and Discussion}\label{discussion}

We have proposed \emph{Genetic Transformer–Assisted Quantum
Neural Networks} (GTQNNs), a hybrid architecture that (i) employs a transformer
to compress high-dimensional data, (ii) processes the compressed features with a
shallow variational quantum circuit, and (iii) uses NSGA-II to co-optimize
classification accuracy and hardware cost.  
Experiments on Iris, Breast-Cancer Wisconsin, MNIST, and Heart-Disease
demonstrate consistent accuracy gains over state-of-the-art quantum models while
halving the number of entangling gates.

The present study shows that GTQNNs can navigate the accuracy–versus–hardware trade-off
that dominates NISQ–era machine learning.  A transformer encoder performs an
\emph{in-network} dimensionality-reduction step, feeding a compact feature
vector to a shallow variational circuit whose design is refined with NSGA-II.
Across four benchmarks GTQNNs match or exceed the best published QNN
accuracies while significantly cutting quantum gate usage.

A hybrid Fisher–information analysis deepens this picture.
The leading eigenvectors of the Fisher matrix increasingly concentrate on the
\emph{quantum} coordinates as the qubit budget grows, indicating that
(i)~the transformer has already compressed the classical features and
(ii)~the remaining curvature---and therefore representational power---resides
mostly in the QNN subspace.  This clear separation of roles keeps the model
well away from the barren-plateau regime.
  
Our experiments restrict the quantum ansatz to single-qubit $R_{y}$ rotations
plus a fixed pattern of entangling gates.  Employing \emph{richer} gate
families—for example, leveraging
widely used IBM Qiskit constructs such as the ZZFeature map, or
incorporating geometry-aware interaction patterns—could expand the expressive
power of the QNN layer without substantially increasing circuit depth.
Integrating these gate sets into the genetic search therefore constitutes a
promising direction for future work.

Our evaluation focuses on small- to medium-sized data sets; scaling GTQNNs
to ImageNet-scale problems will require further investigation into
transformer parameter sharing and cost-effective quantum circuit design. Finally, integrating more-sophisticated evolutionary operators—e.g.\
grammar-based mutations or learned crossover policies—may uncover circuit
families that generalise better across tasks. Incorporating realistic noise models and hardware connectivity
should sharpen the Pareto front even further.

In summary, GTQNNs offer a practical route toward quantum-enhanced machine
learning on near-term devices by tightly coupling classical feature learning
with quantum-native decision making and by explicitly optimising for the
hardware constraints that dominate today’s quantum landscape.





\section*{Declarations}
{\bf Conflict of interest} The author declares no competing interests.
 
\noindent {\bf Availability of data and materials} The datasets used in paper are publicly available in the Python packages.  The code will be available upon request.


\begin{thebibliography}{128}

\bibitem{abbas2021power}
A.~Abbas, D.~Sutter, C.~Zoufal, A.~Lucchi, A.~Figalli, and S.~Woerner.
\newblock The power of quantum neural networks.
\newblock {\em Nature Computational Science}, 1(6):403--409, 2021.
\newblock \href{https://doi.org/10.1038/s43588-021-00084-1}{DOI:\,10.1038/s43588-021-00084-1}.

\bibitem{alam2022deepqmlp}
M.~Alam and S.~Ghosh.
\newblock DeepQMLP: A scalable quantum–classical hybrid deep neural network architecture for classification.
\newblock In {\em Proc.\ 35th Int.\ Conf.\ VLSI Design \& 21st Int.\ Conf.\ Embedded Systems (VLSID)}, pages 275--280. IEEE, 2022.

\bibitem{chen2024}
H.-Y.~Chen, Y.-J.~Chang, S.-W.~Liao, and C.-R.~Chang.
\newblock Deep Q-learning with hybrid quantum neural network on solving maze problems.
\newblock Quantum Machine Intelligence, 2024, 6:2 
\newblock https://doi.org/10.1007/s42484-023-00137-w

\bibitem{liu2024training}
C.-Y.~Liu, E.~J.~Kuo, C.~H.~A.~Lin \emph{et al.}
\newblock Training classical neural networks by quantum machine learning.
\newblock 2024 IEEE International Conference on Quantum Computing and Engineering (QCE), 2024, Volume: 2, Pages: 34-38
\newblock \href{https://doi.org/10.1109/QCE60285.2024.10248}{DOI 10.1109/QCE60285.2024.10248}
\newblock arXiv:2402.16465, 2024.

\bibitem{maccormack2022branching}
I.~MacCormack, C.~Delaney, A.~Galda, N.~Aggarwal, and P.~Narang.
\newblock Branching quantum convolutional neural networks.
\newblock {\em Physical Review Research}, 4(1):013117, 2022.
\newblock \href{https://doi.org/10.1103/PhysRevResearch.4.013117}{DOI:\,10.1103/PhysRevResearch.4.013117}.

\bibitem{pointing2024simplicity}
J.~Pointing.
\newblock Do quantum neural networks have simplicity bias?
\newblock {\em arXiv preprint} arXiv:2407.03266, 2024.

\bibitem{schuld2021kernel}
M.~Schuld.
\newblock Supervised quantum machine learning models are kernel methods.
\newblock {\em arXiv preprint} arXiv:2101.11020, 2021.

\bibitem{amari1998natural}
S.-I. Amari.
\newblock Natural gradient works efficiently in learning.
\newblock {\em Neural computation}, 10(2):251--276, 1998.
\newblock \texttt{\href{http://dx.doi.org/10.1162/089976698300017746}{DOI:\,10.1162/089976698300017746}}.

\bibitem{kunstner2019limitations}
F.~Kunstner, P.~Hennig, and L.~Balles.
\newblock Limitations of the empirical Fisher approximation for natural-gradient descent.
\newblock In {\em Advances in Neural Information Processing Systems~32}, pages 4156--4167, 2019.

\bibitem{frieden_2004}
B.~R. Frieden.
\newblock {\em Science from Fisher Information: A Unification}.
\newblock Cambridge University Press, 2004.
\newblock \texttt{\href{http://dx.doi.org/10.1017/CBO9780511616907}{DOI:\,10.1017/CBO9780511616907}}.

\bibitem{rissanen1996fisher}
J.~J. {Rissanen}.
\newblock Fisher information and stochastic complexity.
\newblock {\em IEEE Transactions on Information Theory}, 42(1):40--47, 1996.
\newblock \texttt{\href{http://dx.doi.org/10.1109/18.481776}{DOI:\,10.1109/18.481776}}.


\bibitem{Ji2020}
W.~Ji, D.~Liu, Y.~Meng, and Y.~Xue.  
\newblock A review of genetic-based evolutionary algorithms in SVM parameter optimisation.  
\newblock {\em Evolutionary Intelligence}, 14:1389--1414, 2021.  
\newblock \texttt{\href{https://doi.org/10.1007/s12065-020-00439-z}{DOI:\,10.1007/s12065-020-00439-z}}.

\bibitem{Lahoz-Beltra2016}
R.~Lahoz-Beltra.  
\newblock Quantum genetic algorithms for computer scientists.  
\newblock {\em Computers}, 5(4):24, 2016.  
\newblock \texttt{\href{https://doi.org/10.3390/computers5040024}{DOI:\,10.3390/computers5040024}}.

\bibitem{Acampora2021}
G.~Acampora and A.~Vitiello.  
\newblock Implementing evolutionary optimisation on actual quantum processors.  
\newblock {\em Information Sciences}, 575:542--562, 2021.  
\newblock \texttt{\href{https://doi.org/10.1016/j.ins.2021.06.049}{https://doi.org/10.1016/j.ins.2021.06.049}}.

\bibitem{ARG21}
S.~Altares-L\'{o}pez, A.~Ribeiro, and J.~J. Garc\'{\i}a-Ripoll.  
\newblock Automatic design of quantum feature maps.  
\newblock {\em Quantum Science and Technology}, 6(4):045010, 2021.  
\newblock \texttt{\href{https://iopscience.iop.org/article/10.1088/2058-9565/ac1ab1}{https://iopscience.iop.org/article/10.1088/2058-9565/ac1ab1}}.

\bibitem{Chivilikhin2020}
D.~Chivilikhin, A.~Samarin, V.~Ulyantsev, I.~Iorsh, A.~R. Oganov, and O.~Kyriienko.  
\newblock MoG-VQE: Multi-objective genetic variational quantum eigensolver.  
\newblock {\em arXiv preprint} arXiv:2007.04424, 2020.

\bibitem{Deb2002}
K.~Deb, S.~Agrawal, A.~Pratap, and T.~Meyarivan.  
\newblock A fast and elitist multi-objective genetic algorithm: NSGA-II.  
\newblock {\em IEEE Transactions on Evolutionary Computation}, 6(2):182--197, 2002.  
\newblock \texttt{\href{https://doi.org/10.1109/4235.996017}{DOI:\,10.1109/4235.996017}}.

\bibitem{Miettinen1999}
K.~Miettinen.  
\newblock {\em Nonlinear Multiobjective Optimization}.  
\newblock Kluwer Academic Publishers, Boston, 1998.

\bibitem{SDB14}
E.B.\ Alaia, I.H.\ Dridi, H.\ Bouchriha and P.\ Borne,
\newblock {\em Genetic algorithm with pareto front selection for multi-criteria optimization of multi-depots and multi-vehicle pickup and delivery problems with time windows},
\newblock 2014 15th International Conference on Sciences and Techniques of Automatic Control and Computer Engineering, pp. 488-493, (2014).

\bibitem{baran2021}
B.\ Bar{\'{a}}n, A.\ Carballude and M.\ Villagra,
\newblock {\em Neighbor Optimization of N-Dimensional Quantum Circuits}, 
\newblock SN Computer Science 2, 2021 

\bibitem{Cerezo2022}
M.~Cerezo \emph{et al.}
\newblock Challenges and opportunities in quantum machine learning.
\newblock \emph{Nat.\ Comput.\ Sci.} \textbf{2}, 567 (2022).

\bibitem{Beer2020}
K.~Beer \emph{et al.}
\newblock Training deep quantum neural networks.
\newblock \emph{Nat.\ Commun.} \textbf{11}, 808 (2020).

\bibitem{Havlicek2019}
V.~Havlíček \emph{et al.}
\newblock Supervised learning with quantum-enhanced feature spaces.
\newblock \emph{Nature} \textbf{567}, 209–212 (2019).

\bibitem{Garcia2022}
D.~P. García, J.~Cruz-Benito, F.~J. García-Peñalvo.
\newblock Systematic literature review: quantum machine learning and its applications.
\newblock \emph{Computer Science Review}: 51, 100619(2024).

\bibitem{Sim2019}
S.~Sim, P.~D. Johnson, A.~Aspuru-Guzik.
\newblock Expressibility and entangling capability of parameterized quantum circuits.
\newblock \emph{Adv.\ Quantum Technol.} \textbf{2}, 1900070 (2019).

\bibitem{Zhang2023}
S.~X. Zhang \emph{et al.}
\newblock TensorCircuit: a quantum software framework for the NISQ era.
\newblock \emph{Quantum} \textbf{7}, 912 (2023).

\bibitem{Farhi2018}
E.~Farhi, H.~Neven.
\newblock Classification with quantum neural networks on near-term processors.
\newblock \emph{arXiv}:1802.06002 (2018).

\bibitem{Sharma2022}
K.~Sharma \emph{et al.}
\newblock Trainability of dissipative perceptron-based quantum neural networks.
\newblock \emph{Phys.\ Rev.\ Lett.} \textbf{128}, 180505 (2022).

\bibitem{Sagingalieva2023}
A.~Sagingalieva \emph{et al.}
\newblock Hybrid quantum neural network for drug response prediction.
\newblock \emph{Cancers} \textbf{15}, 2705 (2023).

\bibitem{Singh2024}
U.~Singh, A.~Z. Goldberg, K.~Heshami.
\newblock Coherent feed-forward quantum neural network.
\newblock \emph{Quantum Machine Intelligence} \textbf{6}, 89 (2024).

\bibitem{Schuld2021}
M.~Schuld.
\newblock Supervised quantum machine learning models are kernel methods.
\newblock \emph{arXiv}:2101.11020 (2021).

\bibitem{Vaswani2017}
A.~Vaswani \emph{et~al.}
\newblock Attention is all you need.
\newblock In \emph{Advances in Neural Information Processing Systems~30}, pages 5998--6008, 2017.

\bibitem{Devlin2019}
J.~Devlin, M.-W. Chang, K.~Lee, and K.~Toutanova.
\newblock BERT: Pre-training of deep bidirectional transformers for language understanding.
\newblock In \emph{Proceedings of NAACL}, pages 4171--4186, 2019.
\newblock https://doi.org/10.18653/v1/N19-1423

\bibitem{Dosovitskiy2021}
A.~Dosovitskiy \emph{et~al.}
\newblock An image is worth $16\\times16$ words: Transformers for image recognition at scale.
\newblock \emph{International Conference on Learning Representations}, 2021.
\newblock https://arxiv.org/abs/2010.11929

\bibitem{ASUComputing2023}
Jennewein, Douglas M. et al. 
\newblock The Sol Supercomputer at Arizona State University.
\newblock PEARC '23: Practice and Experience in Advanced Research Computing 2023: Computing for the Common Good
Pages 296 - 301, 2023.
\newblock https://doi.org/10.1145/3569951.3597573

\bibitem{Wang2025}
H.~Wang 
\newblock Several fitness functions and entanglement gates in quantum kernel generation.
\newblock \emph{Quantum Mach. Intell.}, 7, 7 (2025).
\newblock https://doi.org/10.1007/s42484-024-00233-5

\bibitem{Wang2025-1}
H.~Wang 
\newblock A novel feature selection method based on quantum support vector machine.
\newblock \emph{Physica Scripta}, 99 056006 (2024).  , Volume 99, Number 5
\newblock https://doi.org/DOI 10.1088/1402-4896/ad36ef

\bibitem{jMetalPy}
Antonio Benítez-Hidalgo, Antonio J. Nebro, José García-Nieto, Izaskun Oregi, Javier Del Ser,
\newblock jMetalPy: A Python framework for multi-objective optimization with metaheuristics,
\newblock \emph{Swarm and Evolutionary Computation}, 51, 100598, 2019
\newblock https://doi.org/10.1016/j.swevo.2019.100598

\bibitem{Schuld}
M.\ Schuld and F.\ Petruccione.
\newblock {\em Machine Learning with Quantum Computers}.
\newblock Springer International Publishing, Cham, 2021.
\newblock \texttt{\href{https://link.springer.com/book/10.1007/978-3-030-83098-4}{https://link.springer.com/book/10.1007/978-3-030-83098-4}}.

\bibitem{NielsonChuang}
M.\,A.\ Nielsen and I.\,L.\ Chuang.
\newblock {\em Quantum Computation and Quantum Information} (10th-anniv.\ ed.).
\newblock Cambridge University Press, Cambridge, 2011.

\bibitem{qml1}
J.\ Biamonte, P.\ Wittek, N.\ Pancotti, P.\ Rebentrost, N.\ Wiebe, and S.\ Lloyd.
\newblock Quantum machine learning.
\newblock {\em Nature}, 549(7671):195–202, 2017.
\newblock \texttt{\href{https://doi.org/10.1038/nature23474}{DOI:\,10.1038/nature23474}}.

\bibitem{qml2}
M.\ Schuld, I.\ Sinayskiy, and F.\ Petruccione.
\newblock An introduction to quantum machine learning.
\newblock {\em Contemporary Physics}, 56(2):172–185, 2015.
\newblock \texttt{\href{https://doi.org/10.1080/00107514.2014.964942}{DOI:\,10.1080/00107514.2014.964942}}.

\bibitem{qml16}
V.\ Havlíček, A.\,D. Córcoles, K.\ Temme \emph{et al.}
\newblock Supervised learning with quantum-enhanced feature spaces.
\newblock {\em Nature}, 567(7747):209–212, 2019.
\newblock \texttt{\href{https://doi.org/10.1038/s41586-019-0980-2}{DOI:\,10.1038/s41586-019-0980-2}}.

\bibitem{qml15}
D.\,P. García, J.\ Cruz-Benito, and F.\,J. García-Peñalvo.
\newblock Systematic literature review: Quantum machine learning and its applications.
\newblock {\em arXiv:2201.04093}, 2022.
\newblock \texttt{\href{https://doi.org/10.48550/arXiv.2201.04093}{DOI:\,10.48550/arXiv.2201.04093}}.

\bibitem{qml6}
D.\,N. Diep.
\newblock Some quantum neural networks.
\newblock {\em International Journal of Theoretical Physics}, 59:1179–1190, 2020.
\newblock \texttt{\href{https://doi.org/10.1007/s10773-020-04397-1}{DOI:\,10.1007/s10773-020-04397-1}}.

\bibitem{qml9}
A.\ Chalumuri, R.\ Kune, and B.\,S. Manoj.
\newblock A hybrid classical-quantum approach for multi-class classification.
\newblock {\em Quantum Information Processing}, 20:119, 2021.
\newblock \texttt{\href{https://doi.org/10.1007/s11128-021-03029-9}{DOI:\,10.1007/s11128-021-03029-9}}.

\bibitem{qml11}
B.\,Q. Chen and X.\,F. Niu.
\newblock Quantum Neural Network with Improved Quantum Learning Algorithm.
\newblock {\em International Journal of Theoretical Physics}, 59:1978–1990, 2020.
\newblock \texttt{\href{https://doi.org/10.1007/s10773-020-04470-9}{DOI:\,10.1007/s10773-020-04470-9}}.

\bibitem{qml12}
F.\ Tacchino, S.\ Mangini, P.\,K. Barkoutsos \emph{et al.}
\newblock Variational Learning for Quantum Artificial Neural Networks.
\newblock {\em IEEE Transactions on Quantum Engineering}, 2:1–10, 2021.
\newblock \texttt{\href{https://doi.org/10.1109/TQE.2021.3062494}{DOI:\,10.1109/TQE.2021.3062494}}.

\bibitem{qml13}
J.\ Wang, Y.\ Chen, R.\ Chakraborty, and S.\,X. Yu.
\newblock Quantum gradient descent algorithms.
\newblock {\em arXiv:1911.12207}, 2019.
\newblock \texttt{\href{https://doi.org/10.48550/arXiv.1911.12207}{DOI:\,10.48550/arXiv.1911.12207}}.

\bibitem{qml14}
Y.\ Li, R.\,G. Zhou, R.\ Xu, J.\ Luo, and W.\ Hu.
\newblock A quantum deep convolutional neural network for image recognition.
\newblock {\em Quantum Science and Technology}, 5:044003, 2020.
\newblock \texttt{\href{https://doi.org/10.1088/2058-9565/ab9f93}{DOI:\,10.1088/2058-9565/ab9f93}}.

\bibitem{qml10}
S.\,L. Wu \emph{et al.}
\newblock Application of quantum ensemble learning to particle-physics analysis at the LHC.
\newblock {\em Physical Review Research}, 3:033221, 2021.
\newblock \texttt{\href{https://doi.org/10.1103/PhysRevResearch.3.033221}{DOI:\,10.1103/PhysRevResearch.3.033221}}.

\bibitem{Wan2017Sep}
K.\,H. Wan, O.\ Dahlsten, H.\ Kristjánsson, R.\ Gardner, and M.\,S. Kim.
\newblock Quantum generalisation of feed-forward neural networks.
\newblock {\em npj Quantum Information}, 3:36, 2017.
\newblock \texttt{\href{https://doi.org/10.1038/s41534-017-0032-4}{DOI:\,10.1038/s41534-017-0032-4}}.

\bibitem{Beer2020Feb}
K.\ Beer, D.\ Bondarenko, T.\ Farrelly \emph{et al.}
\newblock Training deep quantum neural networks.
\newblock {\em Nature Communications}, 11:808, 2020.
\newblock \texttt{\href{https://doi.org/10.1038/s41467-020-14454-2}{DOI:\,10.1038/s41467-020-14454-2}}.

\bibitem{Cong2019Dec}
I.\ Cong, S.\ Choi, and M.\,D. Lukin.
\newblock Quantum convolutional neural networks.
\newblock {\em Nature Physics}, 15:1273–1278, 2019.
\newblock \texttt{\href{https://doi.org/10.1038/s41567-019-0648-8}{DOI:\,10.1038/s41567-019-0648-8}}.

\bibitem{Sharma2022May}
K.\ Sharma, M.\ Cerezo, L.\ Cincio, and P.\,J. Coles.
\newblock Trainability of dissipative perceptron-based quantum neural networks.
\newblock {\em Physical Review Letters}, 128:180505, 2022.
\newblock \texttt{\href{https://doi.org/10.1103/PhysRevLett.128.180505}{DOI:\,10.1103/PhysRevLett.128.180505}}.

\bibitem{Zhou2023May}
M.\,G. Zhou, Z.\,P. Liu, H.\,L. Yin \emph{et al.}
\newblock Quantum Neural Network for Quantum Neural Computing.
\newblock {\em Research}, 6:0134, 2023.
\newblock \texttt{\href{https://doi.org/10.34133/research.0134}{DOI:\,10.34133/research.0134}}.

\bibitem{Preskill2018Aug}
J.\ Preskill.
\newblock Quantum computing in the NISQ era and beyond.
\newblock {\em Quantum}, 2:79, 2018.
\newblock \texttt{\href{https://doi.org/10.22331/q-2018-08-06-79}{DOI:\,10.22331/q-2018-08-06-79}}.

\bibitem{Cherrat2024}
E.\,A. Cherrat, I. Kerenidis, N. Mathur, J. Landman, M. Strahm, and Y.\ Y. Li.
\newblock Quantum vision transformers.
\newblock {\em Quantum}, 8:1265, 2024.
\newblock \texttt{\href{https://doi.org/10.22331/q-2024-02-22-1265}{DOI:\,10.22331/q-2024-02-22-1265}}.

\bibitem{Ma2024}
H.\ Ma, H.\ Shang, and J.\ Yang.
\newblock Quantum embedding method with transformer neural network quantum states for strongly correlated materials.
\newblock {\em npj Computational Materials}, 10:220, 2024.
\newblock \texttt{\href{https://doi.org/10.1038/s41524-024-01406-3}{DOI:\,10.1038/s41524-024-01406-3}}.

\bibitem{Zhang2025arxiv}
H.\ Zhang and Q.\ Zhao.
\newblock A survey of quantum transformers: Technical approaches, challenges and outlooks.
\newblock {\em arXiv:2504.03192}, 2025.
\newblock \texttt{\href{https://arxiv.org/abs/2504.03192}{arXiv:2504.03192}}.

\bibitem{Li2025quantumS}
Li, G., Zhao, X. , Wang, X. 
\newblock Quantum self-attention neural networks for text classification. 
\newblock Sci. China Inf. Sci. 67, 142501 (2024). 
\newblock https://doi.org/10.1007/s11432-023-3879-7


\bibitem{Baevski2020}
A.~Baevski, Y.~Zhou, A.~Mohamed, and M.~Auli.
\newblock wav2vec~2.0: A framework for self-supervised learning of speech representations.
\newblock In \emph{Advances in Neural Information Processing Systems~33}, pages 12449--12460, 2020.
\newblock https://arxiv.org/abs/2006.11477

\bibitem{ADB14}
    {E.B.\ Alaia, I.H.\ Dridi, H.\ Bouchriha and P.\ Borne},
\newblock  {\em Genetic algorithm with pareto front selection for multi-criteria optimization of
    multi-depots and multi- vehicle pickup and delivery problems with time windows},
\newblock 2014 15th International Conference on Sciences and Techniques of Automatic Control
    and Computer Engineering, pp. 488-493, (2014).

\end{thebibliography}
\end{document}